\documentclass[10pt,journal,compsoc]{IEEEtran}
\ifCLASSOPTIONcompsoc
  \usepackage[nocompress]{cite}
\else
  \usepackage{cite}
\fi
\ifCLASSINFOpdf
\else
\fi
\usepackage[cmex10]{amsmath}
\usepackage{amssymb}
\usepackage[linesnumbered,noline,boxed,commentsnumbered, ruled]{algorithm2e}
\usepackage{float}
\usepackage{graphicx}
\usepackage{subfig}
\usepackage{booktabs}
\usepackage{flushend}
\usepackage{multirow}
\usepackage{mathrsfs}
\usepackage{lscape}
\usepackage{bm}
\usepackage{tikz,pgf}
\usepackage{doi}
\usepackage{amsmath}
\usetikzlibrary{arrows,shapes,positioning,shadows,trees}

\begin{document}
%
\title{Three-Dimensional Mesh Steganography \\and Steganalysis: A Review}
%
%
%
%

\author{Hang~Zhou,
        Weiming~Zhang,
        Kejiang~Chen,
        Weixiang~Li,
        and~Nenghai~Yu
\IEEEcompsocitemizethanks{\IEEEcompsocthanksitem This work was supported in part by the Natural Science Foundation of China under Grant U1636201, 62002334 and 62072421, by Anhui Science Foundation of China under Grant 2008085QF296 and by the Fundamental Research Funds for the Central Universities under Grant WK2100000018.

\IEEEcompsocthanksitem H. Zhou is with the School of Computing Science, Simon Fraser University, Burnaby, Canada (Email: zhouhang2991@gmail.com).
\IEEEcompsocthanksitem W. Zhang, K. Chen, W. Li and N. Yu are with the CAS Key Laboratory of Electromagnetic Space Information, University of Science and Technology of China, Hefei, 230026, China (E-mail: zhangwm@ustc.edu.cn, chenkj@ustc.edu.cn, wxli6049@mail.ustc.edu.cn, ynh@ustc.edu.cn).
\IEEEcompsocthanksitem Corresponding authors: Weiming Zhang and Kejiang Chen.}}

\IEEEtitleabstractindextext{%
\begin{abstract}
Three-dimensional (3-D) meshes are commonly used to represent virtual surfaces and volumes. Over the past decade, 3-D meshes have emerged in industrial, medical, and entertainment applications, being of large practical significance for 3-D mesh steganography and steganalysis. In this article, we provide a systematic survey of the literature on 3-D mesh steganography and steganalysis. Compared with an earlier survey~\cite{girdhar2017comprehensive}, we propose a new taxonomy of steganographic algorithms with four categories: 1) two-state domain, 2) LSB domain, 3) permutation domain, and 4) transform domain. Regarding steganalysis algorithms, we divide them into two categories: 1) universal steganalysis and 2) specific steganalysis. For each category, the history of technical developments and the current technological level are introduced and discussed. Finally, we highlight some promising future research directions and challenges in improving the performance of 3-D mesh steganography and steganalysis.
\end{abstract}

\begin{IEEEkeywords}
3-D, polygonal mesh, information hiding, steganography, steganalysis, survey, review
\end{IEEEkeywords}}

\maketitle

\IEEEdisplaynontitleabstractindextext

%
\IEEEpeerreviewmaketitle

\IEEEraisesectionheading{\section{Introduction}\label{sec:introduction}}
\subsection{Motivation}

%
%
%
%

\IEEEPARstart{I}{n} this paper, we report a systematic review of papers on 3-D mesh steganography and steganalysis published in conferences and journals related to computer graphics and security.
We provide objective reports on the types of steganographic and steganalytic methods encountered in the literature. At the same time, we also quantitatively evaluate these papers from the perspective of security evaluations.
Our goal in this work is to understand the assessment procedure in the 3-D mesh steganography and steganalysis methods as a whole.
The significance of this research has been well recognized by the growing body of work on how to improve the anti-steganalysis ability on the steganographer side and how to improve the steganalysis ability on the steganalyzer side. In this paper, we contribute to previous work by providing some standard evaluation metrics, an overall summary and an understanding of related papers that have not been subject to this kind of systematic assessment in the past.


Different from image steganography that embeds data by modifying pixel values, 3-D mesh steganography modifies vertex coordinates or vertex order to embed data. 
In the latest literature analysis of 3-D steganography and steganalysis by Girdhar and Kumar~\cite{girdhar2017comprehensive}, steganography is divided into three categories (geometrical domain, topological domain and representation domain), the robustness of the algorithms against attacks is examined, and steganalysis is briefly introduced. Their paper is an important contribution but is not a comprehensive survey nor reflects the entire 3-D steganography and steganalysis community. For example, the geometrical domain can still be split into the two-state domain and LSB
 domain. Moreover, the concepts of ``steganography'' and ``watermarking'' are used interchangeably. Watermarking pursues robustness and is used to protect copyright ownership and reduce the counterfeiting of digital multimedia, while steganography pursues undetectability and is used for covert communication. 
They focus mainly on analyzing the robustness of the existing methods, while the undetectability of steganography is a more important property, resulting from the requirement of its real application: covert communication. 
Compared with \cite{girdhar2017comprehensive}, we have included the missing related literature in our survey.

In this paper, we strive to provide readers with a more comprehensive survey, a clear taxonomy and several standard evaluation metrics in terms of both robustness and undetectability. 
From the perspective of reversible or not, we classify data hiding as either reversible data hiding or steganography, and regarding the structure of 3-D data, we mainly consider the 3-D mesh and RGBD image. Here, we consider only 3-D meshes as carriers and steganographic techniques. We also group the steganographic techniques into several domains (two-state domain, LSB domain, permutation domain and transform domain) but in a subdivided manner, excluding those with small embedding capacities. Addtionally, we divide 3-D mesh steganalysis into two aspects (universal steganalysis and specific steganalysis). Our analysis of the existing methods reveals massive weaknesses and strengths, from which we can learn lessons for future work. Therefore, we not only describe the current methods but also show how to improve them.

In summary, the contributions of our paper are fourfold:
\begin{itemize}
\item We objectively report the current evaluation metrics used in the 3-D mesh steganography and steganalysis community.
\item We introduce in detail the embedding capacity, computational complexity and security analysis of steganographic methods.
\item We introduce in detail the complexity and security analysis of steganalytic methods and explore the effectiveness of submodels and the generalizability of trained models.
\item We summarize the current challenges in this field and propose possible directions for researchers to address them in future work.
\end{itemize}

\subsection{History}

The word steganography comes from New Latin \textit{steganographia} and combines the Greek stegan\'os ($\sigma\tau\varepsilon\gamma\alpha\upsilon\acute{o}\varsigma$), which means ``covered or concealed'', and -graphia ($\gamma\rho\alpha\varphi\acute{\eta}$), which means ``writing''~\cite{steganography}. The first recorded use of the term was in 1499 by Johannes Trithemius in his work, disguised as a magic book, \textit{Steganographia}, a monograph on cryptography and steganography.

Generally, the technique of hiding secret messages in innocuous objects (called covers) dates back to ancient times~\cite{johnson1998exploring}. For example, hidden messages may be displayed in invisible ink between visible lines of private letters. Some implementations of steganography that lack shared secrets are a form of security achieved through obscurity, and steganographic schemes that rely on keys follow Kerckhoffs's principle~\cite{fridrich2004searching}. The advantage of steganography over cryptography is that the expected secret message will not attract attention because of scrutiny. Plainly visible encrypted messages, no matter how indestructible they are, will arouse interest~\cite{steganography2}.

More formally, the goal of steganography is to allow communication between a sender and receiver through secret communication channels, in which Alice embeds messages in an innocuous-looking cover object with a specified steganographic method, making it impossible for potential eavesdroppers to detect their presence.
Recently, according to some news reports, before the 9/11 attack, Al-Qaeda and other terrorist organizations used steganography for covert communication~\cite{terror,terror2,terror3}.
Thus far, there have been many research papers on digital image steganography, such as~\cite{hussain2018image, cheddad2010digital,chandramouli2003image} and books such as~\cite{johnson2001information,wayner2009disappearing,fridrich2009steganography,cox2007digital}.

Steganography and steganalysis are counterpart problems that usually occur in pairs and have attracted worldwide attention.
Steganalysis, from the opponent's perspective, aims to detect the existence of confidential data hidden in digital media. Its primary demand is to accurately decide whether confidential data are hidden in the test object or not. More precisely, it is possible to even determine the steganographic method type, estimate the message length, and extract secret data. 
Research on image steganalysis includes survey papers~\cite{luo2008review,li2011survey} and books~\cite{bhme2010advanced,schaathun2012machine}.

While the early steganography and steganalysis methods mostly address images, audio files or videos, the usage of 3-D geometry as the host object has attracted people's attention over the past several years. The price of 3-D hardware is lower than ever before, which has stimulated the widespread usage of 3-D meshes, from the CAM/CAD industry to real-world end-user applications such as virtual reality (VR), web integration, Facebook support, video games, 3-D printing, and animated movies~\cite{frank2018digital}. Therefore, the development of computer graphics has accelerated the production, usage and distribution of 3-D geometric figures, the newest generation of digital media. Moreover, the flexible data structure of 3-D geometry may provide ample room for hosting secret information, making it very suitable as a cover object for steganography. The development of 3-D techniques has facilitated the rapid development of 3-D-related applications. Investigations of 3-D mesh watermarking methods are available in the form of research papers~\cite{wang2008comprehensive,alface20073d} and as book chapters~\cite{nematollahi2017digital}. In addition, a survey on 3-D mesh steganography was initially conducted by Girdhar and Kumar~\cite{girdhar2017comprehensive}.

\subsection{3-D Mesh}
\captionsetup{font={footnotesize}}
\begin{figure}
\begin{center}
  \includegraphics[height=2.40in]{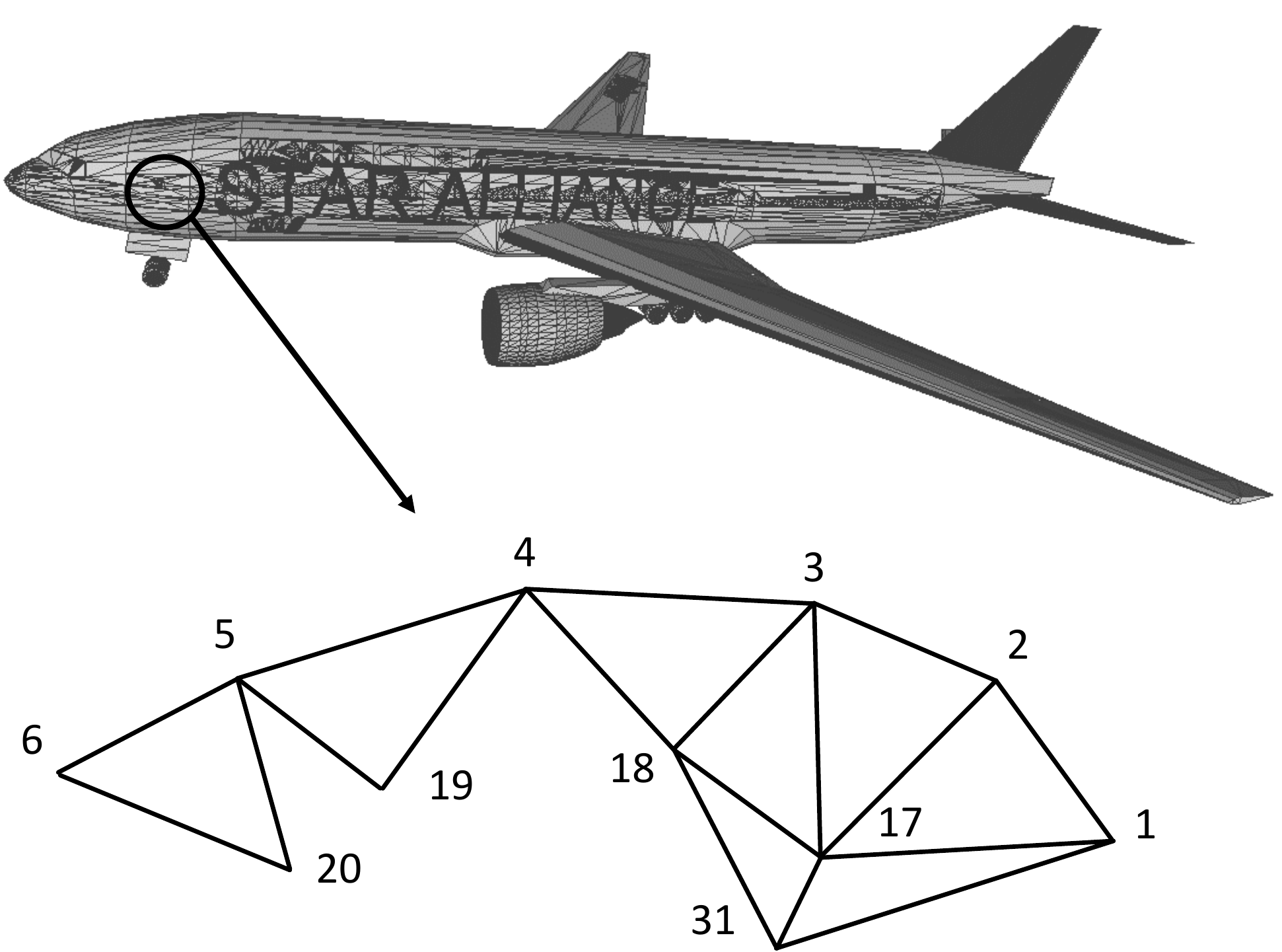}
  \caption{An example of a 3-D mesh (from the Princeton ModelNet dataset). }
  \label{fig:01}
\end{center}
\end{figure}

A mesh is a set of polygonal faces, and the goal is to form an approximation of a real 3-D object. A polygonal mesh has three different combined elements: \emph{vertices}, \emph{edges}, and \emph{faces}; a mesh can also be taken as the combination of \emph{geometry} \emph{connectivity}, where the \emph{geometry} provides the 3-D positions of all its vertices, and \emph{connectivity}, which provides the information hidden between different adjacent vertices. Mathematically, a 3-D polygonal mesh $\mathcal{M}$ containing $V$ vertices and $F$ faces can be formed as a set $\mathcal{M}=\{\mathcal{V},\mathcal{F}\}$, where
\begin{equation}
\label{eqn:01}
\begin{aligned}
\mathcal{V}&=\{v_i\}_{i=1,2,...,V},\\
\mathcal{F}&=\{f_i\}_{i=1,2,...,F},\ f_i\in\mathcal{V}\times\mathcal{V}\times\mathcal{V}\\
\end{aligned}
\end{equation}
in the Cartesian coordinate system and $\mathcal{F}$ is the face set.
In addition, the edge set $\mathcal{E}$ is defined as
\begin{equation}
\mathcal{E}=\{e_i\}_{i=1,2,...,E},\ e_i\in\mathcal{V}\times\mathcal{V}.
\end{equation}

\captionsetup[table]{labelsep=space}
\begin{table}[htbp]
\centering\caption{\label{tab:01}\\
 Indexed face-set data structure for triangle meshes (Figure~\ref{fig:01}). }
 \centering
 \newcommand{\tabincell}[2]{\begin{tabular}{@{}#1@{}}#2\end{tabular}}
   \begin{tabular}{ccccc|cc}
    \toprule
    & \multicolumn{4}{c}{Vertex list} & \multicolumn{2}{c}{Face list}\\
    \cline{1-7}
     & \tabincell{c}{Index of\\vertex} & \tabincell{c}{\textit{x}-axis} & \tabincell{c}{\textit{y}-axis} & \tabincell{c}{\textit{z}-axis} & \tabincell{c}{Index of\\face} & \tabincell{c}{Elements in\\each face} \\
    \midrule
    & 1 & $x_{1}$ & $y_{1}$ & $z_{1}$ & 1 & (17, 1, 2) \\
    & 2 & $x_{2}$ & $y_{2}$ & $z_{2}$ & 2 & (3, 2, 17) \\
    & 3 & $x_{3}$ & $y_{3}$ & $z_{3}$ & 3 & (4, 3, 18) \\
    & 4 & $x_{4}$ & $y_{4}$ & $z_{4}$ & 4 & (5, 4, 19) \\
    & 5 & $x_{5}$ & $y_{5}$ & $z_{5}$ & 5 & (6, 5, 20) \\
    & 6 & $x_{6}$ & $y_{6}$ & $z_{6}$ & ... & ... \\
    & ... & ... & ... & ... & 16 & (31, 17, 1) \\
    & 17 & $x_{17}$ & $y_{17}$ & $z_{17}$ & 17 & (18, 17, 31) \\
    & 18 & $x_{18}$ & $y_{18}$ & $z_{18}$ & ... & ... \\
    & 19 & $x_{19}$ & $y_{19}$ & $z_{19}$ & 241 & (17, 18, 3) \\
    & 20 & $x_{20}$ & $y_{20}$ & $z_{20}$ & ... & ... \\
    & ... & ... & ... & ... & ... & ... \\
    & 31 & $x_{31}$ & $y_{31}$ & $z_{31}$ & ... & ... \\
    & ... & ... & ... & ... & ... & ... \\
    \bottomrule
  \end{tabular}
\end{table}

The geometric embedding of a triangular mesh into $\mathbb{R}^{3}$ is determined by connecting the 3-D position $\mathbf{p}_i$ to each vertex $v_i\in \mathcal{V}$~\cite{botsch2010polygon}:
\begin{equation}
\begin{aligned}
\mathcal{P}&=\left\{\mathbf{p}_{1}, \ldots, \mathbf{p}_{V}\right\}, \\ \mathbf{p}_{i}:&=\mathbf{p}\left(v_{i}\right)=[x\left(v_{i}\right),y\left(v_{i}\right),z\left(v_{i}\right)]^T \in \mathbb{R}^{3}.
\end{aligned}
\end{equation}

The list of mesh faces is often determined by some algorithms aiming to facilitate the speed of geometric and topological operations on a given mesh.
Fig.~\ref{fig:01} shows an example of a 3-D mesh, and Table~\ref{tab:01} shows the corresponding file formats. As shown in the enlarged view, the \emph{degree} of a face is the number of its component edges, and the \emph{valence} of a vertex is defined as the number of its incident edges. Faces are usually composed of triangles (triangle meshes), quadrilaterals (quads), or other simple convex polygons (n-gons). Since the triangle mesh is the current mainstream mesh, this paper considers only triangle meshes.

\subsection{Outline}
The structure of this paper is as follows. Section~\ref{2} introduces the basic concepts, including the basic model of steganography and steganalysis, some standard evaluation metrics for the quantitative assessment of security, and the 3-D mesh structure. Section~\ref{3} presents the 3-D mesh steganographic techniques. Four main methods are described: two-state steganography, LSB steganography, permutation steganography and transform steganography. Section~\ref{4} introduces the 3-D mesh steganalysis technology, which includes universal steganalysis and specific steganalysis. Section~\ref{5} reveals the experimental results. Section~\ref{6} discusses open problems and some interesting research topics. Section~\ref{7} offers the conclusions of our work.

\section{Fundamental Concepts}\label{2}
\subsection{Basic Model}
\begin{figure}
\begin{center}
  \includegraphics[height=1.10in]{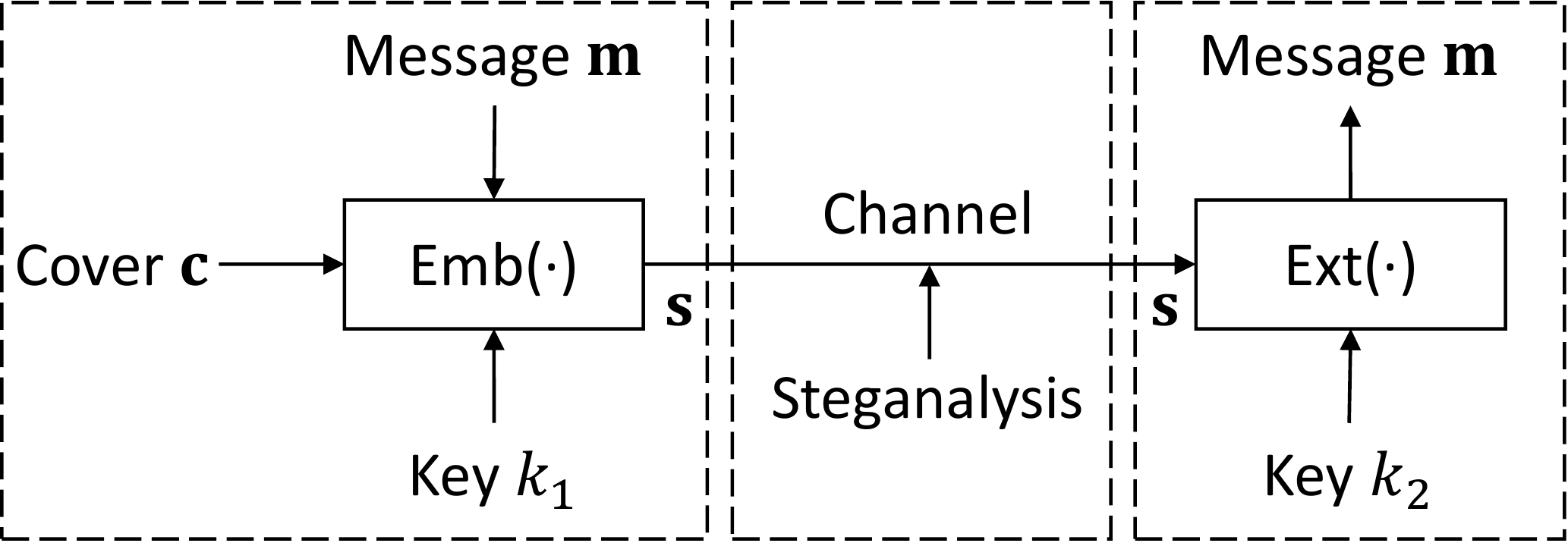}
  \caption{A diagram of steganographic communication and steganalysis. }
  \label{fig:com}
\end{center}
\end{figure}

As shown in Fig.~\ref{fig:com}, the problem of steganography and steganalysis is usually modeled as a prisoner's problem~\cite{simmons1984prisoners} related to three parties. In this problem, Alice and Bob are regarded as two prisoners who jointly work out an escape plan while Wendy, a warden, oversees their communications. With the data embedding function $\operatorname{Emb}(\cdot)$, Alice utilizes the secret key $k_1$ to hide the secret information $\mathbf{m}$ in a cover object $\mathbf{c}$ and generates an innocuous-looking stego object $\mathbf{s}$~\cite{kerckhoffs1883cryptographic}:
\begin{equation}
\begin{aligned}
\operatorname{Emb}(\mathbf{c}, \mathbf{m}, k_1)=\mathbf{s}.
\end{aligned}
\end{equation}

On the receiving side, the object represented by $\mathbf{s}$ obtained by Bob is subjected to the data extraction method $\operatorname{Ext}(\cdot)$, which is used to extract the embedded data $\mathbf{m}$ with the key $k_2$:
\begin{equation}
\operatorname{Ext}(\mathbf{s},k_2)=\mathbf{m}.
\end{equation}
Although in some papers the symmetric key steganographic scheme is utilized, the most common assumption is to adopt the private key steganographic scheme of $k_1=k_2$ in a steganographic system.
If Wendy can distinguish $\mathbf{s}$ from $\mathbf{c}$, the steganographic scheme is regarded as invalid. It should be noted that this example is utilized only to interpret the fundamental concept of steganography and steganalysis and does not fully explain the actual implementation.

\captionsetup{font={footnotesize}}

\subsection{Evaluation Metrics}
To fairly evaluate the performance of various steganographic and steganalytic methods, it is vital to define some standards that most people can accept. In addition, some evaluation criteria can help us to improve the techniques in the right direction.

\subsubsection{Criteria for Steganography}
Three general requirements, i.e., security, capacity and robustness, are utilized to evaluate the steganographic performance:
\begin{itemize}
\item \textbf{Security.} If the existence of secret information can be estimated only with a probability no higher than a random guess in the steganalytic system, steganography can be considered perfectly secure in this steganalytic system. In fact, the security level is evaluated by the anti-steganalysis performance. The definition of security is discussed further in subsection~\ref{sec:steganalysis}.
\item \textbf{Capacity.} To be useful in transmitting secret messages, the hiding capacity provided by steganography should be as high as possible. This capacity can be provided by an absolute metric (such as the size of the secret message) or a relative payload (called the embedding rate, e.g., bits per vertex (bpv)).
\item \textbf{Robustness.} Although most steganographic methods do not aspire to be robust, in some practical applications, due to the limited network traffic, bandwidth and processing capacity of smart devices, communication channels are lossy, resulting in reduced performance of the transmitted media.
\end{itemize}

%
%

\subsubsection{Criteria and Typical Classifiers for Steganalysis}\label{sec:steganalysis}
The main goal of steganalysis is to determine whether a suspicious medium is embedded with secret data, in other words, to determine whether the test medium belongs to the cover class or the stego class. If a certain steganalytic method is utilized to detect suspicious media, there are four possible results:
\begin{itemize}
\item True positive (TP), which means that the stego medium is correctly classified as stego.
\item False negative (FN), which means that the stego medium is incorrectly classified as cover.
\item True negative (TN), which means that the cover medium is correctly classified as cover.
\item False positive (FP), which means that the cover medium is incorrectly classified as stego.
\end{itemize}

\textbf{Confusion Matrix.} The steganalytic results of mixing cover and stego data can form a $2\times 2$ confusion matrix~\cite{fawcett2004roc}, which represents the configuration of the instances in the collection. Based on this, several evaluation metrics can be defined:
\begin{equation}
\begin{aligned}
\mathrm{TP} \text { Rate } &=\frac{\mathrm{TPs}}{\mathrm{TPs}+\mathrm{FNs}}, \\
\mathrm{FP} \text { Rate } &=\frac{\mathrm{FPs}}{\mathrm{TNs}+\mathrm{FPs}}, \\
\text { Accuracy } &=\frac{\mathrm{TPs}+\mathrm{TNs}}{\mathrm{TPs}+\mathrm{FNs}+\mathrm{TNs}+\mathrm{FPs}}, \\
\text { Precision } &=\frac{\mathrm{TPs}}{\mathrm{TPs}+\mathrm{FPs}}.
\end{aligned}
\end{equation}

\textbf{Receiver Operating Characteristic (ROC) Curve.} The performance of a steganalytic classifier can be visualized by the ROC curve~\cite{fawcett2004roc}, where the true positive rate is plotted on the vertical axis and the false positive rate is plotted on the horizontal axis. If the area under the curve (AUC) is larger, the performance of the steganalytic method is better.

Below, we introduce two typical supervised classifiers for training feature vectors extracted from the data.

\textbf{Support Vector Machine.}
A support vector machine (SVM) constructs a set of hyperplanes in a high-dimensional space, which is used for classification, regression and other tasks such as outlier detection. A fine separation is achieved by the hyperplane whose distance (i.e., margin) from the nearest training data of any class is the largest. In general, the larger the margin is, the lower the generalization error of the classifier. The linear, polynomial and Gaussian kernels vary when making the hyperplane decision boundary between the classes. 
The kernel functions are utilized to map the original features into a higher-dimensional space to create a linear dataset. Usually, linear and polynomial kernels are less time consuming and provide less accuracy than Gaussian kernels.



However, as the dimension of the feature space and the number of training samples increase, the complexity and memory requirements of SVMs also increase rapidly. Therefore, to deal with high-dimensional steganalytic features, an ensemble classifier for steganalysis is utilized.

\textbf{Ensemble Classifier for Steganalysis.}
Ensemble learning is a way of producing diverse base classifiers from which a new classifier is derived that performs better than any individual classifier. In the task of steganalysis, an ensemble classifier for steganalysis~\cite{kodovsky2011ensemble} is built as random forests by fusing decisions of weak and unstable base learners into a Fisher linear discriminant. Notably, this is a commonly used tool for steganalysis, since the computational cost of an SVM is much higher than that of an ensemble classifier when dealing with more than $500$-D features (e.g., 686-D SPAM~\cite{DBLP:journals/tifs/PevnyBF10} and 34671-D SRM~\cite{fridrich2012rich}). Moreover, ensemble classifiers yield lower detection errors than do SVMs. 

Specifically, for each payload, a separate FLD-ensemble is trained on the original features and on the stego features.
The testing error is evaluated using the minimal total error probability under equal priors, i.e.,
\begin{equation}
\label{eqn:wegs}
P_\textrm{E}=\min_{P_{\textrm{FP}}}\frac{1}{2}(P_{\textrm{FP}}+P_{\textrm{FN}}),
\end{equation}
achieved on a test set averaged over ten 50/50 splits of the database. The symbols $P_{\textrm{FP}}$ and $P_{\textrm{FN}}$ stand for the false positive and false negative rates, respectively. 

\section{3-D Mesh Steganography Techniques}\label{3}

In this section, we categorize the steganographic techniques into four domains, i.e., two-state domain, LSB domain, permutation domain and transform domain, and elaborate on the development of each.

Before introducing each technique, we list some typical 3-D mesh digital operations below, which can be regarded as attacks, as they may invalidate the correct extraction of embedded watermark messages:
\begin{itemize}
\item \textbf{Affine transform} (including translation, rotation and scaling). These operations are basic techniques used to understand a 3-D object by moving it or the camera, and they can be expressed by homogeneous transformation matrices.
\item \textbf{Vertex reordering.} Reordering does not change the topology of the mesh; it changes only the storage layout of the vertices and gives them new indices. It is commonly used for mesh optimization efficiency~\cite{DBLP:conf/imr/ShontzK08}, cache coherency~\cite{DBLP:journals/ewc/SastryKSK14}, etc.
\item \textbf{Noise addition.} Measured mesh models often contain noise, introduced by the scanning devices and digitization processes.
\item \textbf{Smoothing.} The aim of this operation is to generally remove certain high-frequency information in the mesh.
\item \textbf{Simplification.} This is carried out to transform a given 3-D mesh into another complexity-reduced mesh with fewer faces, edges, and vertices.
\end{itemize}

%
%
%
%

\subsection{Two-State Domain}
In the early research on 3-D mesh data hiding, steganography and watermarking were regarded as the same technique by researchers.
A large number of steganography schemes for meshes create multiple two-state domains and embed messages by aligning together the state information and message bit.

\subsubsection{MEP-based method}
\captionsetup{font={footnotesize}}
\begin{figure}
\begin{minipage}[t]{0.5\linewidth}
\centering
\subfloat[]{
  \label{fig:02a}
    \includegraphics[width=1.50in]{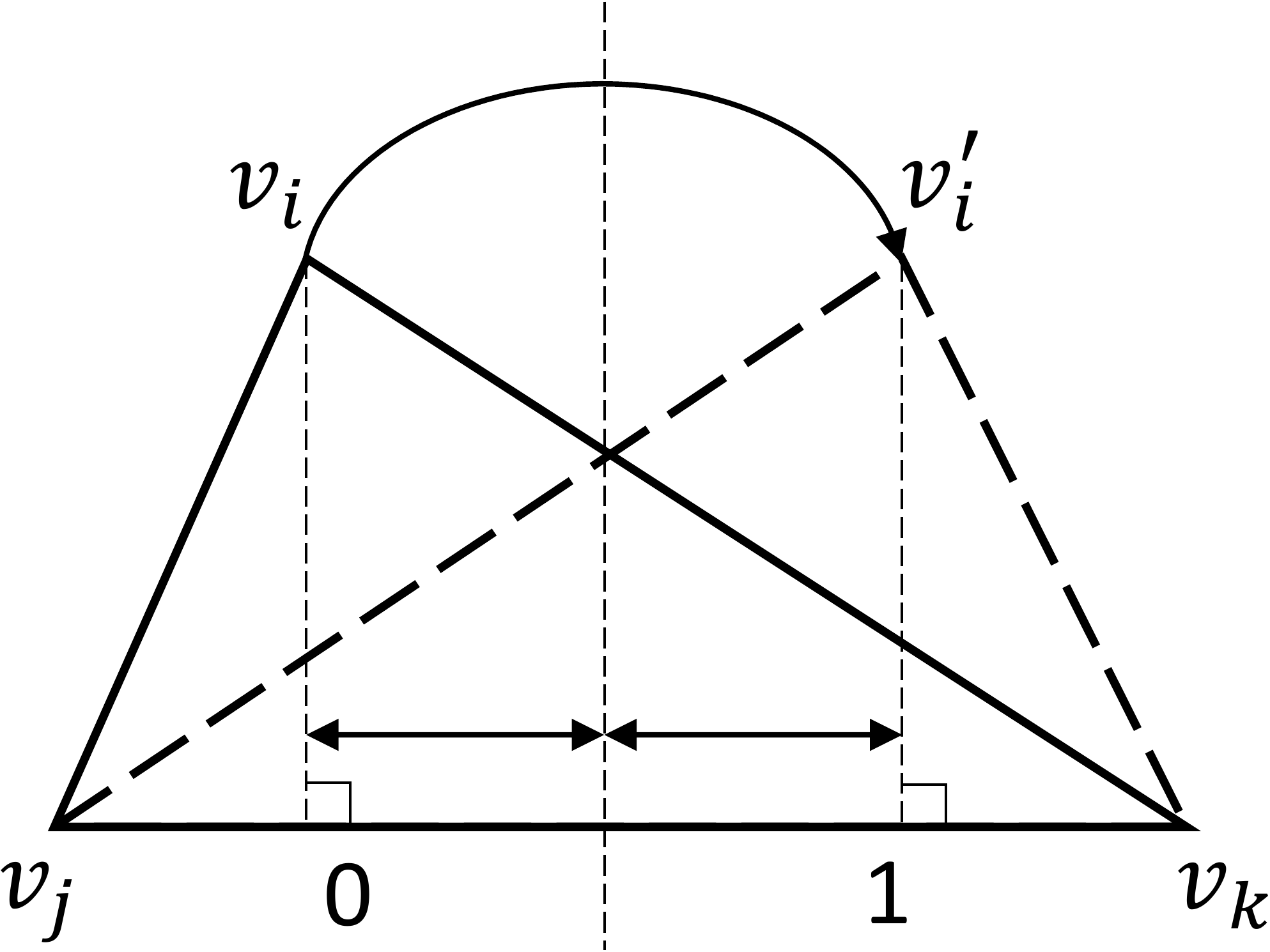}
    }
\centering
\subfloat[]{
  \label{fig:02b}
    \includegraphics[width=1.50in]{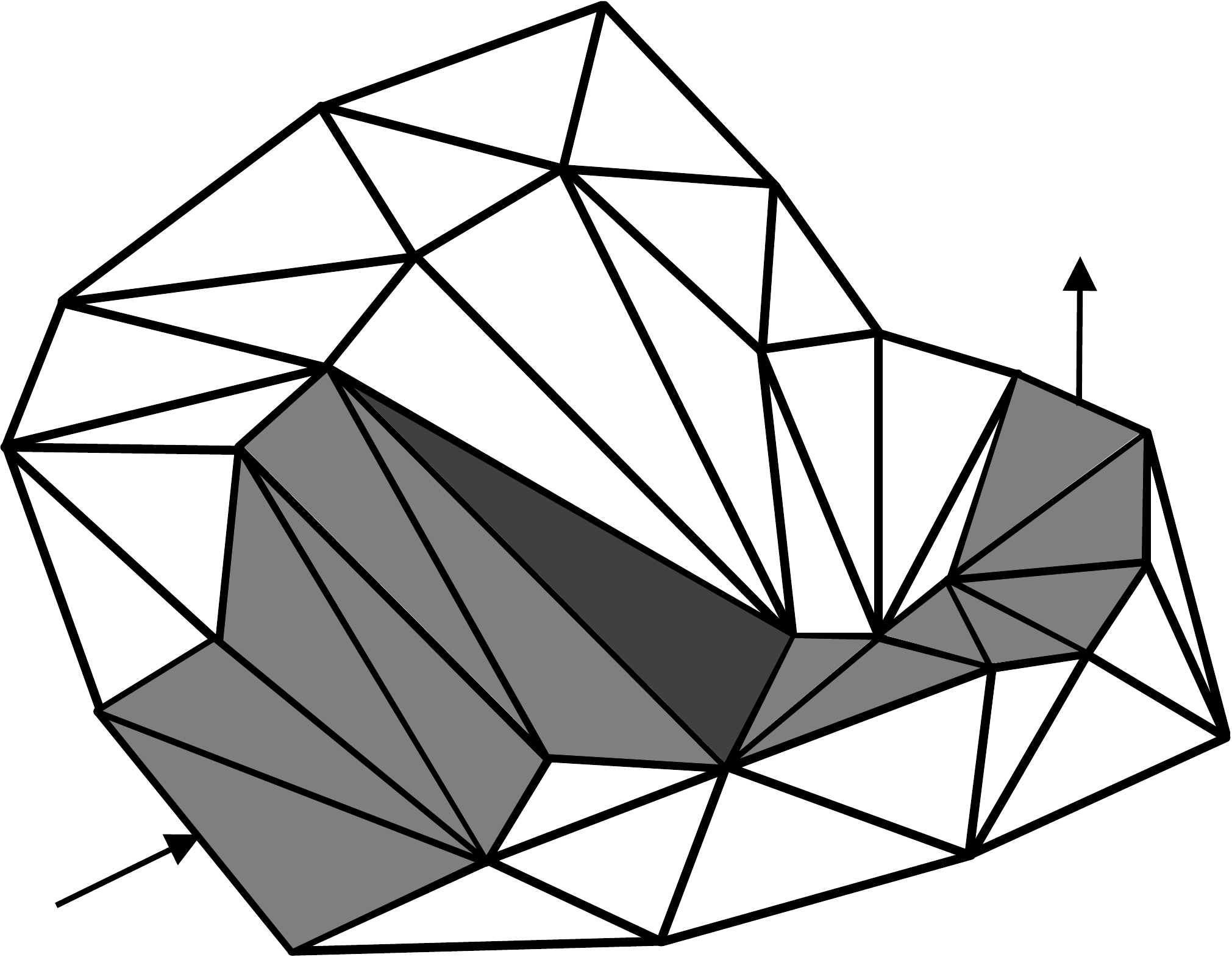}
    }
\end{minipage}
\caption{(a) Diagram of the MEP. The triangle is viewed from the topology perspective, with the entry edge ($\overline{v_jv_k}$) related to the current message bit to be inserted and the two candidate exit edges ($\overline{v_iv_j}$ and $\overline{v_iv_k}$). The exit edges are ordered in a clockwise manner. (b) A polygonal mesh, shown with the TSPS path (gray). Each face represents a one-bit message, with the MEP shown in black. Figure from~\cite{cayre2003data}. }
\label{fig:02}
\end{figure}

Cayre and Macq~\cite{cayre2003data} proposed a 3-D mesh steganography scheme. Geometrically, it is a quantization index modulation scheme that extends to the edge of a triangle. As shown in Fig.~\ref{fig:02}\subref{fig:02a}, the macro embedding procedure (MEP) is a spatial substitutive procedure. The global core is to treat the triangle as a two-state geometrical object, meaning that by orthogonally projecting the position of the vertex $v_i$ onto the edge $\overline{v_jv_k}$, the edge can be partitioned into ``0'' and ``1'' states. Depending on the message bits, $v_i$ remains unchanged or moves to $v_i'$. Fig.~\ref{fig:02}\subref{fig:02b} shows the establishment of the triangle processing list. The so-called triangle strip peeling sequence (TSPS) algorithm inserts message bits as it moves along the mesh.

In summary, the method is robust to affine transformations and vertex reordering. Its theoretical upper-bound capacity is 1 bit per vertex. However, the algorithm has neither a large enough capacity nor computational efficiency because it has visible embedding modifications and requires too much preprocessing time to create the stencil. Because of the large modifications, it cannot resist steganalysis.

\subsubsection{MLEP-based method}

\captionsetup{font={footnotesize}}
\begin{figure}
\begin{minipage}[t]{0.5\linewidth}
\centering
\subfloat[]{
  \label{fig:03a}
    \includegraphics[width=1.50in]{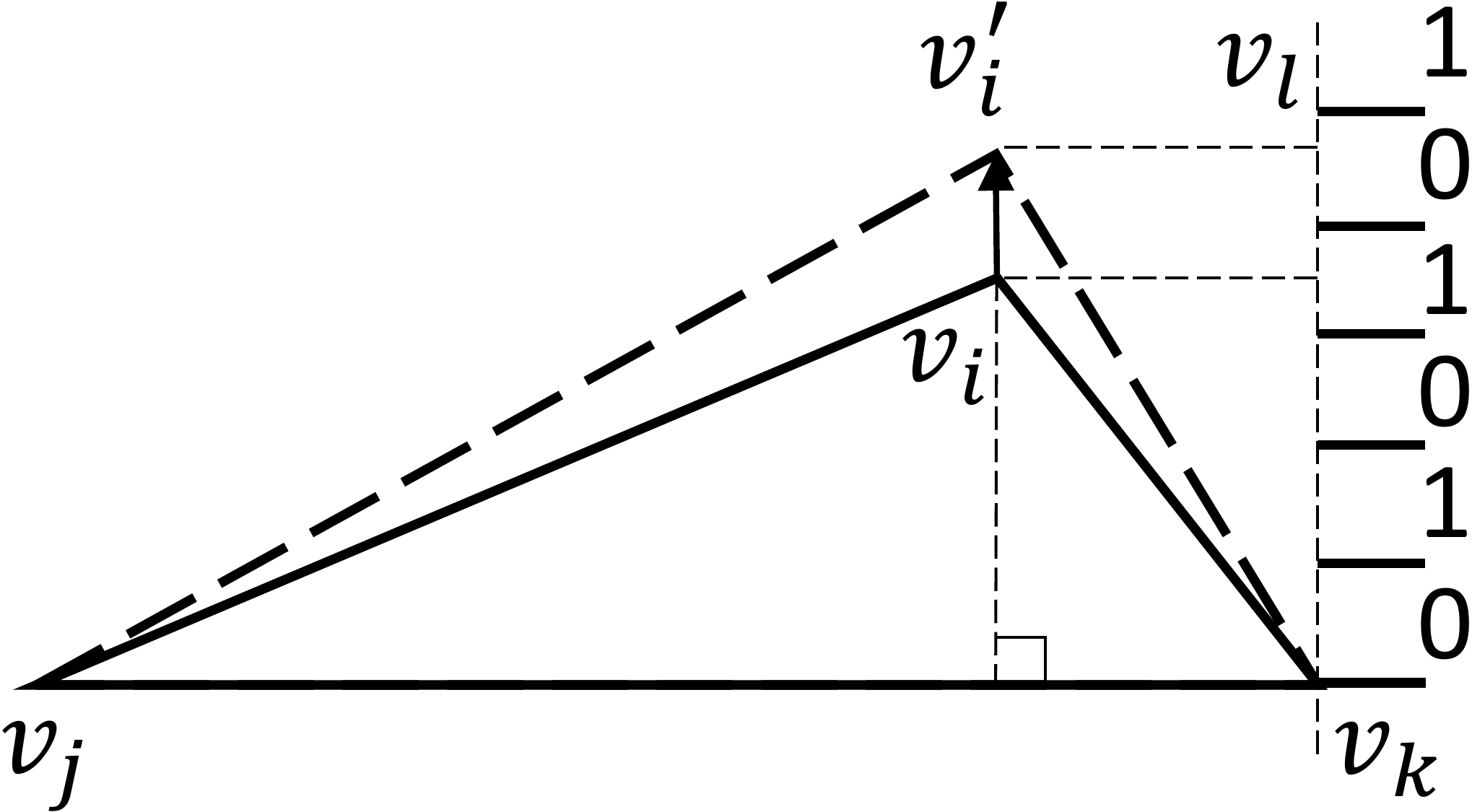}
    }
\centering
\subfloat[]{
  \label{fig:03b}
    \includegraphics[width=1.75in]{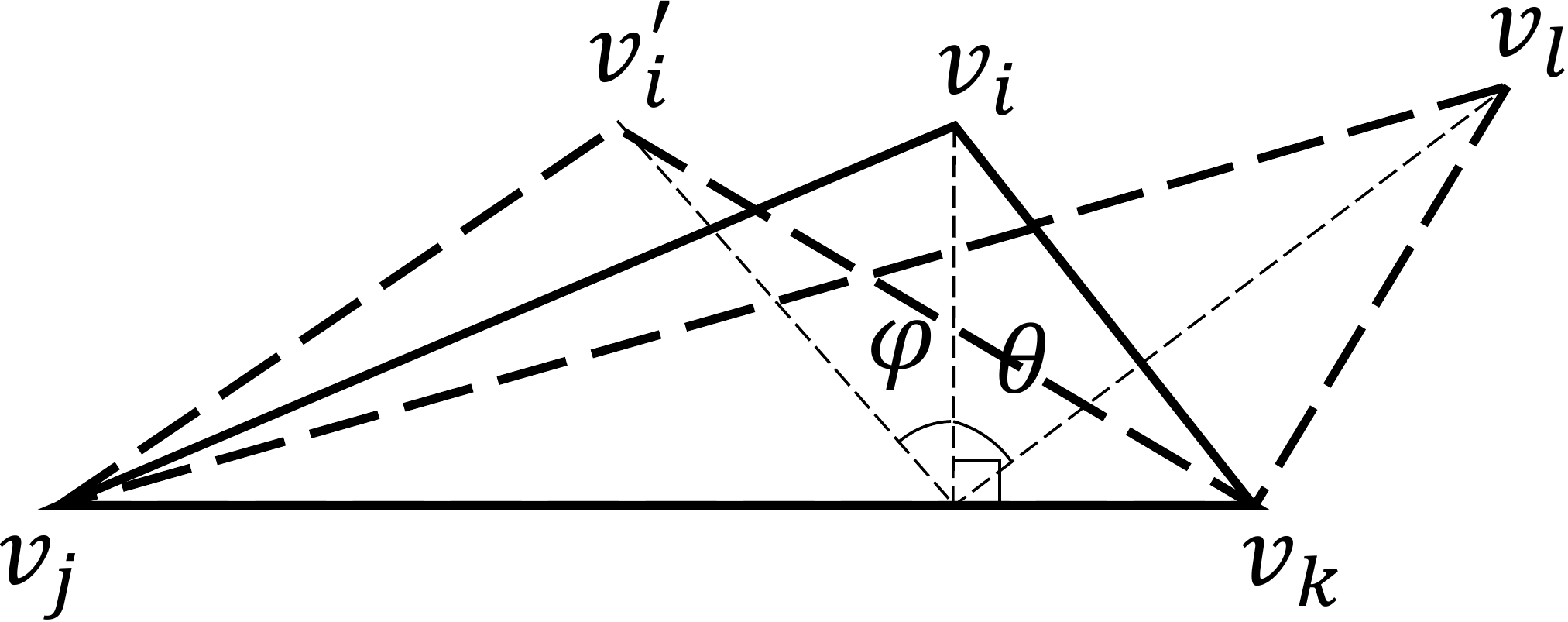}
    }
\end{minipage}
\caption{(a) Extending domain of the MLEP. (b) Rotating domain of the MLEP. Figure from~\cite{wang2005efficient}.}
\label{fig:03}
\end{figure}

Wang and Cheng~\cite{wang2005efficient} proposed an efficient steganographic technique based on a substitutive blind procedure in the spatial domain. To quickly obtain the processing order of vertices, they created a hierarchical data structure based on the kd-tree and an advanced jump strategy. They also proposed the multilevel embed procedure (MLEP), which includes sliding, extending, and rotating domains to embed messages based on the geometric properties of the message vertices.

Fig.~\ref{fig:03}\subref{fig:03a} is a diagram of the extension domain. Let a vertex $v_l$ and the line defined by $v_l$ and $v_k$ be orthogonal to the line $\overline{v_jv_k}$. The state of the triangle depends on the orthogonally projected location of the vertex $v_i$ onto the virtual edge $\overline{v_lv_k}$. Depending on the message bits, $v_i$ remains unchanged or moves to $v_i'$. Fig.~\ref{fig:03}\subref{fig:03b} is a diagram of the rotation domain. The message is embedded in the angle between two triangular planes. $v_l$ is the reference vertex obtained by the barycenter of the initial triangle. The message is embedded by adding or subtracting the angle $\phi$.

In summary, this method is also robust to affine transformations and vertex reordering. It increases the capacity up to 6 bits per vertex and reduces the calculation time, but the embedding capacity is still very low. Moreover, it also cannot resist steganalysis. 

\subsubsection{Point-sampled-geometry-based method}
\captionsetup{font={footnotesize}}
\begin{figure}
\begin{center}
  \includegraphics[height=1.25in]{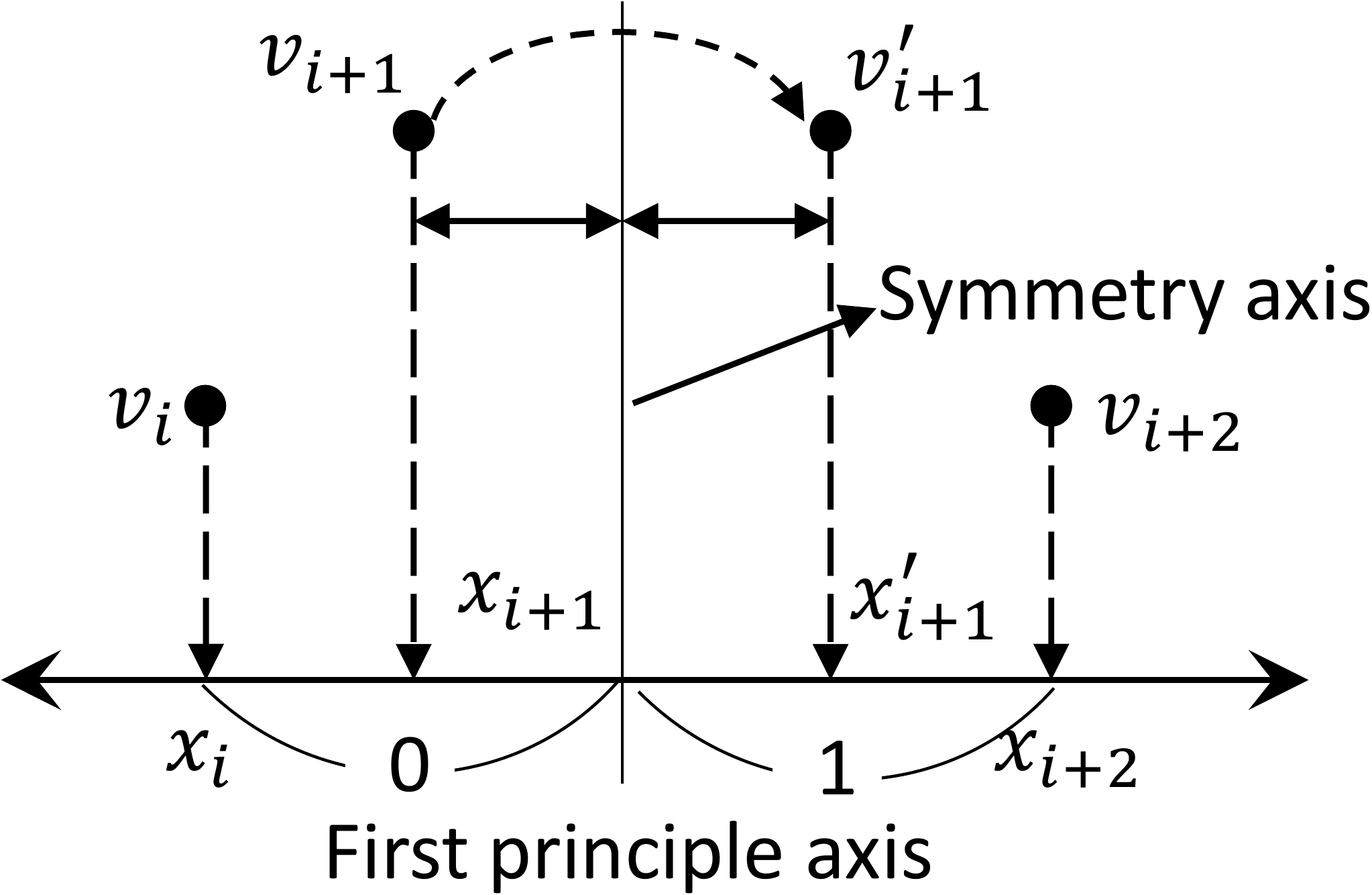}
  \caption{An interval $\overline{x_ix_{i+2}}$ is divided into two subintervals for embedding data. Figure from~\cite{wang2006steganography}.}
  \label{fig:05}
\end{center}
\end{figure}

Wang et al.~\cite{wang2006steganography} proposed the first steganographic method for point-sampled geometry. It is based on principal component analysis (PCA) of the vertex permutation-invariant method, which can convert the original vertex coordinates into a new coordinate system. The method builds a list of intervals for each of $xyz$-axes and embeds a bit in each interval by changing the position of the vertex. As shown in Fig.~\ref{fig:05},
each interval (consisting of two values $x_i$ and $x_{i+2}$ on the $x$-axis) is regarded as a two-state object. The authors used the $x$-coordinate $x_{i+1}$ of the vertex $v_{i+1}$ to define the state of the interval and partitioned the line segment $\overline{x_{i}x_{i+2}}$ into two sets $S_0$ and $S_1$. If $x_{i+1}\in S_0$, the interval is considered to be in the ``0'' state; otherwise, it is $x_{i+1}\in S_1$, and the interval is in the ``1'' state. To embed data, if the embedded message is equal to the state, then no modification is required; otherwise, the subinterval boundary must be utilized as the symmetry axis to move $v_{i+1}$ toward $v_{i+1}'$.

In addition, they located the list of macro embedding primitives and embedded up to 6 bits at each macro embedding primitive to increase the capacity. In summary, the upper-bound capacity of the method is $6/2=3$ bpv. Additionally, this method is robust to affine transformations and vertex reordering. The method is easy to attack by PCA transform-targeted steganalysis, as stated in subsection~\ref{pca}.

\subsubsection{Multilayer-based method}
\captionsetup{font={footnotesize}}
\begin{figure}
\begin{center}
  \includegraphics[height=1.75in]{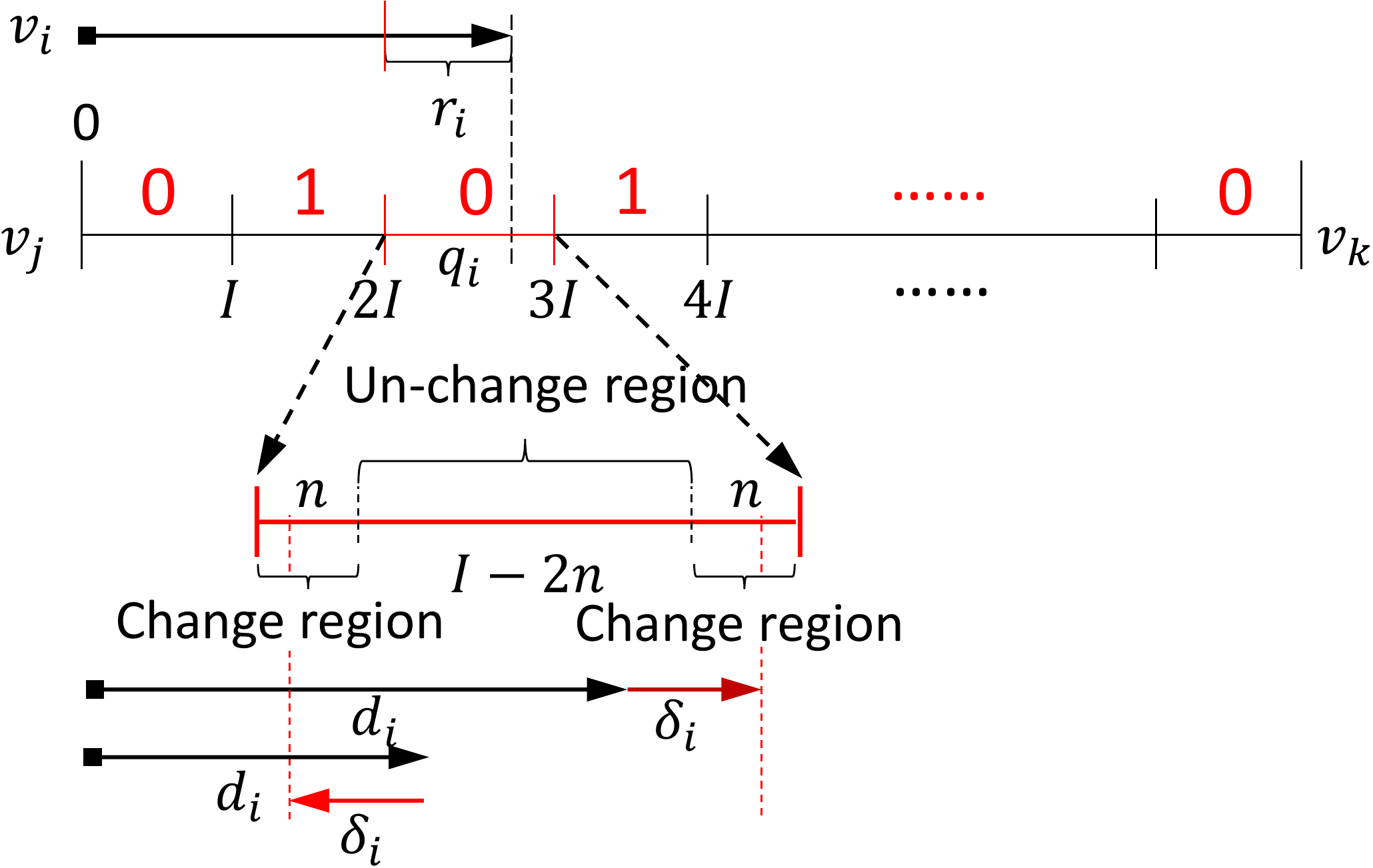}
  \caption{Diagram of the state region $q_i$ and its changed and unchanged regions. During the data-embedding period, vertices are moved into change regions while no vertices are moved into unchanged regions. Figure from~\cite{chao2009high}.}
  \label{fig:07}
\end{center}
\end{figure}
As in~\cite{wang2006steganography}, Chao et al.~\cite{chao2009high} transform the coordinate system through PCA. The two most extreme vertices that fall on the first principal axis are regarded as the end vertices $v_j$ and $v_k$.
The interval $\overline{v_jv_k}$ is evenly partitioned into two-state region sets in an interleaved manner (for example, $010101...$), expressed as $S_0$ and $S_1$, as shown in Fig.~\ref{fig:07}.
The region, in which vertex $v_i$ is located, can be calculated using the following equations:
\begin{equation}
\label{eqn:05}
\begin{aligned}
d_i/I&=q_i...r_i,\\
  q_i&=\lceil d_i/I\rceil,\\
  r_i&=d_i\%I,
\end{aligned}
\end{equation}
where $...$ is the modulo operation, $d_i$ is the projection distance between vertex $v_i$ and end vertex $v_j$ of the $x$ component.
$I$ indicates the width of the interval in the state region.
$q_i$ represents the state region to which the vertex $v_i$ belongs, and $r_i$ represents the position of the vertex in the state region $q_i$.
There are two cases of embedded message bits:

\textbf{Case 1. }$q_i\in S_k$ and $bit(i)=k,k=0$ or 1: No modification.

\textbf{Case 2. }$q_i\in S_k$ and $bit(i)\neq k,k=0$ or 1: Move the vertex $v_i$ to the adjacent change region in $q_i$. If $d_i\%I\leq I/2$, shift $v_i$ to the range $\left[0,n\right]$; if not, shift $v_i$ to the range $\left[I-n, I\right]$. The offset $\delta_i=\min(r_i,I-r_i)(1-2n/I)$, and the new position is acquired by
\begin{equation}
\label{eqn:06}
d_i'=
\left\{
\begin{aligned}
d_i-\delta_i, \qquad \text{if}\ r_i\leq I/2\\
d_i+\delta_i, \qquad \text{if}\ r_i>I/2
\end{aligned}
\right..
\end{equation}

In addition, by moving the state regions in even layers by $\frac{1}{2}I$, one can achieve multilayered high-capacity steganography. In the first layer, $n=\frac{1}{4}I$, and the change region is $[0,\frac{1}{4}I]\cup[\frac{3}{4}I,I]$. Iteratively, for the second layer, $n=\frac{3}{8}I$, and the change region is $[0,\frac{3}{8}I]\cup[\frac{5}{8}I,I]$. Thus, in the $n$-th layer, the change region is $[0,\frac{2^n-1}{2^{n+1}}I]\cup[\frac{2^n+1}{2^{n+1}}I,I]$. Since the precision of a single floating point is $2^{-23}$, the distance between the two boundaries has to be larger than the precision, i.e., $\frac{2^n+1}{2^{n+1}}I-\frac{2^n-1}{2^{n+1}}I\geq 2^{-23}$. Thus, the limitation of $n\leq\log_{2}I+23$. 

In summary, when $I=1$, considering the $x,y,z$-coordinates, the method has a theoretical upper-bound capacity of 69 bpv, and it can resist vertex reordering and affine transform attacks. However, it can be easily attacked via PCA transform-targeted steganalysis.

\subsubsection{Static-arithmetic-coding-based method}


Itier and Puech~\cite{itier2017high} constructed Hamiltonian paths on the entire vertex graph without using connectivity information. The Hamiltonian path is a uniquely traversed path of all the vertices starting from the one selected by the secret key. For each stage, the nearest neighbor vertex $v_{i+1}$ of the present vertex $v_i$ is selected. As a matter of fact, the message is embedded with its synchronization with the message sequence guaranteed. Once the vertex $v_{i+1}$ is added to the path, the message can be embedded by changing the relative location of the vertex $v_{i+1}$ to its predecessor $v_i$. To achieve a high embedding capacity, the vertex moves along three coordinate components in the spherical coordinate system.

In addition, the algorithm divides the edge vector into multiple intervals for each coordinate, and these intervals correspond to different message bits. Specifically, static arithmetic coding is utilized to embed messages, which first slices subintervals while taking into account the value distribution of all the messages and then associates the embedded bits with the corresponding subinterval. Then, the vertex is moved to the new subsegment within the segment to embed the message bit. The check phase is performed to retain the created Hamiltonian path and message synchronization process during steganography. The method can achieve a maximum capacity of 24 bits per vertex.

Later, Li et al.~\cite{li2017rethinking} reconsidered the method to improve its resistance to steganalysis. The original steganography method~\cite{itier2017high} embeds data into all three coordinate components of the edge vector in the spherical coordinate system. However, changes in polar and azimuthal coordinates have a greater impact on steganalytic features (such as LFS76~\cite{li2017steganalysis}) than in the radial coordinate system. Therefore, they improved the method by resisting against steganalysis through modification of only the radial component represented by the edge vector in the spherical coordinate system. Therefore, the upper-bound embedding capacity is only 8 bits per vertex.

In summary, the method of Itier and Puech~\cite{itier2017high} has a maximum capacity of 24 bpv but cannot resist steganalysis. The method of Li et al.~\cite{li2017rethinking} has a maximum capacity of 8 bpv and can resist steganalysis effectively.

\subsubsection{Statistical-embedding-based method}
Cho et al.~\cite{cho2006oblivious} proposed a blind watermarking scheme that uses the distribution of vertex norms to embed data.
First, the vertices are divided into multiple bins according to the distances from the vertices to the center $\mathbf{o}$ of the mesh:
\begin{equation}
\label{eqn:02}
\mathbf{o}=\frac{\sum_{v_i\in \mathcal{V}}A(v_i)\mathbf{p}_i}{\sum_{v_i\in \mathcal{V}}A(v_i)},
\end{equation}
where $A(v_i)$ represents the area of all faces containing vertex $v_i$.
For a given vertex $v_i$ and its distance to the center $\rho_{i}=\left\|\mathbf{p}_i-\mathbf{o}\right\|$, we obtain $\rho_{min}=\min_{v_i}(\rho_{i})$ and $\rho_{max}=\max_{v_i}(\rho_{i})$ and evenly group vertices according to the segmentations between $\rho_{min}$ and $\rho_{max}$. For a normal mesh, the distribution of the statistical variable of the bin is usually close to a uniform distribution, for example, the expected average value of the statistical variable is $1/2$. Therefore, if the message is 0 or 1, the watermark is embedded by moving the statistical variable to $1/2+\epsilon$ or $1/2-\epsilon$. In short, the embedding capacity is directly related to the number of bins.

Bors and Luo~\cite{bors2013optimized} extended the statistical watermark embedding method to preservation of the mesh surface.
The surface preservation function includes the distance from the vertices shifted to the original surface by watermarking, the distance to the surface of the watermarked object, and the actual vertex shift, and the sum of these Euclidean distances are minimized using Levenberg-Marquardt optimization~\cite{marquardt1963algorithm,levenberg1944method} in a spherical coordinate system.

In summary, since the method is based on statistical shifting, it is robust to additive-noise attacks, Laplacian-smoothing attacks and mesh-simplification attacks. It is worth mentioning that there have been some steganalytic methods for detecting watermarks~\cite{yang2014steganalytic,yang20163d,yang2013information}.
In addition, the optimization process hinders the speed of watermark embedding. Due to the limited number of positions, the embedding capacity is still very low. Moreover, because of the large modification, it cannot
resist steganalysis.

\subsection{LSB Domain}
LSB steganography first embeds data on the LSB layer and then embeds it on the LSB layer of the remaining bitplanes of the object iteratively until all data are embedded.

\subsubsection{Gaussian-curvature-constraint-based method}
Yang et al.~\cite{yang2013linear} designed a data hiding algorithm that balances a high embedding capacity and low embedding distortion.
A simplified Gaussian curvature $\kappa_i$, as the smoothness of the mesh at vertex $v_i$, is defined as:
\begin{equation}
\label{eqn:07}
\kappa_G(v_i)=2\pi-\sum_{v_j\in\mathcal{N}_1(v_i)}\theta_j,
\end{equation}
where $\mathcal{N}_1(v_i)$ is the one-ring neighbor of vertex $v_i$ and $\theta_j$ represents the angles of the incident triangles at vertex $v_i$. $\left|\kappa_G(v_i)\right|$ could reflect the smoothness of the local region of the $i$-th vertex. Based on $\kappa_G(v_i)$, a vector of quantization levels for each vertex is calculated. Since visual distortion is closely related to normal degradation, the least significant bit (LSB) replacement is used~\cite{DBLP:journals/spl/Mielikainen06a} (which embeds data by replacing the lowest bit with the message bit) to inform the capacity under a given distortion tolerance.

Each vertex coordinate is in the 32-bit IEEE 754 single precision standard format. Apart from the top bit, which indicates whether the coordinate is positive or negative, the remaining 31 bits can be embedded with the messages. Considering the $x,y,z$ coordinates altogether, each vertex has a maximum 93-bit capacity.

In summary, this method can achieve a high capacity (93 bpv) and low distortion, but when the amount of embedded noise becomes larger, the error will greatly increase. This method is robust to the vertex reordering attack but is fragile against PCA transform-targeted steganalysis.

%

\subsubsection{Adaptive-steganography-based method}

Zhou et al.~\cite{zhou2018distortion} proposed an adaptive steganography technique to resist steganalysis. Unlike previous schemes (belonging to the nonadaptive mode) that shift the vertex coordinates to embed messages without considering steganalysis, this technique provides vertices with various costs for determining the probability of modification.

Considering the storage form of the uncompressed mesh, each vertex coordinate is usually specified as a 32-bit single precision pattern, where the significant precision number is 23 bits~\cite{chao2009high}. These coordinates are converted into multiple binary bitplanes~\cite{jiang2017reversible}. Considering that steganography in a low-level bitplane causes fewer artifacts in the overall coordinates, the data are first embedded into the low-level bitplane and then iteratively embedded into the high-level bitplane. These bitplanes are embedded by LSB replacement, except for the highest bitplane determined by the message.

The operation of the top biplane is as follows. Since the steganographic security performance is primarily dependent on several effective steganalytic features, the various submodels of steganalytic features are independently evaluated, and the vertex normal feature is found to have the best discriminability over the cover and stego meshes. To compete with these subdetectors, the authors designed vertex distortions based on vertex normals. The cost value $\rho_i$ was designed as:
\begin{equation}
\label{eqn:08}
\rho_i=\frac{1}{\ln(\left\|\mathbf{n}(v_i)-\mathbf{n}(v_i')\right\|_2+1)+\epsilon},
\end{equation}
where $\mathbf{n}(v_i')$ comes from the Laplacian-smoothed mesh. Using syndrome-trellis codes (STCs)~\cite{filler2011minimizing} or steganographic polar codes (SPCs)~\cite{li2020design}, steganography can be well implemented. A larger distortion cost means a smaller modification probability. A better designed distortion function could effectively withstand steganalysis.

In summary, the method has a maximum capacity of 69 bpv, but it is fragile against the existing attacks, including affine transform, vertex reordering, noise addition, smoothing and simplification. However, it is strong enough to contend against existing steganalysis when the embedding payload is low.

\subsection{Permutation Domain}

\captionsetup{font={footnotesize}}
\begin{figure}
\begin{center}
  \includegraphics[height=1.25in]{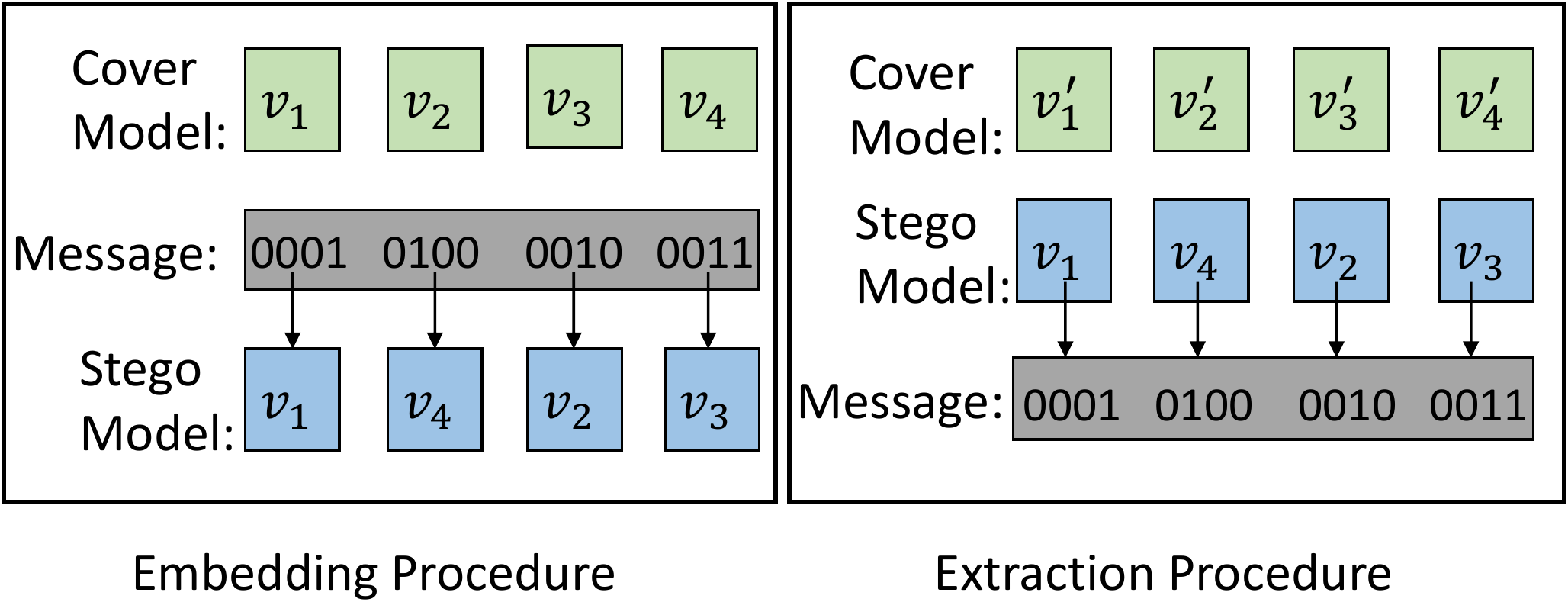}
  \caption{Schematic diagram of steganography and message extraction of vertex index embedding by rearranging vertex indexes. }
  \label{fig:06}
\end{center}
\end{figure}

Permutation steganography~\cite{artz2001digital} hides data in the order of the set elements. We can map each permutation of the set to an integer and encode the message into the cover object by altering the element order in the set. The 3-D mesh contains a group of rearrangeable vertices and triangles, which provides space for permutation-based steganography without changing the geometry of the mesh.

Given $n$ elements, assuming that they are rearrangeable, the embedded data can be encoded by arranging elements related to the known reference order. With $n!$ possible orders, the standard permutation steganography is able to encode no more than $\log_2{(n!)}=\mathcal{O}(n\log{n})$ bits of the optimal message capacity at the cost of a computational complexity of $\Omega(n^2\log^2n\log{\log{n}})$.

The specific implementation of the optimal permutation steganography algorithm is described below. Let $\mathbf{m}$ represent the secret data. $\mathbf{m}$ comprises multiple consecutive $0$s and $1$s and can be regarded as a long integer $m$. The core purpose is to obtain the replacement $\pi$ related to the value of $m$ on $\mathcal{V}$ recursively. For the $i$-th iteration step, the $i$-th element of the permutation is determined as the $m/b_i$-th residual element of a reference ordering of $\mathcal{P}$, in which $b_i=(i-1)!$ is regarded as the factorial basis; in addition, $m$ is updated using $m\%b_i$. Through this processing, the message is embedded by arranging $\pi$. To accurately recover $m$ from $\pi$ on the receiver side, some methods for the canonical traversal of meshes are usually utilized to determine the unique reference order of $\mathcal{P}$~\cite{rossignac1999edgebreaker}. This simple method is illustrated in Fig.~\ref{fig:06}.

In summary, owning to the fact that the reference ordering is obtained by traversing via the mesh compression algorithm called ``Edgebreaker''~\cite{rossignac1999edgebreaker} based on the mesh connectivity only, permutation steganography is robust and resistant to geometric affine transformation attacks.
However, due to the integer algorithm, this algorithm has a very large runtime complexity, which makes it difficult to use the standard permutation steganography method for a large number of elements. Therefore, various variants of the method are put forward to find a feasible trade-off between embedding capacity and computational complexity. Because the adjacent correlation is broken, this method is fragile to permutation-targeted steganalysis, as stated in subsection~\ref{permutation}.

\subsubsection{Order-encoding-based method}
Bogomjakov et al.~\cite{bogomjakov2008distortion} proposed a permutation steganography algorithm, which can be efficiently implemented. The idea is to maximize the length of the bitstream while encoding each embedded message with the larger index of an embedding value in the reference ordering as much as possible. Specifically, instead of dividing the two long integers, this method utilizes the next $k=\left\lfloor\log_2{i}\right\rfloor$-bit data to select the $(n-i)$-th element in the permutation $\pi$. To enlarge the capacity, Bogomjakov et al. used a special trick to encode one more bit of data: when the next $k+1$-bit data point is selected, if the decimal integer is smaller than $i$ but greater or equal to $2^k$, the element is able to encode $k+1$ bits. Otherwise, the element can encode only a $k$-bit message and use the decimal integer to choose the correct element.
Obviously, all the bitstreams of maximum length have the most significant bit ``1''. The expected capacity (in bits) is calculated by
\begin{equation}
\label{eqn:03}
E_1(C_i)=(k_i+1)\frac{i-2^{k_i}}{2^{k_i}}+(k_i)\frac{2^{k_i}-i}{2^{k_i}}.
\end{equation}

In summary, the encoder and decoder have a computational complexity of $\mathcal{O}(n)$, being more efficient than the optimal permutation steganography method. The upper bound of the embedding capacity is $\frac{1}{V}\sum_{i=1}^{V}\left\lfloor\log_2{i}\right\rfloor$ bpv. The compact and simple computation of the method is efficient in implementation and loses only one bit per vertex compared with the optimal one regarding capacity. In addition, this method is fragile against permutation-targeted steganalysis.

\subsubsection{Enhanced-order-encoding-based method}
Huang et al.~\cite{huang2009toward} proposed a permutation steganographic algorithm with high efficiency, which can increase the capacity to an amount closer to the optimal case while maintaining the same time complexity. The next $k$-bit data are chosen, and if their decimal integer is less than $i-2^k$, then the element can encode the $k+1$-bit message; otherwise, only the $k$-bit message can be encoded.
The expected capacity (in bits) is calculated as
\begin{equation}
\label{eqn:04}
E_2(C_i)=(k_i+1)\frac{i-2^{k_i}}{2^{k_i+1}}+(k_i)\frac{2^{k_i+1}-(i-2^{k_i})}{2^{k_i+1}}.
\end{equation}

Because $E_2(C)=\sum E_2(C_i)>E_1(C)=\sum E_1(C_i)$, the expected capacity of the algorithm is closer to the optimum capacity than~\cite{bogomjakov2008distortion}, i.e., the lower bound is $\frac{1}{V}\sum_{i=1}^{V}\left\lfloor\log_2{i}\right\rfloor$ bpv. In addition, the same runtime complexity of $\mathcal{O}(n)$ is achieved. Moreover, the method is fragile to permutation-targeted steganalysis.

\subsubsection{Binary-tree-based method}
Tu et al.~\cite{tu2010improved} improved the method of Bogomjakov et al.~\cite{bogomjakov2008distortion} by adopting a complete binary tree to embed and extract data, which doubles the probability of encoding the additional bit-length message. This method does not directly interpret the primitive indexes in the reference order as encoding/decoding bits but makes some modifications to the mapping. The number of different $k+1$ bits is twice the difference between $n$ and the highest $2^k$ less than $n$ (i.e., $(i-2^k)\times 2$). If the primitive is less than $2(i-2^k)$, the method will peek at the next $k+1$-bit data. Otherwise, only the $k$-bit data can be encoded, and the offset $i-2^k$ is added to the value of these original bits, which is then interpreted as an index.

In summary, compared with the algorithm of \cite{bogomjakov2008distortion}, this method doubles the chance of encoding additional bits and increases the average capacity by 0.63 bits per vertex. In short, this scheme has the same minimum and maximum capacity, and the average embedding rate is still one bit per vertex less than the optimum. In addition, the method has the same embedding capacity (the upper bound is $\frac{1}{V}\sum_{i=1}^{V}\left\lfloor\log_2{i}\right\rfloor$ bpv) and computational complexity $\mathcal{O}(n)$ as those of~\cite{huang2009toward} and is fragile against permutation-targeted steganalysis.

\subsubsection{Coding-tree-based method}
Tu and Tai~\cite{tu2010permutation} established a left-biased binary coding tree for embedding bitstrings into primitives. The core operation of the method constitutes two parts: the left-skewed subtree, which extends the bitstring, and the right subtree, which is a complete binary tree.

In summary, the minimum capacity of the method remains the same as that of the complete binary tree. The scheme has a computational complexity of $\mathcal{O}(n\log{(n)})$. In addition, this method increases the embedding capacity of each vertex by 0.63 bits compared with \cite{bogomjakov2008distortion}, i.e., the upper bound is $\frac{1}{V}\sum_{i=1}^{V}\left\lfloor\log_2{i}\right\rfloor+0.63$ bpv. Moreover, the method is fragile to permutation-targeted steganalysis.

\subsubsection{Maximum-expected-level-tree-based method}

\captionsetup[table]{labelsep=space}
\begin{table*}[htbp]
\centering\caption{\label{tab:02}\\
Comparison of steganographic methods in terms of distortion, capacity and security. ``$\surd$'' indicates that the steganographic method is secure against all steganalysis methods, and ``$\times$'' indicates that the method is fragile to at least one steganalysis method. }
 \centering
 \newcommand{\tabincell}[2]{\begin{tabular}{@{}#1@{}}#2\end{tabular}}
   \begin{tabular}{c|c|cc}
    \toprule
    Category (domain) & Method & Distortion & Security\\
    \midrule
    \multirow{6}{*}{Two-state}
    & MEP~\cite{cayre2003data}     & Large  & $\times$ \\
    & MLEP~\cite{wang2005efficient} & Medium & $\times$\\
    & Point-sampled geometry~\cite{wang2006steganography}& Medium  & $\times$ \\
    & Multilayer~\cite{chao2009high}      & Medium  & $\times$ \\
    & Static arithmetic coding~\cite{itier2017high,li2017rethinking} & Medium & $\surd$\\
    & Statistical embedding~\cite{cho2006oblivious,bors2013optimized}   & Medium & $\times$\\
    \midrule
    \multirow{2}{*}{LSB}
    & Gaussian curvature constraint~\cite{yang2013linear}    & Small  & $\times$\\
    & Adaptive steganography~\cite{zhou2018distortion}    & Small  & $\surd$\\
    \midrule
    \multirow{6}{*}{Permutation}
    & Order encoding~\cite{bogomjakov2008distortion} & Null  & $\times$ \\
    & Enhanced order encoding~\cite{huang2009toward} & Null  & $\times$ \\
    & Binary tree~\cite{tu2010improved}     & Null  & $\times$ \\
    & Coding tree~\cite{tu2010permutation}  & Null & $\times$ \\
    & Maximum expected level tree~\cite{tu2012high} & Null  & $\times$ \\
    & One-ring neighborhood~\cite{wang2019breaking} & Null  & $\surd$ \\
    \midrule
    \multirow{2}{*}{Transform}
    & Mesh spectrum analysis~\cite{ohbuchi2001watermarking}    & Large  & $\times$ \\
    & Wavelet transform~\cite{kanai1998digital}    & Large  & $\times$ \\
    \bottomrule
  \end{tabular}
\end{table*}

When embedding and extracting a certain vertex, Tu et al.~\cite{tu2012high} established a maximum expected level tree for the remaining vertices in the reference order. At each level, the number of leaf vertices in the subtree is determined by the probability of the next message to be embedded. In this way, the distance between the root node and the leaf vertex can be as long as possible. Messages are represented by traversed paths. Given the message to be embedded, the message probability model of the 0-bit and 1-bit run-length histograms needs to be updated after each embedding or extraction. In the message extraction process, one needs to extract the message histogram from the stego model first and then extract the message from the histogram.

In summary, the capacity of this method is related to the run-length histograms of the embedded data, and it has a lower-bound capacity of $\frac{1}{V}\sum_{i=1}^{V}\left\lfloor\log_2{i}\right\rfloor+0.63$ bpv. The computational complexity of the method varies from $\mathcal{O}(n^2)$ to $\mathcal{O}(n\log{n})$ since it is directly correlative to the height of the constructed maximum expected level tree. Therefore, this method is analytically slower. In addition, this method is fragile against permutation-targeted steganalysis.

\subsubsection{One-ring-neighborhood-based method}
Previous permutation-based methods utilize the encoding capabilities of vertex and face lists to embed data. Although the vertex ordering is initially configured regularly and the order of the vertices is directly related to the surface normals, the permutation operation will destroy the correlation of the adjacent vertices and obfuscate the triangle normals.

To avoid bringing about global changes, Wang et al.~\cite{wang2019breaking} proposed to embed secret messages into the local neighbors of each vertex. In each embedding round, the 1-ring neighbor of the current vertex is utilized to carry the next few bits. In some cases, the vertices of the neighbor have already been utilized; thus, among all unused vertices, the method selects the vertex that appears first in the reference order and assigns an alias to it. Then, all the vertices not picked by $i$ are cascaded in clockwise order. According to the next $\left\lfloor\log_2{i} \right\rfloor$ bit value, the corresponding vertex is selected.

In summary, the embedding capacity of the method is much smaller than $\frac{1}{V}\sum_{i=1}^{V}\left\lfloor\log_2{i}\right\rfloor$ bpv, but its security level is high since it can withstand permutation-targeted steganalysis and universal steganalysis.

\captionsetup[table]{labelsep=space}
\begin{table*}[htbp]
\centering\caption{\label{tab:02-2}\\
Comparison of steganographic methods in terms of robustness. ``$\surd$'' indicates that the steganographic method is robust against the specified attack (affine transform, vertex reordering, noise addition, smoothing and simplification), and ``$\times$'' indicates that the method is fragile to the attack. }
 \centering
 \newcommand{\tabincell}[2]{\begin{tabular}{@{}#1@{}}#2\end{tabular}}
   \begin{tabular}{c|c|ccccc}
    \toprule
    \tabincell{c}{Category\\(domain)} & Method & Affine transform & Vertex reordering & Noise addition & Smoothing & Simplification\\
    \midrule
    \multirow{6}{*}{Two-state}
    & MEP~\cite{cayre2003data}     & $\surd$  & $\surd$  & $\times$ & $\times$ & $\times$ \\
    & MLEP~\cite{wang2005efficient} & $\surd$  & $\surd$  & $\times$ & $\times$ & $\times$ \\
    & Point-sampled geometry~\cite{wang2006steganography}& $\surd$ & $\surd$ & $\times$ & $\times$ & $\times$ \\
    & Multilayer~\cite{chao2009high}      & $\surd$ & $\surd$ & $\times$ & $\times$ & $\times$ \\
    & Static arithmetic coding~\cite{itier2017high,li2017rethinking} & $\surd$ & $\surd$ & $\times$ & $\times$ & $\times$ \\
    & Statistical embedding~\cite{cho2006oblivious,bors2013optimized}   & $\surd$ & $\surd$ & $\surd$ & $\surd$ & $\surd$ \\
    \midrule
    \multirow{2}{*}{LSB}
    & Gaussian curvature constraint~\cite{yang2013linear}    & $\times$ & $\surd$ & $\times$ & $\times$ & $\times$ \\
    & Adaptive steganography~\cite{zhou2018distortion}    & $\times$ & $\times$ & $\times$ & $\times$ & $\times$ \\
    \midrule
    \multirow{6}{*}{Permutation}
    & Order encoding~\cite{bogomjakov2008distortion} & $\surd$ & $\times$ & $\surd$ & $\surd$ & $\times$ \\
    & Enhanced order encoding~\cite{huang2009toward} & $\surd$ & $\times$ & $\surd$ & $\surd$ & $\times$ \\
    & Binary tree~\cite{tu2010improved}     & $\surd$ & $\times$ & $\surd$ & $\surd$ & $\times$ \\
    & Coding tree~\cite{tu2010permutation}  & $\surd$ & $\times$ & $\surd$ & $\surd$ & $\times$ \\
    & Maximum expected level tree~\cite{tu2012high} & $\surd$ & $\times$ & $\surd$ & $\surd$ & $\times$ \\
    & One-ring neighborhood~\cite{wang2019breaking} & $\surd$ & $\times$ & $\surd$ & $\surd$ & $\times$ \\
    \midrule
    \multirow{2}{*}{Transform}
    & Mesh spectrum analysis~\cite{ohbuchi2001watermarking}    & $\surd$  & $\surd$ & $\surd$ & $\surd$ & $\times$ \\
    & Wavelet transform~\cite{kanai1998digital}    & $\surd$  & $\surd$ & $\times$ & $\times$ & $\times$ \\
    \bottomrule
  \end{tabular}
\end{table*}

\subsection{Transform Domain}
Many transform-domain-based methods~\cite{kanai1998digital,praun1999robust,ohbuchi2001watermarking,wang2011robust,zafeiriou2005blind,wang2008new} are watermarking-based algorithms. In some steganographic applications, to avoid varying attacks in the lossy communication channels, robustness is a necessary property. Therefore, we summarize the transform-domain-based methods below.

\subsubsection{Mesh-spectrum-based method}
Ohbuchi~\cite{ohbuchi2001watermarking} proposed a spectral watermarking algorithm based on mesh spectral analysis. The mesh spectrum is obtained from a Laplacian matrix derived from connectivity of a  3-D mesh. The watermarking method embeds data into the mesh shape by modifying its mesh spectral coefficients. An inverse transformation converts the watermarked spectral coefficients back into the original mesh whose vertex coordinates are slightly altered. Since only the low-frequency end of the spectrum is modulated, the watermark is less perceptible and the watermarked mesh can become resilient against attacks including similarity transformation, random noise addition and smoothing.

\subsubsection{Wavelet-transform-based method}

\captionsetup{font={footnotesize}}
\begin{figure}
\begin{center}
  \includegraphics[height=1.20in]{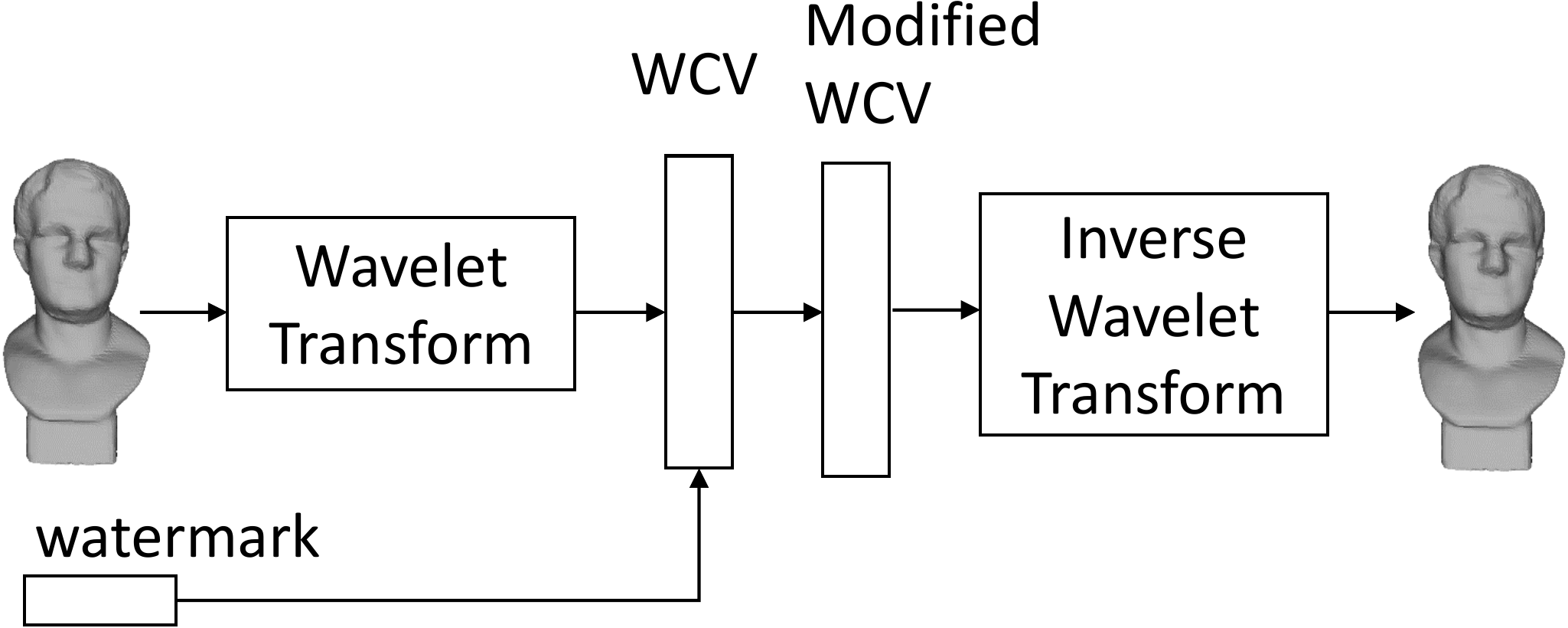}
  \caption{Outline of the embedding process of the wavelet-based watermarking method. WCV represents the wavelet coefficient vector. }
  \label{fig:04}
\end{center}
\end{figure}

Kanai et al.~\cite{kanai1998digital} proposed a nonblind watermarking algorithm, which takes advantage of the wavelet transform and the multiresolution representation of the mesh. As shown in Fig.~\ref{fig:04}, watermarks are embedded in the vectors of several large wavelet coefficients acquired in different resolution levels, which makes the embedded watermarks imperceptible and invariant to affine transformation. This also makes it reliable for controlling geometric errors caused by the watermarking.

In summary, the method has neither a large enough capacity nor computational efficiency. In addition, the method is a nonblind watermarking method, which requires the original mesh to detect the watermark. Because of the large modification, it cannot resist steganalysis.

\tikzset{
  basic/.style  = {draw, text width=2.5cm, drop shadow, font=\sffamily, rectangle},
  root/.style   = {basic, rounded corners=2pt, thin, align=center, fill=white, text width=4.0cm},
  level-1/.style = {basic, rounded corners=4pt, thin, align=center, fill=white, text width=2.3cm},
  level-2/.style = {basic, rounded corners=6pt, thin, align=center, fill=white, text width=2.3cm},
  level-3/.style = {basic, thin, align=center, fill=white, text width=1.7cm},
  level-4/.style = {basic, thin, align=center, fill=white, text width=1.9cm},
  level-5/.style = {basic, thin, align=center, fill=white, text width=1.1cm},
  level-6/.style = {basic, thin, align=center, fill=white, text width=2.0cm},
  level-7/.style = {basic, thin, align=center, fill=white, text width=2.1cm},
}

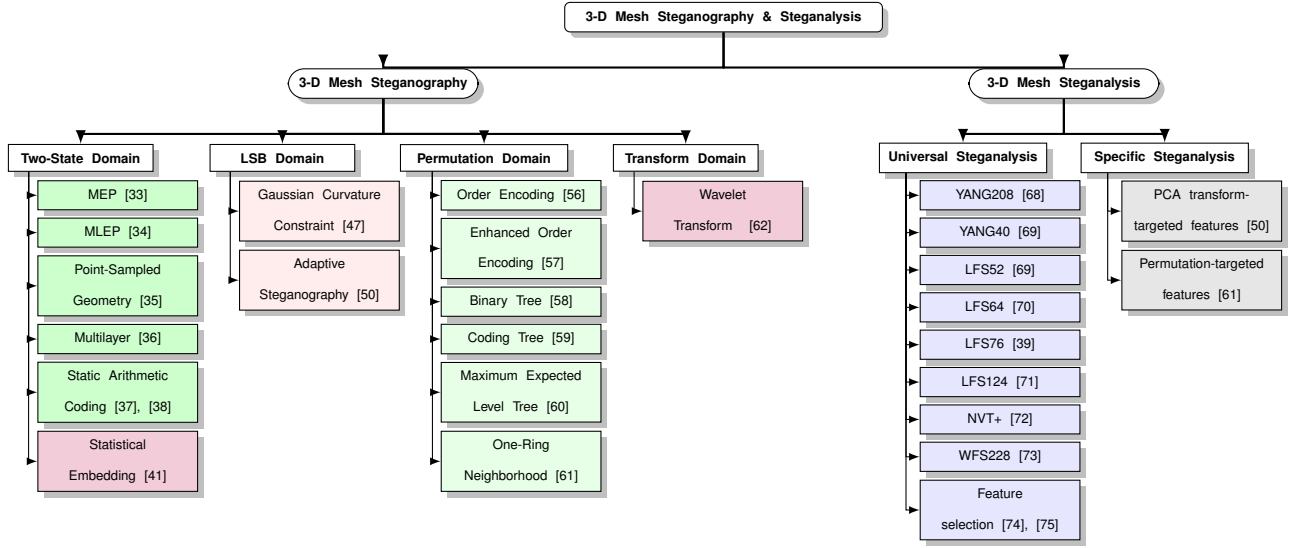
\begin{figure*}
    \centering
\begin{tikzpicture}[
  level 1/.style={sibling distance=27em, level distance=4em},
  level 2/.style={sibling distance=8em, level distance=4em},
  edge from parent/.style={->,solid,black,thick,sloped,draw},
  edge from parent path={(\tikzparentnode.south) -- (\tikzchildnode.north)},
  >=latex, node distance=1.2cm, edge from parent fork down]

\node[root] {\tiny\textbf{3-D Mesh Steganography \& Steganalysis}}
  child {node[level-2, yshift=13pt] (c1) {\tiny\textbf{3-D Mesh Steganography}}
          child {node[level-3, yshift=10pt] (c11) {\tiny\textbf{Two-State Domain}}}
          child {node[level-3, yshift=10pt] (c12) {\tiny\textbf{LSB Domain}}}
          child {node[level-6, yshift=10pt] (c13) {\tiny\textbf{Permutation Domain}}}
          child {node[level-3, yshift=10pt] (c14) {\tiny\textbf{Transform Domain}}}}
  child { node[level-2, yshift=13pt] (c2) {\tiny\textbf{3-D Mesh Steganalysis}}
          child {node[level-6, yshift=10pt] (c21) {\tiny\textbf{Universal Steganalysis}}}
          child {node[level-6, yshift=10pt] (c22) {\tiny\textbf{Specific Steganalysis}}}};

\begin{scope}[every node/.style={level-4}]
\node [below of = c21, xshift=14pt, yshift=20pt, fill=blue!10] (c211) {\tiny YANG208~\cite{yang2014mesh}};
\node [below of = c211, yshift=20pt, fill=blue!10] (c212) {\tiny YANG40~\cite{li20163d}};
\node [below of = c212, yshift=20pt, fill=blue!10] (c213) {\tiny LFS52~\cite{li20163d}};
\node [below of = c213, yshift=20pt, fill=blue!10] (c214) {\tiny LFS64~\cite{kim2017improved}};
\node [below of = c214, yshift=20pt, fill=blue!10] (c215) {\tiny LFS76~\cite{li2017steganalysis}};
\node [below of = c215, yshift=20pt, fill=blue!10] (c216) {\tiny LFS124~\cite{li20183d}};
\node [below of = c216, yshift=20pt, fill=blue!10] (c217) {\tiny NVT+~\cite{zhou2019feature}};
\node [below of = c217, yshift=20pt, fill=blue!10] (c218) {\tiny WFS228~\cite{li2020steganalysis}};
\node [below of = c218, yshift=14pt, fill=blue!10] (c219) {\tiny Feature selection~\cite{li2016selection,li2018selection}};

\end{scope}

\begin{scope}[every node/.style={level-4}]

\node [below of = c11, xshift=14pt, yshift=20pt, fill=green!20] (c111) {\tiny MEP~\cite{cayre2003data}};
\node [below of = c111, yshift=20pt, fill=green!20] (c112) {\tiny MLEP~\cite{wang2005efficient}};
\node [below of = c112, yshift=14pt, fill=green!20] (c113) {\tiny Point-Sampled Geometry~\cite{wang2006steganography}};
\node [below of = c113, yshift=14pt, fill=green!20] (c114) {\tiny Multilayer~\cite{chao2009high}};
\node [below of = c114, yshift=14pt, fill=green!20] (c115) {\tiny Static Arithmetic Coding~\cite{itier2017high,li2017rethinking}};
\node [below of = c115, yshift=8pt, fill=purple!20] (c116) {\tiny Statistical Embedding~\cite{bors2013optimized}};

\node [below of = c12, xshift=14pt, yshift=14pt, fill=pink!30] (c121) {\tiny Gaussian Curvature Constraint~\cite{yang2013linear}};
\node [below of = c121, yshift=8pt, fill=pink!30] (c122) {\tiny Adaptive Steganography~\cite{zhou2018distortion}};

\node [below of = c13, xshift=14pt, yshift=20pt, fill=green!10] (c131) {\tiny Order Encoding~\cite{bogomjakov2008distortion}};
\node [below of = c131, yshift=14pt, fill=green!10] (c132) {\tiny Enhanced Order Encoding~\cite{huang2009toward}};
\node [below of = c132, yshift=14pt, fill=green!10] (c133) {\tiny Binary Tree~\cite{tu2010improved}};
\node [below of = c133, yshift=20pt, fill=green!10] (c134) {\tiny Coding Tree~\cite{tu2010permutation}};
\node [below of = c134, yshift=14pt, fill=green!10] (c135) {\tiny Maximum Expected Level Tree~\cite{tu2012high}};
\node [below of = c135, yshift=8pt, fill=green!10] (c136) {\tiny One-Ring Neighborhood~\cite{wang2019breaking}};

\node [below of = c14, xshift=14pt, yshift=14pt, fill=purple!20] (c141) {\tiny Wavelet Transform ~\cite{kanai1998digital}};

\node [below of = c22, xshift=14pt, yshift=14pt, fill=black!10] (c221) {\tiny PCA transform-targeted features~\cite{zhou2018distortion}};
\node [below of = c221, yshift=8pt, fill=black!10] (c222) {\tiny Permutation-targeted features~\cite{wang2019breaking}};
\end{scope}

\foreach \value in {1,...,6}
  \draw[->] (c11.195) |- (c11\value.west);

\foreach \value in {1,2}
  \draw[->] (c12.195) |- (c12\value.west);

\foreach \value in {1,...,6}
  \draw[->] (c13.195) |- (c13\value.west);

\foreach \value in {1}
  \draw[->] (c14.195) |- (c14\value.west);

\foreach \value in {1,...,9}
  \draw[->] (c21.195) |- (c21\value.west);

\foreach \value in {1,2}
  \draw[->] (c22.195) |- (c22\value.west);

\end{tikzpicture}
    \caption{Our proposed taxonomy for 3-D mesh steganography and steganalysis.}
    \label{fig:my_label}
\end{figure*}

\subsection{Summary}

Table~\ref{tab:02} presents and compares the previously discussed steganographic methods in terms of distortion and security. Specifically, when the steganographic method is secure against all steganalysis methods, then we call it secure; otherwise, the method is not secure. 
Capacities are not compared, because some methods have tight upper bounds, some have approximated upper bounds and some do not have upper bounds. For two-state domain and LSB domain, most of the methods have tight upper bounds, which imply that they cannot embed more than the upper bounds. For permutation domain, the existing upper-bound capacities are the estimated ones, which can be approximated but can be barely reached, thus they are not tight. 
Table~\ref{tab:02-2} presents the robustness of different steganographic methods against various digital attacks such as affine transformation, vertex reordering, noise addition, smoothing and simplification.
The steganographic papers we have reviewed are classified in the taxonomy shown in Fig.~\ref{fig:my_label}.

\section{3-D Mesh Steganalysis Techniques}\label{4}

In this section, we divide the steganalysis techniques into two categories: universal steganalysis and specific steganalysis.
Universal blind steganalysis can detect embedded messages independent of steganographic algorithms and is
more frequently used in practical applications. This technique is very important because it is flexible and can quickly adapt to new unknown steganographic methods~\cite{fridrich2002practical}.
Specific steganalysis is designed for a specified type of steganographic method.
The development history of each group is elaborated below.

\subsection{Universal Steganalysis}

\begin{figure}
\begin{center}
  \includegraphics[height=1.25in]{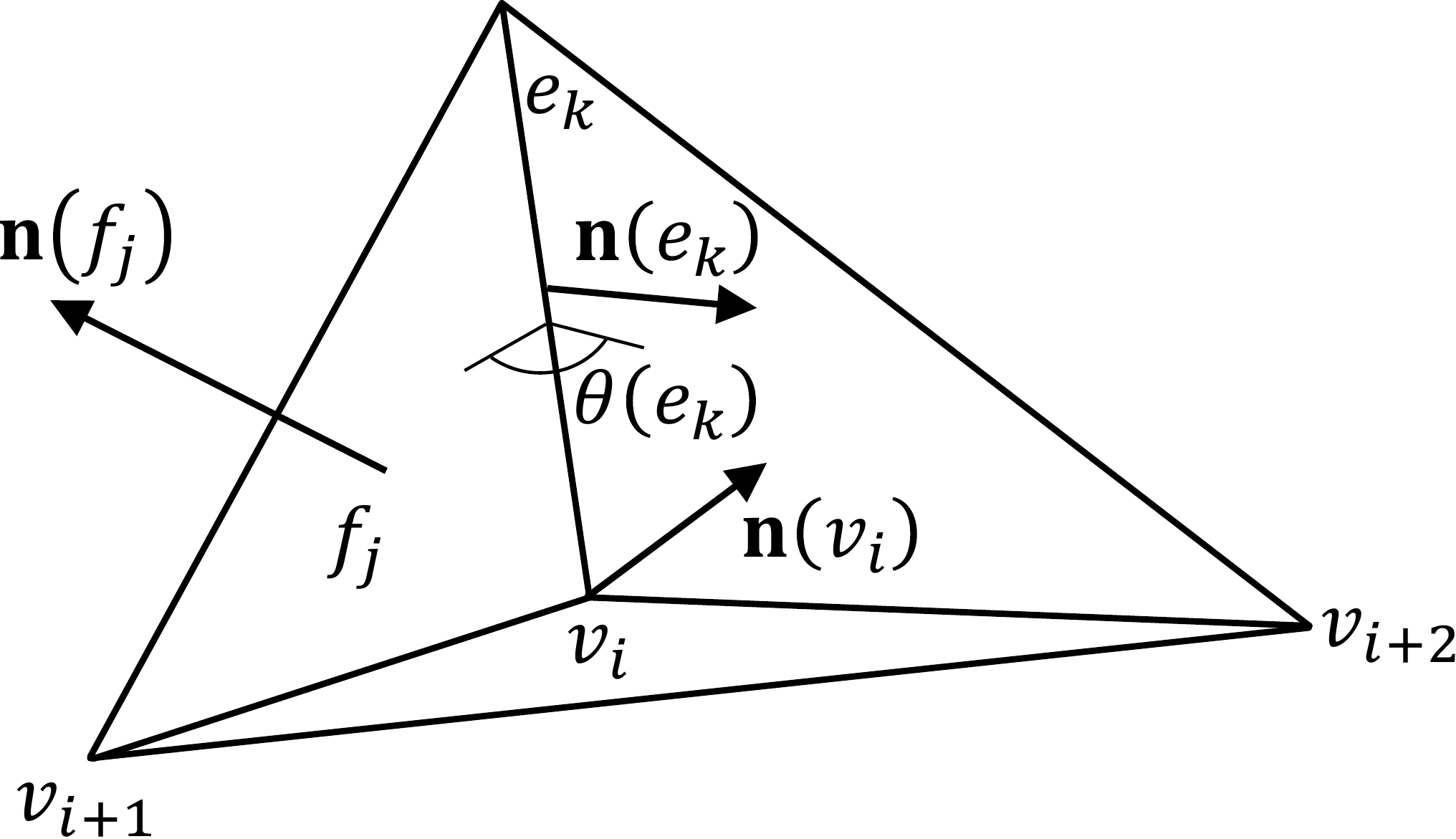}
  \caption{Notations of elements in a local region, which include vertices ($v_i$, $v_{i+1}$, $v_{i+2}$), edges ($e_k$), faces ($f_j$), normal vectors ($\mathbf{n}(v_i)$, $\mathbf{n}(f_j)$, $\mathbf{n}(e_k)$) and dihedral angles ($\theta(e_k)$). }
  \label{fig:08}
\end{center}
\end{figure}

Universal steganalysis aims to detect steganographic artifacts by designing features based on the differences between the mesh object and its smoothed object. The steganalysis performance is evaluated by machine learning classifiers~\cite{hearst1998support,kodovsky2012ensemble}.
In the following, we first provide a framework for 3-D mesh steganalysis and then introduce the existing 3-D mesh steganalysis algorithms according to the date of publication. In addition, the notations of elements (vertex, edge, face, normal, etc.) are illustrated in Fig.~\ref{fig:08}.

\subsubsection{Framework of universal steganalysis}

\captionsetup{font={footnotesize}}

\captionsetup{font={footnotesize}}
\begin{figure*}
\begin{center}
  \includegraphics[height=1.55in]{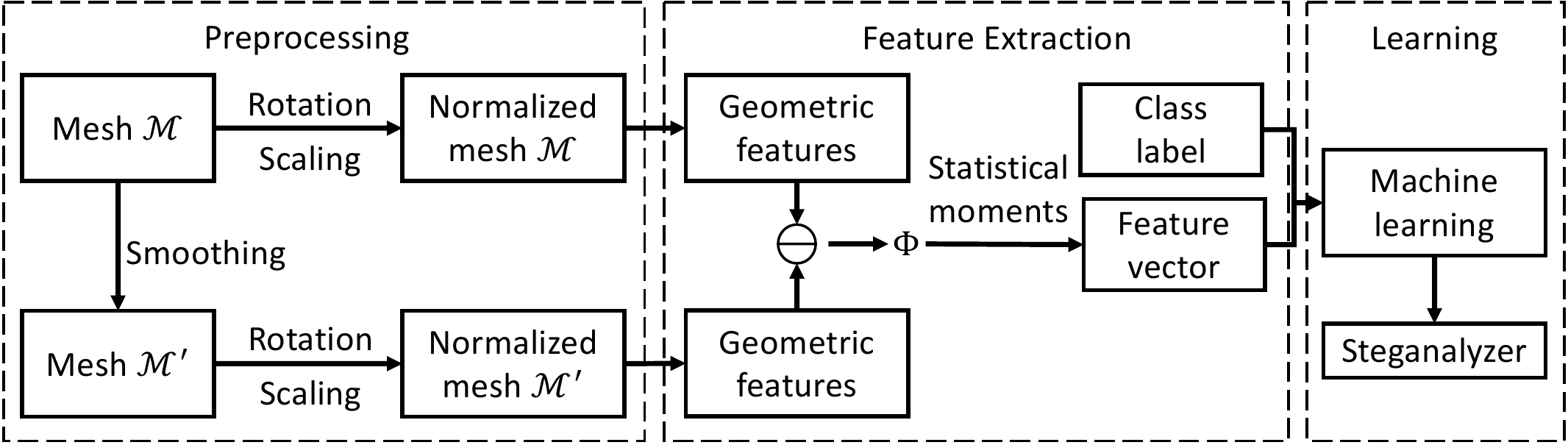}
  \caption{The 3-D mesh steganalysis framework based on learning from statistics of residual features and classification. Figure from~\cite{li20183dlevel}.}
  \label{fig:09}
\end{center}
\end{figure*}

Fig.~\ref{fig:09} is a flowchart of the universal 3-D mesh steganalysis. The framework is essentially based on the learning of residual feature statistics and classification, which includes calibrating the original mesh to a canonical version, Laplacian smoothing, extracting features, and mapping features. Before feature extraction, the vertices are preprocessed into a canonical version by rotating the object and aligning its coordinates with the three principal directions collected by PCA. Then, the object is scaled to fit inside a unit cube.

Motivated by the steganalysis of digital images, the difference between the stego image and its smoothed image is more significant than the difference between the cover and its smoothed image \cite{fridrich2002steganalysis,kodovsky2009calibration}; similarly, it is expected that the differences relating to 3-D steganalysis comply with the same rules. By applying a unified Laplacian smoothing process to the original mesh $\mathcal{M}$ for one iteration, we can obtain a smoothed mesh $\mathcal{M}'$, which moves the current vertex $\mathbf{p}_i$ to its one-ring average as follows \cite{taubin1995signal}:
\begin{equation}
\label{eqn:09}
\mathbf{p}_i\leftarrow\mathbf{p}_i+\frac{\tau}{\sum_{v_j\in \mathcal{N}_1(v_i)}w_{ij}}\sum_{v_j\in \mathcal{N}_1(v_i)}w_{ij}(\mathbf{p}_j-\mathbf{p}_i),
\end{equation}
where $\tau$ is a scalar term and $w_{ij}$ is a weighting term. Li et al.~\cite{li20183dlevel} analyzed the effect of $\tau$ on 3-D mesh smoothing and feature extraction.

Usually, ensemble classifiers~\cite{kodovsky2011ensemble} are trained for steganalysis. The core task of designing effective steganalyzers is feature design; hence, we introduce different features below.

\subsubsection{YANG208 features}
Yang and Ivrissimtzis~\cite{yang2014mesh} proposed the first 208-D steganalysis features of 3-D meshes.
The absolute values of the 3-D coordinate difference between $\mathcal{M}$ and $\mathcal{M}'$ are calculated, and then the vector length of each for each vertex vector is determined. For example, the feature of the $x$-component is:
\begin{equation}
\label{eqn:10}
\begin{aligned}
\phi_1(i)&=\left|x(v_i)-x(v_i')\right|.\\
\end{aligned}
\end{equation}

The Laplacian coordinates of $\mathcal{M}$ and $\mathcal{M}'$ are calculated as $\bar{\mathbf{p}}_i=[\bar{x}(v_i),\bar{y}(v_i),\bar{z}(v_i)]^T$ and $\bar{\mathbf{p}}_i'=[\bar{x}(v_i'),\bar{y}(v_i'),\bar{z}(v_i')]^T$, where they are the outcome of the Cartesian coordinates multiplied by the Kirchhoff matrix~\cite{bollobas2013modern} of the 3-D mesh. The other four vectors, including the three absolute values of the difference between each of the three coordinates components of $\mathcal{M}$ and $\mathcal{M}'$ and the $\ell_2$ norm of the coordinates, are calculated by the same absolute differences under Laplacian coordinates. The above computations are made separately on vertices whose valence is less than, equal to or greater than 6, excluding all boundary vertices, which make up 24 features.

Then, the absolute difference between the dihedral angles $\theta(e_i)$ of the adjacent faces in the vertical plane connected by the common edge $e_i$ is calculated:
\begin{equation}
\label{eqn:12}
\phi_2(i)=\left|\theta(e_i)-\theta(e_i')\right|.
\end{equation}

For mesh faces, the change in local surface direction is acquired by measuring the angle between the surface normal $\mathbf{n}(f_i)$ of the original object and the corresponding $\mathbf{n}(f_i)'$ of the smoothed 3-D mesh:
\begin{equation}
\label{eqn:13}
\phi_3(i)=\arccos\frac{\mathbf{n}(f_i)\cdot\mathbf{n}(f_i')}
{\left\|\mathbf{n}(f_i)\right\|\cdot\left\|\mathbf{n}(f_i')\right\|}.
\end{equation}

Based on each of the above 26 vectors, the components of the eight feature vectors are calculated, thereby acquiring a vector of 208 dimensions expressed as $\Phi_{208}$. Suppose $\phi$ is one of these 26 vectors and the first four components are constructed from the differences between the adjacent histogram bins of $\phi$. The remaining four components are the mean, variance, skewness, and kurtosis of logarithm $\log(|\phi|+\epsilon)$. 

In summary, though the computational cost of YANG208 is not large, the feature number is relatively large, and its discriminability is weaker than that of other features (see below), which implies that a few features in YANG208 are noneffective. Moreover, the classification accuracy on different payloads is inconsistent, which indicates that YANG208 is not robust enough.

\subsubsection{YANG40 features}
Li and Bors~\cite{li20163d} proposed YANG40, which consists of 40-D features, i.e., the most effective features in YANG208. The first 6 features are the absolute distance calculated along the 3-D axis between the vertex positions of $\mathcal{M}$ and $\mathcal{M}'$, considered in both the Cartesian and Laplacian coordinate systems.

Moreover, the two variations measured by the $\ell_2$ norm of the vertex vectors (from the mesh center to the vertex position) are calculated in the Cartesian and Laplacian coordinate systems, respectively.

$\phi_2$ and $\phi_3$ from YANG208 are regarded as two additional feature vectors. From each of the above 10 vectors, 4 components (mean, variance, skewness and kurtosis) are calculated. These components form a vector of dimension 40, which is represented as $\Phi_{40}$.

In summary, compared with YANG208, YANG40 does not calculate feature vectors separately based on vertex groups with different valence, thereby reducing the overall feature dimension while still maintaining good steganalysis performance.

\subsubsection{LFS52 Features}
Li and Bors~\cite{li20163d} proposed features based on the 52-D local feature set (LFS), including YANG40 and 12 local shape features. The first designed feature is calculated as the angle between the normal vectors of two specified vertices, in which the vertex normal vector is calculated as the weighted sum of the normal vectors of the triangles related to the vertex:
\begin{equation}
\label{eqn:15}
\mathbf{n}(v_i)=\frac{\sum_{f_j\in\mathcal{N}_1(\mathbf{v}_i)}{A(f_j)\mathbf{n}(f_j)}}
{\left\|\sum_{f_j\in\mathcal{N}_1(v_i)}{A(f_j)\mathbf{n}(f_j)}\right\|},
\end{equation}
where $\mathcal{N}_1(v_i)$ is the one-ring neighboring face of $v_i$ and $A(f_j)$ is the area of triangle $f_j$. Therefore, the absolute value of the angle between the two vertex normals $\mathbf{n}(v_i)$ and $\mathbf{n}(v'_i)$ is regarded as the designed feature.

The local shape curvature measures the surface smoothness of the 3-D mesh, where the principal curvatures $\kappa_1(v_i)$ and $\kappa_2(v_i)$ reflect the bending degree of the local surface in the orthogonal direction at vertex $v_i$.

The Gaussian curvature~\cite{rugis2006scale} is calculated as the product of the minimum and maximum principal curvatures:
$\kappa_G(v_i)=\kappa_1(v_i)\cdot\kappa_2(v_i)$. The second feature is calculated as the absolute difference between the two Gaussian curvatures $\kappa_G(v_i)$ and $\kappa_G(v'_i)$.

The curvature ratio is obtained by taking the ratio of the minimum to the maximum principal curvature:
\begin{equation}
\label{eqn:18}
\kappa_r(v_i)=\frac{\min{(\left|\kappa_1(v_i)\right|, \left|\kappa_2(v_i)\right|})}
{\max{(\left|\kappa_1(v_i)\right|, \left|\kappa_2(v_i)\right|})},
\end{equation}
and the corresponding feature is obtained by taking the absolute differences of the two curvature ratios $\kappa_r(v_i)$ and $\kappa_r(v'_i)$.

In summary, compared with the features based on coordinates and face normals, features containing vertex normals and curvatures own better discriminability.

\subsubsection{LFS64 features}
Kim et al.~\cite{kim2017improved} extended LFS52 and considered the edge normal vector, mean curvature and total curvature together, forming a 64-D feature vector. The edge normal vector $\mathbf{n}({e_i})$ is defined as the weighted sum of the triangle normal vectors connected by a common edge:
\begin{equation}
\label{eqn:20}
\mathbf{n}({e_i})=\frac{\sum_{f_j\in\mathcal{N}_1(e_i)}{A(f_j)\mathbf{n}(f_j)}}
{\left\|\sum_{f_j\in\mathcal{N}_1(e_i)}{A(f_j)\mathbf{n}(f_j)}\right\|},
\end{equation}
and the absolute value of the angles between the two edge normal vectors $\mathbf{n}(e_i)$ and $\mathbf{n}(e'_i)$ is calculated, which is regarded as the feature.

The mean curvature $\kappa_m(v_i)=(\kappa_1(v_i)+\kappa_2(v_i))/2$ and the total curvature $\kappa_t(v_i)=\left|\kappa_1(v_i)\right|+\left|\kappa_2(v_i)\right|$ contribute two additional features. The difference between $\kappa_m(v_i)$ and $\kappa_m(v'_i)$ and that between $\kappa_t(v_i)$ and $\kappa_t(v'_i)$ are regarded as new features.

In summary, the method outperforms LFS52 greatly because of its three features: edge normal vector, mean curvature and total curvature.

\subsubsection{LFS76 features}
Li and Bors~\cite{li2017steganalysis} extended LFS52 and proposed features based on the spherical coordinates $(R,\theta,\varphi)$, thus forming a 76-D feature vector. They provided a simple representation of the distance from the mesh center to each vertex position on the sphere, including the Euclidean norm $R$, azimuth $\theta$ and elevation $\varphi$ from a fixed origin.
The absolute difference between each coordinate and its corresponding smoothed coordinate is taken as the new feature.

The edge length in spherical coordinates is considered as the feature extraction element. $e_i$ is the edge that connects $v_{j}$ and $v_{k}$, and the edge vector $\mathbf{e}(e_i)=[\mathbf{p}_{j}, \mathbf{p}_{k}]$ (which will be used in the next subsection). For example, the $R$ component is acquired by
\begin{equation}
\label{eqn:24}
\phi_{4}(i)=\left|\left|R(v_{j})-R(v_{k})\right|-\left|R'(v_{j})-R'(v_{k})\right|\right|.\\
\end{equation}

In summary, in terms of steganalytic discriminability, the 6 new features do not improve the steganalysis performance.

\subsubsection{LFS124 features}
Li et al.~\cite{li20183d} extended LFS76, proposed an extended local feature set using edge vectors for steganalysis, and finally formed a 124-D feature vector.
First, in the Cartesian coordinate system, the absolute difference of the edge length of the 3-D component of the vector is calculated. For example:
\begin{equation}
\label{eqn:25}
\phi_{5}(i)=\left||x(v_{j})-x(v_{k})|-|x(v_{j}')-x(v_{k}')|\right|.\\
\end{equation}
Second, the difference norm between the two vectors $\mathbf{e}(e_i)$ and $\mathbf{e}(e_i)$ is calculated, and another two features made up of the absolute differences between them and the angle between them are obtained.

Based on the Laplacian coordinate system, another six features are calculated in the same manner, all of which constitute 12 features.

In summary, LFS124 is efficient in implementation and performs better than former steganalytic features including LFS52 and LFS76, indicating that the edge vector plays a vital role in steganalysis.

\subsubsection{NVT+ features}
Zhou et al.~\cite{zhou2019feature} proposed 100-D steganalytic features using a tensor voting model, which gathers the local shape context. First, three normal voting tensors (NVTs) based on each face or vertex are extracted.
The normal voting tensor $\mathbf{T}_i$ for vertex $v_i$ is defined as the sum of the weighted covariance matrix of its adjacent faces~\cite{sun2002triangle}:
\begin{equation}
\label{eqn:29}
\mathbf{T}_i=\sum_{f_j\in\mathcal{N}_1(v_i)}\mu_{ij}\mathbf{n}(f_j)\cdot\mathbf{n}(f_j)^T,
\end{equation}
where the weighting term $\mu_{ij}$ depends on the ratio of the area between neighboring faces to the distance between each triangle's barycenter $\mathbf{c}(f_j)$ and its vertex:
\begin{equation}
\label{eqn:30}
\mu_{ij} = \frac{A(f_j)}{\max{(A(\mathcal{N}_1(v_i))}}\exp{\left(-\frac{\left\|\mathbf{c}(f_j)-\mathbf{p}_i\right\|_2}{1/3}\right)}.
\end{equation}
In addition, the unit normals of the triangle $f_i$ define two other NVTs with different neighbors:
\begin{equation}
\label{eqn:31}
\mathbf{T}_i=\sum_{f_j\in \mathcal{N}(f_i)}\mu_{ij}\mathbf{n}(f_j)\cdot\mathbf{n}(f_j)^T.
\end{equation}

Second, three eigenvalues $(\lambda_1,\lambda_2,\lambda_3)$ are obtained from the eigendecomposition of each tensor:
\begin{equation}
\label{eqn:32}
\mathbf{T}=\lambda_1\mathbf{e}_1\mathbf{e}_1^T+\lambda_2\mathbf{e}_2\mathbf{e}_2^T+\lambda_3\mathbf{e}_3\mathbf{e}_3^T,
\end{equation}
where $(\mathbf{e}_1,\mathbf{e}_2,\mathbf{e}_3)$ are the corresponding eigenvectors.
Eigenvalues reflect the shape of patches on local surfaces, such as corners, sharp edges, or planes. Therefore, the absolute value of the difference between eigenvalues is considered as a feature. For example:
\begin{equation}
\label{eqn:33}
\phi_{6}(i)=\left|\lambda_1-\lambda_1'\right|.\\
\end{equation}

The NVT constitutes a total of $9\times 4=36$ features. By combining the NVT features and LFS64 features, the dimension of NVT+ can be as high as 100.

In summary, NVT+ offers an obvious improvement in classification accuracy, which indicates that the distribution of a local region's face normals is an effective indicator for detecting stego meshes. However, the calculation cost of this method is very high because it is very time consuming to calculate each feature of the adjacent face. 

\subsubsection{WFS228 features}
Li and Bors~\cite{li2020steganalysis} proposed using multiresolution 3-D wavelet analysis as a new set of 228-D steganalysis features.
The features are originally designed to detect messages embedded in watermarks based on the 3-D wavelet algorithm, and for most steganographic methods, they are effective for boosting steganalysis.

By using 3-D lazy wavelet decomposition~\cite{lounsbery1997multiresolution} and the Butterfly scheme~\cite{dyn1990butterfly}, wavelet coefficient vectors (WCVs) are used to associate the given mesh representation with different graph resolutions.
Because most of the wavelet-related embedding methods embed messages by modifying both the WCVs and edges obtained from the low-resolution version of the 3-D mesh, it is helpful to find and use these features in 3-D mesh steganalysis. In addition, the vertices of the high-resolution 3-D mesh are acquired by analyzing the larger vertex neighbors of the original 3-D mesh, which shows that the geometric features of the high-resolution 3-D mesh are more sensitive to changes in the original mesh.

In summary, based on these analyses, the authors proposed features based on the edge vector, WCVs and their variants at three resolutions for 3-D mesh steganalysis. Their method outperforms LFS76 by a large margin, yet its computational complexity is also very high because of the search for local vertices.

\subsubsection{Feature-selection-based method}
In real scenarios, training and testing sets are not from the same distributions. This is a challenging task for the existing steganalyzers, called the cover source mismatch (CSM) problem, which is caused by the limited generalizability of steganalyzers.

Li and Bors~\cite{li2016selection,li2018selection} proposed a feature selection method that takes the robustness and correlation of features into consideration to alleviate the mesh steganalysis CSM problem.
Specifically, to test a steganalytic method in the CSM scenario, Li and Bors first applied transformations including mesh simplification and noise addition to the original 3-D meshes and treated the transformed 3-D meshes as the cover meshes for steganography.
Then, by evaluating the effectiveness of separating the cover meshes from the stego meshes between the generated sets of objects, they selected the feature subset. Finally, they used the mutual information criterion and Pearson correlation coefficient to select the appropriate features.

In summary, this method deals with the cover source mismatch problem of 3-D steganalysis and provides several robust features. Its limitation is that the selection of features is restricted to transformations only. A promising improvement would be to experiment on a set of transformed objects
originating from completely different cover sources.

\subsection{Specific Steganalysis}
In this subsection, we discuss two specific steganalysis methods: PCA transform-targeted features and permutation-targeted features.

\subsubsection{PCA transform-targeted features}\label{pca}
The defect of steganography methods based on the PCA transform~\cite{wang2006steganography,chao2009high} is noted by Zhou et al.~\cite{zhou2018distortion}: the preprocessing procedures lead to a location distinction between the cover meshes and the stego meshes, which can be easily attacked by specially-designed detectors.

The vertices falling on the two ends of the first principal axis are taken as the end vertices $v_i$ and $v_j$. Similarly, the vertex falling on the farthest end of the second principal axis is taken as the third end vertex $v_k$. Then, the cover 3-D mesh is transformed to align the unit vectors $\overrightarrow{v_i v_j}$, $\overrightarrow{v_i v_k}$ and $\overrightarrow{v_i v_j}\times \overrightarrow{v_i v_k}$ with the $x$-axis, $y$-axis and $z$-axis, respectively. Therefore, the transformation matrix $\mathbf{T}$ is defined as
\begin{equation}
\label{eqn:34}
\mathbf{T}=[\overrightarrow{v_i v_j},\overrightarrow{v_i v_k},\overrightarrow{v_i v_j}\times \overrightarrow{v_i v_k}].
\end{equation}
Since the first and second principle axes of the stego mesh are near the $x$-axis and $y$-axis, respectively, this operation with behavior disorder will cause attackers to be suspicious, and the one-dimensional feature is calculated as the $\ell_1$ norm between the above two matrices:
\begin{equation}
\label{eqn:35}
\phi_{7}=\left\|\mathbf{T}-\mathbf{I}\right\|_1,
\end{equation}
where $\mathbf{I}$ is the identity matrix.

In summary, although this method is efficient, it is only effective in detecting PCA-transform-based steganographic methods.

\subsubsection{Permutation-targeted features}\label{permutation}
Wang et al.~\cite{wang2019breaking} proposed the first steganalytic method to break permutation steganography. They found that there are significant differences in the topological distance distribution of consecutive mesh elements between cover and stego meshes. They designed effective steganalytic features by measuring the order of the vertex and triangle lists.

As we all know, for clean meshes, the vertex lists are relatively orderly, while for stego meshes, they have high randomness. Therefore, they utilized the distance term $D(\mathcal{P})$ to measure the order of the vertex list $\mathcal{P}$:
\begin{equation}
\label{eqn:xae}
D(\mathcal{P})=\frac{1}{n-1}\sum_{i=1}^{n-1}d(\mathbf{p}_i,\mathbf{p}_{i+1}),
\end{equation}
where $d(\mathbf{p}_i,\mathbf{p}_{i+1})$ is the shortest Euclidean distance between $\mathbf{p}_i$ and $\mathbf{p}_{i+1}$. In most cases, $D(\mathcal{P})$ is very small for cover meshes, where consecutive vertices are close to each other, while for stego meshes, most consecutive vertices are not close to each other, and $D(\mathcal{P})$ may be very large.
The order of the face list is similarly designed.

In summary, this method is universal and does not require any prior knowledge, for instance, the concrete steganographic method and embedding payload.

\subsection{Summary}

\captionsetup[table]{labelsep=space}
\begin{table*}[htbp]
\centering\caption{\label{tab:03}\\
Basic feature elements of all steganalytic methods. Features are extracted by taking the absolute difference between two values or by taking the cosine distance between two vectors. }
 \centering
 \newcommand{\tabincell}[2]{\begin{tabular}{@{}#1@{}}#2\end{tabular}}
   \begin{tabular}{c|c|c|cccccccc}
    \toprule
    Index & Features & Dim & YANG208 & YANG40 & LFS52 & LFS64 & LFS76 & LFS124 & NVT+ & WFS228\\
    \midrule
    1 & \tabincell{c}{3-D coordinates and norm \\with different valence} & 24 & $\surd$ &   &  &  &  &  & &\\
    \midrule
    2 & \tabincell{c}{Face normals} & 1 & $\surd$ & $\surd$ & $\surd$ & $\surd$ & $\surd$ & $\surd$ & $\surd$ & \\
    \midrule
    3 & \tabincell{c}{Dihedral angles} & 1 & $\surd$ & $\surd$ & $\surd$ & $\surd$ & $\surd$ & $\surd$ & $\surd$ & \\
    \midrule
    4 & \tabincell{c}{3-D coordinates and norm} & 8 & & $\surd$  & $\surd$ & $\surd$ & $\surd$ & $\surd$ & $\surd$ & \\
    \midrule
    5 & Vertex normals& 1 & & & $\surd$ & $\surd$ & $\surd$ & $\surd$ & $\surd$ & \\
    \midrule
    6 & Gaussian curvature& 1 & & & $\surd$ & $\surd$ & $\surd$ & $\surd$ & $\surd$ & \\
    \midrule
    7 & Curvature ratio& 1 & & & $\surd$ & $\surd$ & $\surd$ & $\surd$ & $\surd$ & \\
    \midrule
    8 & Edge normal & 1 & &  &  & $\surd$ &  &  & $\surd$ & \\
    \midrule
    9 & Mean curvature& 1 & & &  & $\surd$ &  &  & $\surd$ & \\
    \midrule
    10 & Total curvature& 1 & & &  & $\surd$ &  &  & $\surd$ & \\
    \midrule
    11 & Spherical coordinates& 3 & &  &  &  & $\surd$ & $\surd$ & & \\
    \midrule
    12 & Edge angles of spherical coordinates& 3 & &  &  &  & $\surd$ & $\surd$ & & \\
    \midrule
    13 & \tabincell{c}{Edge vectors} & 12 & &  &  &  &  & $\surd$ & & $\surd$ \\
    \midrule
    14 & Eigenvalues of normal voting tensor & 9 & &  &  &  &  & & $\surd$ & \\
    \midrule
    15 & \tabincell{c}{Edge vectors and WCVs\\in multiresolutions} & 45 & &  &  &  &  &  &  & $\surd$ \\
    \bottomrule
  \end{tabular}
\end{table*}

We summarize all the features in Table~\ref{tab:03}.
The steganalysis papers we reviewed are classified into the taxonomy shown in Fig.~\ref{fig:my_label}.

\section{Experimental Results}\label{5}


\captionsetup{font={footnotesize}}
\begin{figure*}
\begin{minipage}[t]{0.5\linewidth}
\centering
\subfloat[The Princeton Segmentation Benchmark dataset.]{
  \label{fig:09a}
    \includegraphics[width=2.80in]{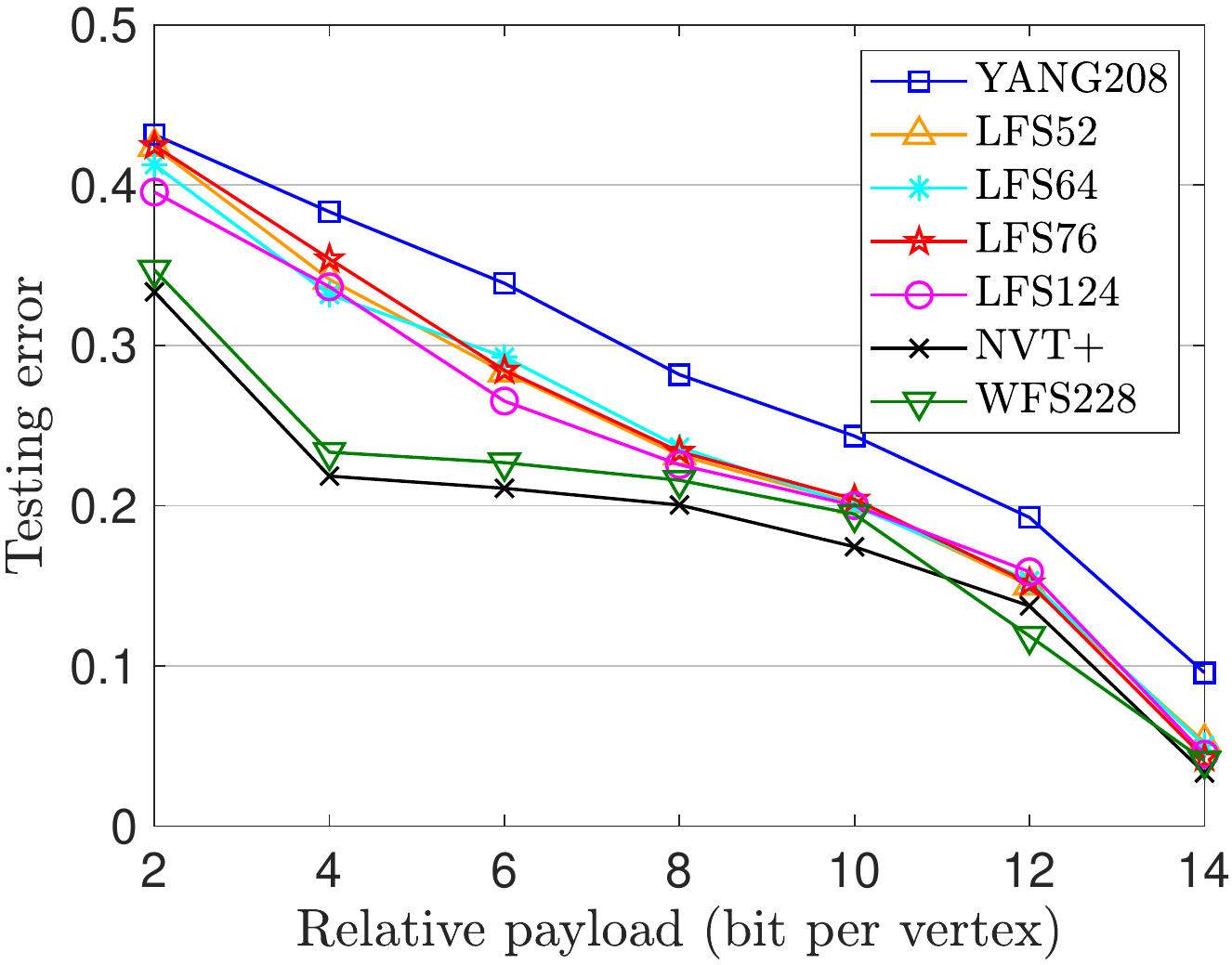}
    }
\end{minipage}%
\begin{minipage}[t]{0.5\linewidth}
\centering
\subfloat[The Princeton ModelNet dataset.]{
    \label{fig:09b}
    \includegraphics[width=2.80in]{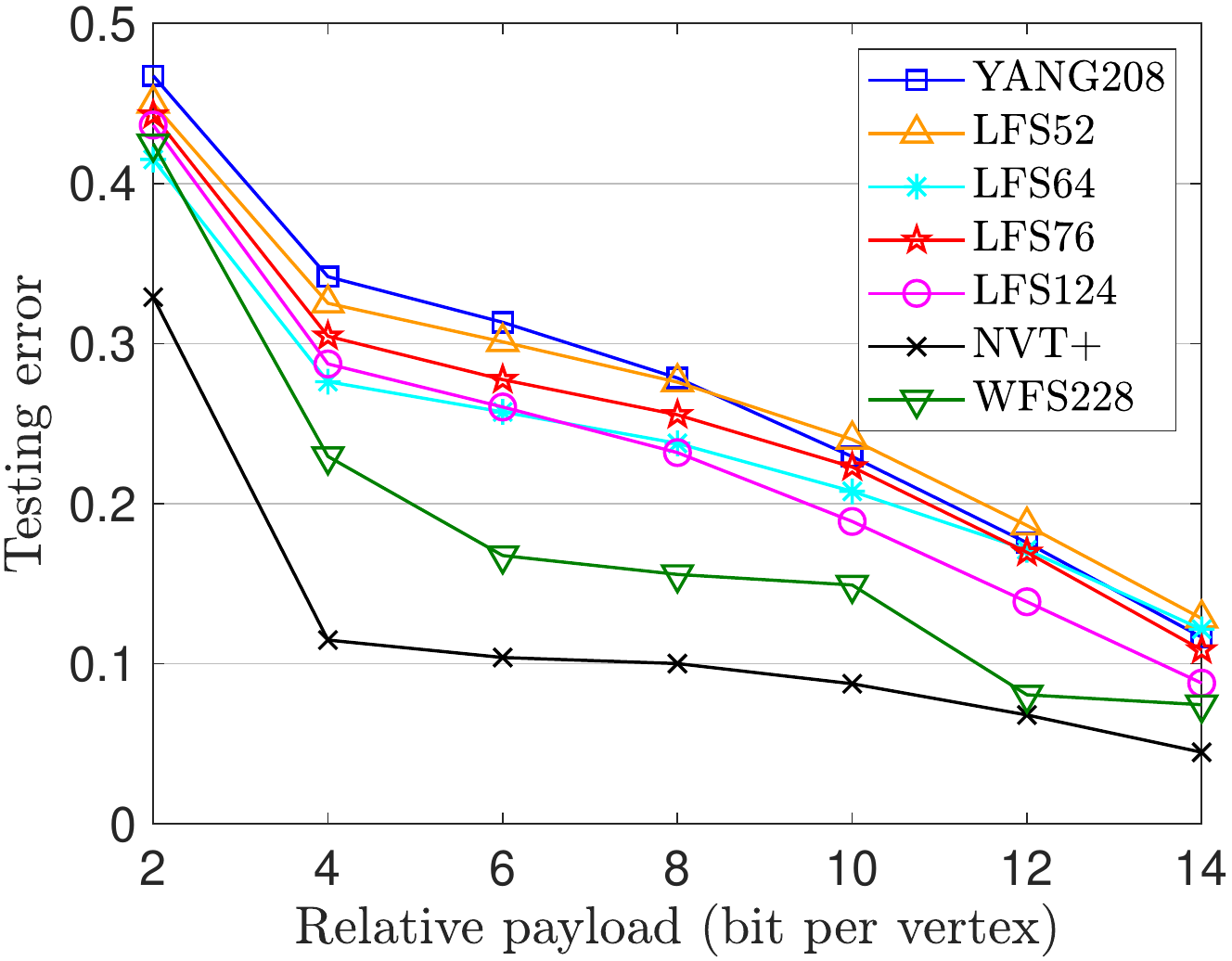}
    }
\end{minipage}
\\
\caption{3-D mesh steganalytic performance comparison.}
\label{fig:steganalysis}
\end{figure*}

\captionsetup{font={footnotesize}}
\begin{figure}
\centering
\subfloat[Testing error vs submodels.]{
  \label{fig:09a}
    \centering\includegraphics[width=3.45in]{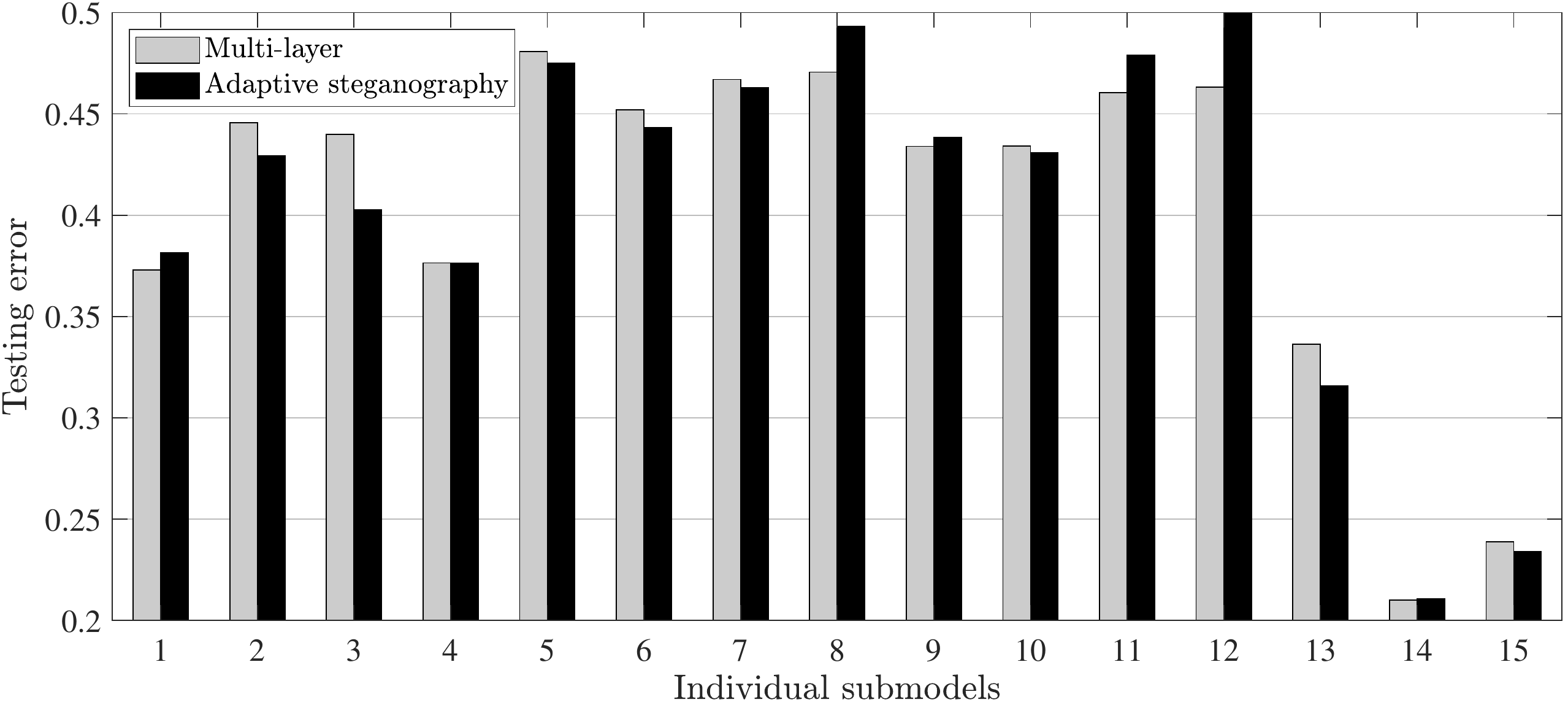}
    }
\\
\centering
\subfloat[Testing time vs submodels.]{
    \label{fig:09b}
    \centering\includegraphics[width=3.45in]{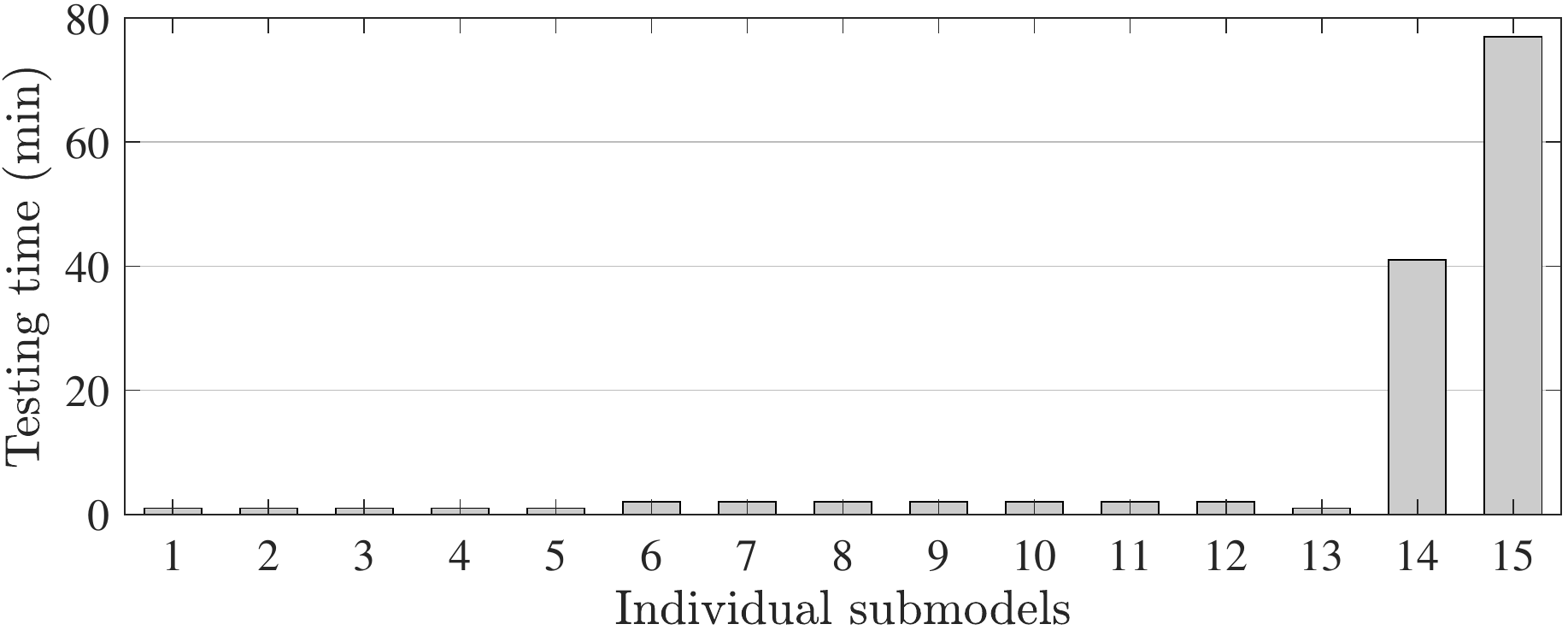}
    }
\\
\caption{(a) Average testing error estimates for the multilayer-based method and adaptive-steganography-based method, both of which are operated with 5 bpv embedding payloads on the Princeton Segmentation Benchmark dataset. (b) The corresponding testing time. }
\label{fig:submodel}
\end{figure}

\captionsetup{font={footnotesize}}
\begin{figure}
\begin{minipage}[t]{0.5\linewidth}
\centering
\subfloat[Testing error of LFS76~\cite{li2017steganalysis} when detecting the steganographic method~\cite{zhou2018distortion} under various payloads. ]{
    \centering\includegraphics[width=3.5in]{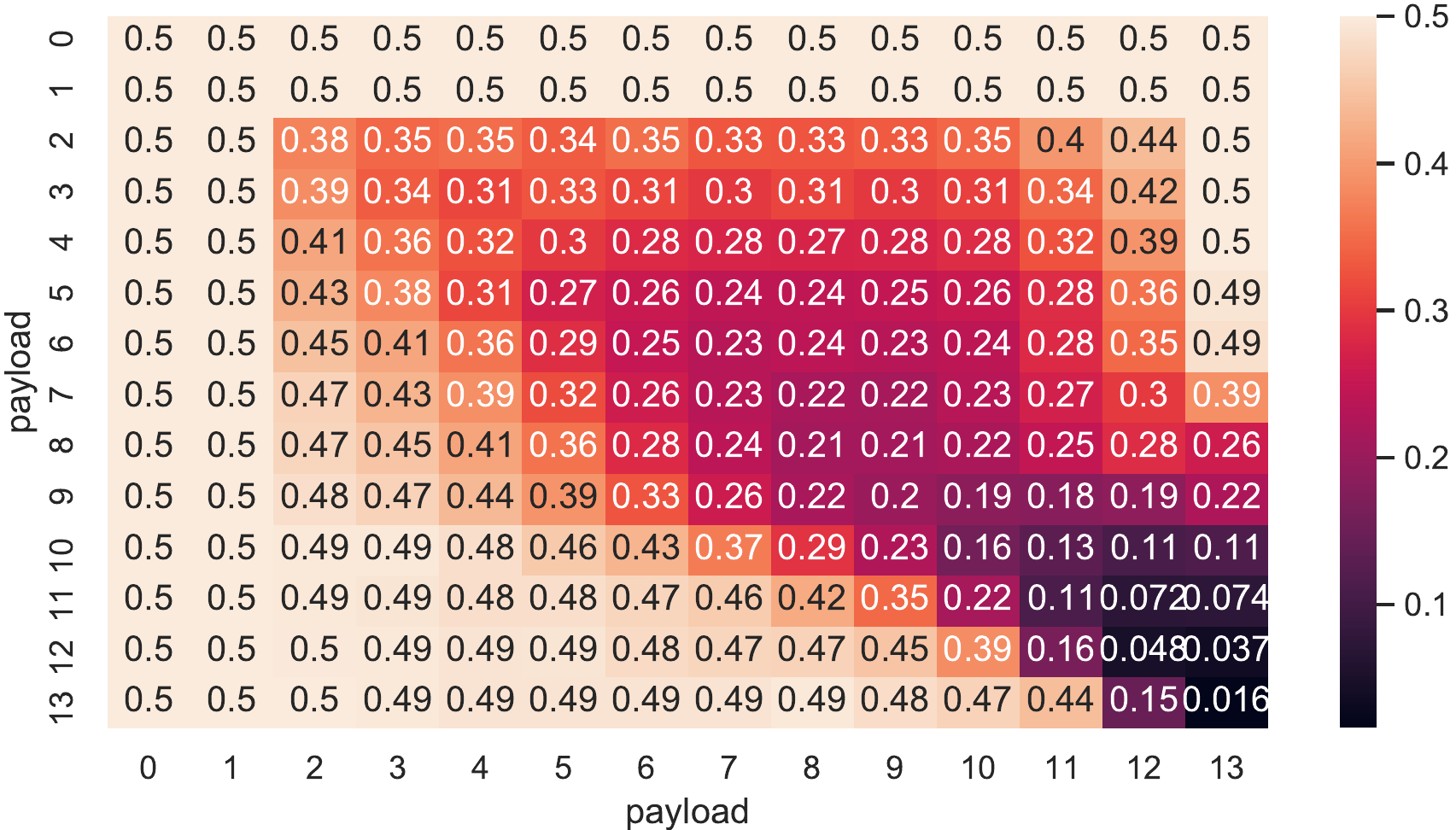}
    }
\end{minipage}
\\
\begin{minipage}[t]{0.5\linewidth}
\subfloat[Testing error of LFS76~\cite{li2017steganalysis} when detecting~\cite{zhou2018distortion} with a 5 bpv payload for two different datasets. ]{
    \centering\includegraphics[width=1.65in]{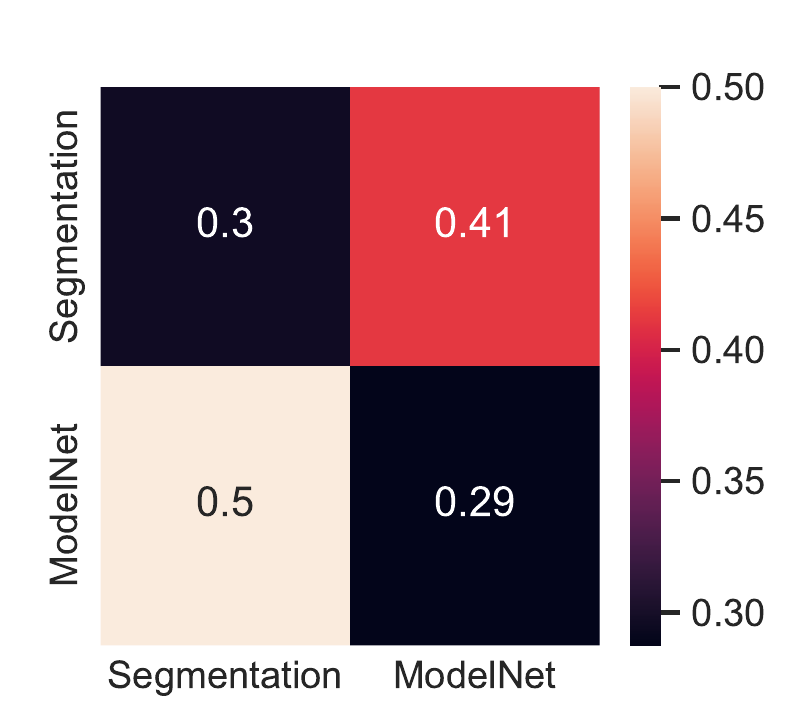}
    }
\end{minipage}
\begin{minipage}[t]{0.5\linewidth}
\subfloat[Testing error of NVT+~\cite{zhou2019feature} when detecting~\cite{zhou2018distortion} with a 5 bpv payload for two different datasets. ]{
    \centering\includegraphics[width=1.65in]{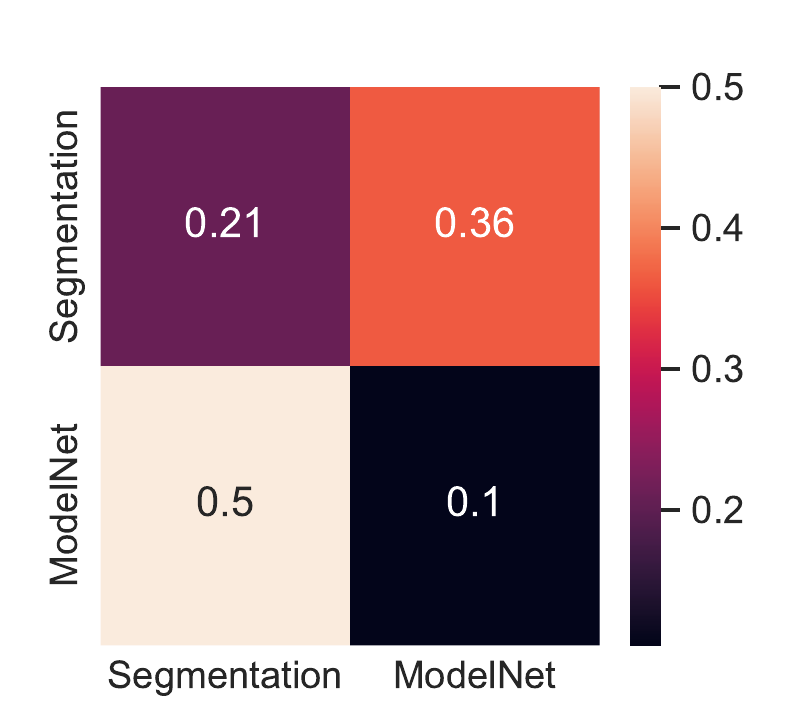}
    }
\end{minipage}
\\
\caption{The generalizability of trained steganalyzers in terms of different embedding payloads and datasets. The lower the testing error is, the stronger the robustness of the detector. }
\label{fig:robustness}
\end{figure}

\subsection{Datasets}

The \textbf{Princeton Segmentation Benchmark}~\footnotemark[1] is a mesh segmentation dataset with 354 objects \cite{chen2009benchmark} splitting into 260 training cover objects and 94 testing cover objects. Given a steganographic algorithm, we generate the corresponding stego object of each cover object. Finally, we have 260 pairs of cover objects and corresponding stego for training and 94 pairs for testing.

The \textbf{Princeton ModelNet}~\footnotemark[2] includes 12,311 mesh objects with 40 categories. 50\% of the mesh objects are taken as the training cover set (6,156), and the remaining are taken as the cover testing set (6,155). Similarly, each object has the corresponding stego object, forming the training set or the testing set. 

\footnotetext[1]{http://segeval.cs.princeton.edu/}
\footnotetext[2]{http://modelnet.cs.princeton.edu/}

\subsection{Comparison among the steganographic algorithms}
In this subsection, a comparison of the steganographic methods is given (see Table~\ref{tab:02} and Table~\ref{tab:02-2}).

First, from the distortion perspective, when embedding low payloads, the two-state domain and transform domain steganography methods modulate vertices by a large margin; thus, they have medium or large geometric distortions. The LSB domain methods first embed messages on the least significant bit; thus, they have low distortions. The permutation domain methods do not change vertex coordinates; thus, they have no distortions.

Second, from the capacity perspective, only the multilayer~\cite{chao2009high}, Gaussian curvature constraint~\cite{yang2013linear}, and adaptive-steganography-based methods have high capacity. Regarding the permutation-domain-based methods, except for the one-ring neighborhood method~\cite{wang2019breaking}, they have similar medium embedding capacities.

Third, from the security perspective, only the static arithmetic coding method~\cite{bors2013optimized}, adaptive steganography~\cite{zhou2018distortion} and the one-ring neighborhood method~\cite{wang2019breaking} can survive all steganalytic detection attempts.

Fourth, from the robustness perspective, the statistical-embedding-based methods~\cite{cho2006oblivious,bors2013optimized} are the most robust ones, as they can withstand all existing attacks. For permutation domain steganography, all are robust against affine transformation, noise addition, and smoothing but are fragile to vertex reordering and simplification. Most of the two-state domain methods can survive the affine transformation, vertex reordering, noise addition, and smoothing attacks but not the simplification attack. 


\subsection{Comparison among the steganalytic algorithms}

In this subsection, comparisons of the universal steganalytic methods are given (see Fig.~\ref{fig:steganalysis}, Fig.~\ref{fig:submodel} and Fig.~\ref{fig:robustness}).

First, to explore the classification performance of each steganalytic method, as shown in Fig.~\ref{fig:steganalysis}, we consider two classic high-capacity steganographic methods~\cite{chao2009high,zhou2018distortion} for generating the two stego mesh sets, respectively. As can be concluded from the two figures, the performance of steganalytic features follows the rule $P_\textrm{E}({\rm NVT+})<P_\textrm{E}({\rm WFS228})<P_\textrm{E}({\rm LFS124})\approx P_\textrm{E}({\rm LFS76})\approx P_\textrm{E}({\rm LFS64})\approx P_\textrm{E}({\rm LFS52})\approx P_\textrm{E}({\rm YANG208})$. In addition, the testing errors on the Princeton ModelNet dataset are lower than those on the Princeton Segmentation Benchmark dataset, which can be attributed to the data sources. The former are crafted by CAD techniques, while the latter are reconstructed from natural 3-D objects, which means that the former have less local complexity and can be easier for steganalyzers to model.

Second, to explore the effectiveness of each subfeature, we train each one grouped by the system of Table~\ref{tab:03} and implement steganalysis, as shown in Fig.~\ref{fig:submodel}~(a). We conclude that eigenvalues of the normal voting tensor are the most effective subfeatures, while edge vectors and WCVs are the second-best steganalytic features. In Fig.~\ref{fig:submodel}~(b), we compare the complexity of each testing procedure and conclude that the feature extraction complexity of both eigenvalues of the normal voting tensors and edge vectors with WCVs make them more time consuming than the other features.

Third, in Fig.~\ref{fig:robustness}, we analyze the generalizability of trained steganalyzers from the perspective of different embedding payloads and datasets to explore the robustness of the source models. Fig.~\ref{fig:robustness}~(a) shows a heatmap of the testing errors achieved by LFS76~\cite{li2017steganalysis} when detecting the steganographic method~\cite{zhou2018distortion} under different payloads. The detectors are trained with one of the cover-stego pairs given a targeted embedding payload listed in the rows and tested against one another listed in the columns. The more distant the payloads of the testing meshes from those of the training meshes are, the weaker the generalizability of the source models. Fig.~\ref{fig:robustness}~(b) and Fig.~\ref{fig:robustness}~(c) show the heatmaps of testing errors achieved by LFS76~\cite{li2017steganalysis} and NVT+~\cite{zhou2019feature} when detecting the steganographic method~\cite{zhou2018distortion} on different datasets, respectively. It can be concluded that the model trained on the Princeton Segmentation Benchmark has better robustness in detecting the Princeton ModelNet than does the reverse setup.

\section{Challenges and Trends}\label{6}
Below, we give several challenges and trends and present some potential solutions.
\subsection{Open Problems for 3-D Mesh Steganography}
Achieving higher steganographic security is the ultimate goal of 3-D mesh steganography. Below are two ideas toward stronger security.
\subsubsection{Combining the permutation domain and LSB domain }
One way to achieve stronger security is to combine the permutation domain and LSB domain by distributing message bits over the two domains. When the number of vertices of a mesh is 5000, the maximum embedding rate of permutation steganography reaches 11 bit per vertex, which can be considered a large embedding rate. The embedding capacity increases, yet it is inevitable to consider the universal steganalysis and permutation-targeted steganalysis together. The key point lies in devising an optimal message allocation scheme that achieves the optimal steganographic security under the same embedding rate.

\subsubsection{Designing spatial steganographic models}
Recent developments of adaptive steganography have verified that by designing nonadditive distortion functions, the steganographic security can be improved further by taking advantage of the mutual impact of modifications among local cover pixels~\cite{li2015strategy,denemark2015improving}. In the work of Zhou et al.~\cite{zhou2018distortion}, they allocate message bits evenly on the $xyz$-axes without considering the embedding effects on each other. It is suggested to design a joint distortion for a triple unit (a vertex consisting of three components), to utilize the DeJoin~\cite{zhang2016decomposing} scheme to allocate message bits and to implement STCs~\cite{filler2011minimizing} to embed data.

Additionally, inspired by the success of adversarial attacks in computer vision, adversarial steganography~\cite{DBLP:conf/ih/ZhangZCLLY18, DBLP:journals/tifs/TangLTBH19, DBLP:conf/ih/BernardPBK19} has been proposed based on generative adversarial networks (GANs) to deceive CNN-based image steganalyzers. It is vital to design adversarial steganography based on deep generative models to deceive mesh steganalyzers.

\subsubsection{Designing steganalysis-resistant permutation steganographic methods}
Wang et al.~\cite{wang2019breaking} proposed a neighborhood embedding scheme that utilizes the next $\left\lfloor\log_2{i} \right\rfloor$ bits in messages to select from the $i$ unpicked vertices of the 1-ring neighbor of the current vertex. It has been verified that $1\leq i\leq 11$ and, in most cases, $i=6$~\cite{jiang2017reversible}, and the average embedding rate is nearly 2 bit per vertex, which is too small for steganography. Moreover, rigorous experiments on steganalytic security are missing from the work of Wang et al.~\cite{wang2019breaking}, and it is expected to design a multiple-ring-neighbor-based permutation steganography to boost the embedding rate.


\subsubsection{Designing 3-D mesh batch steganography methods}
Batch steganography and pooled steganalysis~\cite{ker2006batch} have generalized the problems of data embedding and steganalysis to more than one object. It is speculated that, given images with uniform embedding capacity and a steganalytic scheme satisfying certain assumptions, ``secure'' steganographic capacity is proportional to the square root of the input image number~\cite{ker2007capacity}. It is therefore an interesting and challenging problem for researchers to explore the relationship between capacity and the number of vertices of 3-D meshes and to design a strategy to allocate messages among cover meshes.

\subsubsection{Designing 3-D-printing-material-based robust steganography methods}
Since DNA storage offers substantial information density and exceptional half-life, Koch et al.~\cite{koch2020dna} proposed a ``DNA-of-things'' (DoT) storage architecture to produce materials with immutable memory. They applied
the DoT to 3-D print a Stanford bunny that contained a 45 kB digital DNA blueprint for its synthesis. Specifically, they stored a 1.4 MB video in its DNA in the plexiglass spectacle lenses and retrieved it by excising a tiny piece of the plexiglass and sequencing the embedded DNA. The DoT could be applied to store electronic health records in medical implants, to hide data in everyday objects and to manufacture objects containing their own blueprint, which can be regarded as robust steganography since they can withstand physical-world attacks such as physical damage. However, the extraction side is expensive and time consuming. Therefore, how to design an efficient decoder is a future research direction.

\subsection{Open Problems for 3-D Mesh Steganalysis}

\subsubsection{Designing rich steganalytic features for universal blind steganalysis}

\captionsetup{font={footnotesize}}
\begin{figure}
\begin{center}
  \includegraphics[height=0.90in]{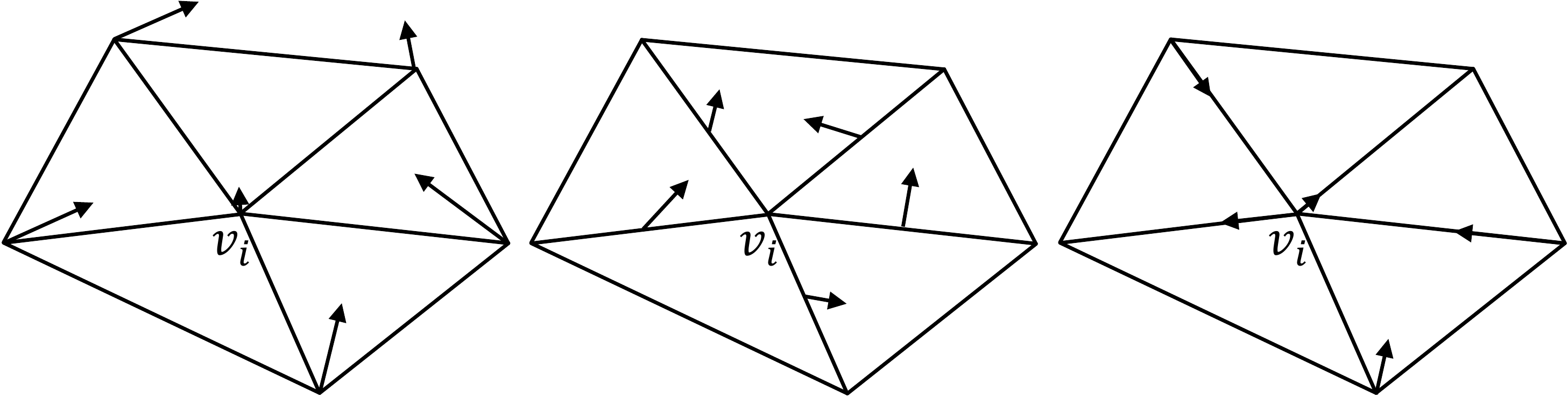}
  \caption{Diagram of local feature set for building tensors: (left) vertex norm, (middle) edge norm and (right) edge vector. }
  \label{fig:12}
\end{center}
\end{figure}

Since the steganalysis performance is poor under the low embedding rate of two-state domain and LSB domain steganography, there is much room for improvement in steganalysis. Inspired by the SRM of Fridrich and Kodovsk{\`y}~\cite{fridrich2012rich}, who designed 34,671-D quantized noise residual features acquired by multiple linear and nonlinear high-pass filters, it is possible to take advantage of more than one mesh smoothing technique, including denoising and fairing~\cite{botsch2010polygon}, to extract steganographic residuals. Advanced surface smoothing methods include anisotropic diffusion flow~\cite{bajaj2003anisotropic}, bilateral filtering~\cite{fleishman2003bilateral}, nonlinear smoothing~\cite{schneider2001geometric,eigensatz2008curvature} and neural-network-based filtering~\cite{zhao2019normalnet}.

In addition, it may be effective to design more features to boost steganalysis. For example, tensors are helpful feature extractors for a set of locally independent vectors. For instance, tensors based on the vertex normal, edge normal, and edge vector can also reflect the local smoothness, as shown in Fig.~\ref{fig:12}. Another option is to consider the $n$-ring-neighborhood-based features ($n>1$) to scale up areas for feature extraction.

Furthermore, the currently used statistical moments for dimensionality reduction may excessively discard informative features. We find that the dimension of features is equal to the number of vertices, edges or faces; thus, we believe that training discriminative models for each of the three and reasonably ensembling them may improve the steganalysis performance.

It is worth mentioning that, in addition to the above points, the runtime of the feature extraction of 3-D meshes is much greater than that of images. Engineering development of 3-D mesh steganalysis should also be considered, such as the use of parallel processing or advanced techniques to resolve the large time consumption of adjacent vertex searching.

\subsubsection{Designing deep-learning-based steganalysis methods}

The existing methods are all handcrafted features. Note that the designs of these features are cumbersome, and it has been recognized that deep learning has superior performance on classification tasks. These deep neural networks (DNNs), such as convolutional neural networks (CNNs)~\cite{krizhevsky2012imagenet}, graph convolutional networks (GCNs)~\cite{hamilton2017inductive} and MeshNet~\cite{feng2019meshnet}, are able to classify testing samples with high accuracy. In addition, CNN-based image steganalyzers such as XuNet~\cite{DBLP:journals/spl/XuWS16}, YeNet~\cite{DBLP:journals/tifs/YeNY17} and SRNet~\cite{DBLP:journals/tifs/BoroumandCF19} that use deep models to classify cover images from stego images are maturing fast. The intrinsic topological property of 3-D meshes conforms to GCNs and MeshNet; thus, one possible solution is to design an end-to-end modified MeshNet with steganalytic network structures and train it with a back-propagation algorithm to improve the 3-D mesh steganalytic performance.

\subsubsection{Designing a finer distance metric to improve the steganalysis of permutation steganography}
Aiming at the detection of permutation steganography, Wang et al.~\cite{wang2019breaking} proposed a theoretical analysis based on the correlation between consecutive mesh elements, but the research is preliminary and quantitative results were not given; thus, there is room for further research. We believe that the distance between two adjacent vertices from the vertex list $\mathcal{P}$ is inadequate for steganalysis and can be improved by calculating the Euclidean distance of adjacent vertices in the permuted order.

\subsubsection{Cover source mismatch problem}
As mentioned before, when a steganalyzer trained on one data source is applied to 3-D meshes from a different source, in general, the detection error will increase because of the mismatch between the two different sources, which is recognized as the CSM problem. In fact, CSM hinders the advancement of steganalysis from the laboratory environment to the real world. One possible solution is to use simple measures, such as by using several steganalyzers trained on several different sources and testing on a steganalyzer trained on the closest source or by increasing the training data diversity.

\section{Conclusions}\label{7}

Three-dimensional (3-D) mesh steganography is an interesting and promising research area, with potential practical applications such as covert communications. In this paper, we gave an overview of 3-D mesh steganography and steganalysis. First, we described our motivation for writing this paper and introduced the development history. Then, we outlined the principle framework of steganography and steganalysis, the evaluation metrics and the structure of 3-D meshes. Afterwards, we introduced the 3-D mesh steganography techniques in detail, including two-state domain, LSB domain, permutation domain and transform domain techniques. Next, the 3-D mesh steganalysis techniques, divided into two categories, were described in detail, i.e., universal steganalysis and specific steganalysis. After that, the experimental performances of these representative methods were compared. Finally, we discussed some valuable problems in the field and provided several interesting directions that may be worth exploring in the future.


\section*{Acknowledgment}

We would like to thank Ali Mahdavi-Amiri from Simon Fraser University and the anonymous reviewers, whose comments helped us improve the paper significantly. 

\bibliographystyle{IEEEtran}
\bibliography{sample-bibliography}

\begin{thebibliography}{100}
\providecommand{\url}[1]{#1}
\csname url@samestyle\endcsname
\providecommand{\newblock}{\relax}
\providecommand{\bibinfo}[2]{#2}
\providecommand{\BIBentrySTDinterwordspacing}{\spaceskip=0pt\relax}
\providecommand{\BIBentryALTinterwordstretchfactor}{4}
\providecommand{\BIBentryALTinterwordspacing}{\spaceskip=\fontdimen2\font plus
\BIBentryALTinterwordstretchfactor\fontdimen3\font minus
  \fontdimen4\font\relax}
\providecommand{\BIBforeignlanguage}[2]{{%
\expandafter\ifx\csname l@#1\endcsname\relax
\typeout{** WARNING: IEEEtran.bst: No hyphenation pattern has been}%
\typeout{** loaded for the language `#1'. Using the pattern for}%
\typeout{** the default language instead.}%
\else
\language=\csname l@#1\endcsname
\fi
#2}}
\providecommand{\BIBdecl}{\relax}
\BIBdecl

\bibitem{girdhar2017comprehensive}
A.~Girdhar and V.~Kumar, ``Comprehensive survey of 3d image steganography
  techniques,'' \emph{IET Image Processing}, vol.~12, no.~1, pp. 1--10, 2017.

\bibitem{steganography}
``Definition of steganography,''
  \url{https://www.merriam-webster.com/dictionary/steganography}, accessed
  February 28, 2020.

\bibitem{johnson1998exploring}
N.~F. Johnson and S.~Jajodia, ``Exploring steganography: Seeing the unseen,''
  \emph{Computer}, vol.~31, no.~2, pp. 26--34, 1998.

\bibitem{fridrich2004searching}
J.~Fridrich, M.~Goljan, and D.~Soukal, ``Searching for the stego-key,'' in
  \emph{Security, Steganography, and Watermarking of Multimedia Contents VI},
  vol. 5306.\hskip 1em plus 0.5em minus 0.4em\relax International Society for
  Optics and Photonics, 2004, pp. 70--82.

\bibitem{steganography2}
O.~Pahati, ``Confounding carnivore: How to protect your online privacy,''
  \url{https://web.archive.org/web/20070716093719/http://www.alternet.org/story/11986/},
  accessed February 28, 2020.

\bibitem{terror}
``Daily telegraph: `terror plot horror hidden in porn film',''
  \url{https://www.adelaidenow.com.au/news/world/terror-plot-horror-hidden-in-porn-film/news-story/49886a10ddeccdfb341cf72c48bbf5d5},
  2012, accessed February 28, 2020.

\bibitem{terror2}
G.~Kolata, ``Veiled messages of terror may lurk in cyberspace,''
  \url{http://www.nytimes.com/2001/10/30/science/veiled-messages-of-terror-may-lurk-in-cyberspace.html},
  2001, accessed February 28, 2020.

\bibitem{terror3}
``Bin laden: Steganography master?''
  \url{https://www.wired.com/2001/02/bin-laden-steganography-master/?currentPage=all},
  accessed February 28, 2020.

\bibitem{hussain2018image}
M.~Hussain, A.~W.~A. Wahab, Y.~I.~B. Idris, A.~T. Ho, and K.-H. Jung, ``Image
  steganography in spatial domain: A survey,'' \emph{Signal Processing: Image
  Communication}, vol.~65, pp. 46--66, 2018.

\bibitem{cheddad2010digital}
A.~Cheddad, J.~Condell, K.~Curran, and P.~Mc~Kevitt, ``Digital image
  steganography: Survey and analysis of current methods,'' \emph{Signal
  Processing}, vol.~90, no.~3, pp. 727--752, 2010.

\bibitem{chandramouli2003image}
R.~Chandramouli, M.~Kharrazi, and N.~Memon, ``Image steganography and
  steganalysis: Concepts and practice,'' in \emph{International Workshop on
  Digital Watermarking}.\hskip 1em plus 0.5em minus 0.4em\relax Springer, 2003,
  pp. 35--49.

\bibitem{johnson2001information}
N.~F. Johnson, Z.~Duric, and S.~Jajodia, \emph{Information hiding:
  steganography and watermarking-attacks and countermeasures: steganography and
  watermarking: attacks and countermeasures}.\hskip 1em plus 0.5em minus
  0.4em\relax Springer Science \& Business Media, 2001, vol.~1.

\bibitem{wayner2009disappearing}
P.~Wayner, \emph{Disappearing cryptography: information hiding: steganography
  and watermarking}.\hskip 1em plus 0.5em minus 0.4em\relax Morgan Kaufmann,
  2009.

\bibitem{fridrich2009steganography}
J.~Fridrich, \emph{Steganography in digital media: principles, algorithms, and
  applications}.\hskip 1em plus 0.5em minus 0.4em\relax Cambridge University
  Press, 2009.

\bibitem{cox2007digital}
I.~Cox, M.~Miller, J.~Bloom, J.~Fridrich, and T.~Kalker, \emph{Digital
  watermarking and steganography}.\hskip 1em plus 0.5em minus 0.4em\relax
  Morgan kaufmann, 2007.

\bibitem{luo2008review}
X.-Y. Luo, D.-S. Wang, P.~Wang, and F.-L. Liu, ``A review on blind detection
  for image steganography,'' \emph{Signal Processing}, vol.~88, no.~9, pp.
  2138--2157, 2008.

\bibitem{li2011survey}
B.~Li, J.~He, J.~Huang, and Y.~Q. Shi, ``A survey on image steganography and
  steganalysis,'' \emph{Journal of Information Hiding and Multimedia Signal
  Processing}, vol.~2, no.~2, pp. 142--172, 2011.

\bibitem{bhme2010advanced}
R.~Bhme, \emph{Advanced statistical steganalysis}.\hskip 1em plus 0.5em minus
  0.4em\relax Springer Publishing Company, Incorporated, 2010.

\bibitem{schaathun2012machine}
H.~G. Schaathun, \emph{Machine learning in image steganalysis}.\hskip 1em plus
  0.5em minus 0.4em\relax Wiley-IEEE Press, 2012.

\bibitem{frank2018digital}
J.~Frank and L.~Mensik, ``Digital construction kit: 3d computer graphics for
  creative and accessible use of museum collections,'' \emph{Biodiversity
  Information Science and Standards}, vol.~2, p. e26023, 2018.

\bibitem{wang2008comprehensive}
K.~Wang, G.~Lavou{\'e}, F.~Denis, and A.~Baskurt, ``A comprehensive survey on
  three-dimensional mesh watermarking,'' \emph{IEEE Transactions on
  Multimedia}, vol.~10, no.~8, pp. 1513--1527, 2008.

\bibitem{alface20073d}
P.~R. Alface and B.~Macq, ``From 3d mesh data hiding to 3d shape blind and
  robust watermarking: A survey,'' in \emph{Transactions on Data Hiding and
  Multimedia Security II}.\hskip 1em plus 0.5em minus 0.4em\relax Springer,
  2007, pp. 91--115.

\bibitem{nematollahi2017digital}
M.~A. Nematollahi, C.~Vorakulpipat, and H.~G. Rosales, \emph{Digital
  watermarking}.\hskip 1em plus 0.5em minus 0.4em\relax Springer, 2017.

\bibitem{botsch2010polygon}
M.~Botsch, L.~Kobbelt, M.~Pauly, P.~Alliez, and B.~L{\'e}vy, \emph{Polygon mesh
  processing}.\hskip 1em plus 0.5em minus 0.4em\relax CRC press, 2010.

\bibitem{simmons1984prisoners}
G.~J. Simmons, ``The prisoners' problem and the subliminal channel,'' in
  \emph{Advances in Cryptology}.\hskip 1em plus 0.5em minus 0.4em\relax
  Springer, 1984, pp. 51--67.

\bibitem{kerckhoffs1883cryptographic}
A.~Kerckhoffs, ``La cryptographic militaire,'' \emph{Journal Des Sciences
  Militaires}, pp. 5--38, 1883.

\bibitem{fawcett2004roc}
T.~Fawcett, ``Roc graphs: Notes and practical considerations for researchers,''
  \emph{Machine Learning}, vol.~31, no.~1, pp. 1--38, 2004.

\bibitem{kodovsky2011ensemble}
J.~Kodovsk{\`y}, J.~Fridrich, and V.~Holub, ``Ensemble classifiers for
  steganalysis of digital media,'' \emph{IEEE Transactions on Information
  Forensics and Security}, vol.~7, no.~2, pp. 432--444, 2011.

\bibitem{DBLP:journals/tifs/PevnyBF10}
T.~Pevn{\'{y}}, P.~Bas, and J.~J. Fridrich, ``Steganalysis by subtractive pixel
  adjacency matrix,'' \emph{IEEE Transactions on Information Forensics and
  Security}, vol.~5, no.~2, pp. 215--224, 2010.

\bibitem{fridrich2012rich}
J.~Fridrich and J.~Kodovsk{\`y}, ``Rich models for steganalysis of digital
  images,'' \emph{IEEE Transactions on Information Forensics and Security},
  vol.~7, no.~3, pp. 868--882, 2012.

\bibitem{DBLP:conf/imr/ShontzK08}
S.~M. Shontz and P.~M. Knupp, ``The effect of vertex reordering on 2d local
  mesh optimization efficiency,'' in \emph{Proceedings of the International
  Meshing Roundtable}.\hskip 1em plus 0.5em minus 0.4em\relax Springer, 2008,
  pp. 107--124.

\bibitem{DBLP:journals/ewc/SastryKSK14}
S.~P. Sastry, E.~Kultursay, S.~M. Shontz, and M.~T. Kandemir, ``Improved cache
  utilization and preconditioner efficiency through use of a space-filling
  curve mesh element- and vertex-reordering technique,'' \emph{Engineering with
  Computers}, vol.~30, no.~4, pp. 535--547, 2014.

\bibitem{cayre2003data}
F.~Cayre and B.~Macq, ``Data hiding on 3-d triangle meshes,'' \emph{IEEE
  Transactions on Signal Processing}, vol.~51, no.~4, pp. 939--949, 2003.

\bibitem{wang2005efficient}
C.-M. Wang and Y.-M. Cheng, ``An efficient information hiding algorithm for
  polygon models,'' in \emph{Computer Graphics Forum}, vol.~24, no.~3.\hskip
  1em plus 0.5em minus 0.4em\relax Wiley Online Library, 2005, pp. 591--600.

\bibitem{wang2006steganography}
C.-M. Wang and P.-C. Wang, ``Steganography on point-sampled geometry,''
  \emph{Computers \& Graphics}, vol.~30, no.~2, pp. 244--254, 2006.

\bibitem{chao2009high}
M.-W. Chao, C.-h. Lin, C.-W. Yu, and T.-Y. Lee, ``A high capacity 3d
  steganography algorithm,'' \emph{IEEE Transactions on Visualization and
  Computer Graphics}, vol.~15, no.~2, pp. 274--284, 2009.

\bibitem{itier2017high}
V.~Itier and W.~Puech, ``High capacity data hiding for 3d point clouds based on
  static arithmetic coding,'' \emph{Multimedia Tools and Applications},
  vol.~76, no.~24, pp. 26\,421--26\,445, 2017.

\bibitem{li2017rethinking}
Z.~Li, S.~Beugnon, W.~Puech, and A.~G. Bors, ``Rethinking the high capacity 3d
  steganography: Increasing its resistance to steganalysis,'' in \emph{IEEE
  International Conference on Image Processing}.\hskip 1em plus 0.5em minus
  0.4em\relax IEEE, 2017, pp. 510--514.

\bibitem{li2017steganalysis}
Z.~Li and A.~G. Bors, ``Steganalysis of 3d objects using statistics of local
  feature sets,'' \emph{Information Sciences}, vol. 415, pp. 85--99, 2017.

\bibitem{cho2006oblivious}
J.-W. Cho, R.~Prost, and H.-Y. Jung, ``An oblivious watermarking for 3-d
  polygonal meshes using distribution of vertex norms,'' \emph{IEEE
  Transactions on Signal Processing}, vol.~55, no.~1, pp. 142--155, 2006.

\bibitem{bors2013optimized}
A.~G. Bors and M.~Luo, ``Optimized 3d watermarking for minimal surface
  distortion,'' \emph{IEEE Transactions on Image Processing}, vol.~22, no.~5,
  pp. 1822--1835, 2013.

\bibitem{marquardt1963algorithm}
D.~W. Marquardt, ``An algorithm for least-squares estimation of nonlinear
  parameters,'' \emph{Journal of the Society for Industrial and Applied
  Mathematics}, vol.~11, no.~2, pp. 431--441, 1963.

\bibitem{levenberg1944method}
K.~Levenberg, ``A method for the solution of certain non-linear problems in
  least squares,'' \emph{Quarterly of Applied Mathematics}, vol.~2, no.~2, pp.
  164--168, 1944.

\bibitem{yang2014steganalytic}
Y.~Yang, R.~Pintus, H.~Rushmeier, and I.~Ivrissimtzis, ``A steganalytic
  algorithm for 3d polygonal meshes,'' in \emph{IEEE International Conference
  on Image Processing}.\hskip 1em plus 0.5em minus 0.4em\relax IEEE, 2014, pp.
  4782--4786.

\bibitem{yang20163d}
------, ``A 3d steganalytic algorithm and steganalysis-resistant
  watermarking,'' \emph{IEEE Transactions on Visualization and Computer
  Graphics}, vol.~23, no.~2, pp. 1002--1013, 2016.

\bibitem{yang2013information}
Y.~Yang, ``Information analysis for steganography and steganalysis in 3d
  polygonal meshes,'' Ph.D. dissertation, Durham University, 2013.

\bibitem{yang2013linear}
Y.~Yang, N.~Peyerimhoff, and I.~Ivrissimtzis, ``Linear correlations between
  spatial and normal noise in triangle meshes,'' \emph{IEEE Transactions on
  Visualization and Computer Graphics}, vol.~19, no.~1, pp. 45--55, 2013.

\bibitem{DBLP:journals/spl/Mielikainen06a}
J.~Mielik{\"{a}}inen, ``{LSB} matching revisited,'' \emph{{IEEE} Signal
  Processing Letters}, vol.~13, no.~5, pp. 285--287, 2006.

\bibitem{li2017high}
N.~Li, J.~Hu, R.~Sun, S.~Wang, and Z.~Luo, ``A high-capacity 3d steganography
  algorithm with adjustable distortion,'' \emph{IEEE Access}, vol.~5, pp.
  24\,457--24\,466, 2017.

\bibitem{zhou2018distortion}
H.~Zhou, K.~Chen, W.~Zhang, Y.~Yao, and N.~Yu, ``Distortion design for secure
  adaptive 3-d mesh steganography,'' \emph{IEEE Transactions on Multimedia},
  vol.~21, no.~6, pp. 1384--1398, 2018.

\bibitem{jiang2017reversible}
R.~Jiang, H.~Zhou, W.~Zhang, and N.~Yu, ``Reversible data hiding in encrypted
  three-dimensional mesh models,'' \emph{IEEE Transactions on Multimedia},
  vol.~20, no.~1, pp. 55--67, 2017.

\bibitem{filler2011minimizing}
T.~Filler, J.~Judas, and J.~Fridrich, ``Minimizing additive distortion in
  steganography using syndrome-trellis codes,'' \emph{IEEE Transactions on
  Information Forensics and Security}, vol.~6, no.~3, pp. 920--935, 2011.

\bibitem{li2020design}
W.~Li, W.~Zhang, L.~Li, H.~Zhou, and N.~Yu, ``Designing near-optimal
  steganographic codes in practice based on polar codes,'' \emph{IEEE
  Transactions on Communications}, 2020, doi: 10.1109/TCOMM.2020.2982624.

\bibitem{artz2001digital}
D.~Artz, ``Digital steganography: Hiding data within data,'' \emph{IEEE
  Internet Computing}, vol.~5, no.~3, pp. 75--80, 2001.

\bibitem{rossignac1999edgebreaker}
J.~Rossignac, ``Edgebreaker: Connectivity compression for triangle meshes,''
  \emph{IEEE Transactions on Visualization and Computer Graphics}, vol.~5,
  no.~1, pp. 47--61, 1999.

\bibitem{bogomjakov2008distortion}
A.~Bogomjakov, C.~Gotsman, and M.~Isenburg, ``Distortion-free steganography for
  polygonal meshes,'' in \emph{Computer Graphics Forum}, vol.~27, no.~2.\hskip
  1em plus 0.5em minus 0.4em\relax Wiley Online Library, 2008, pp. 637--642.

\bibitem{huang2009toward}
N.-C. Huang, M.-T. Li, and C.-M. Wang, ``Toward optimal embedding capacity for
  permutation steganography,'' \emph{IEEE Signal Processing Letters}, vol.~16,
  no.~9, pp. 802--805, 2009.

\bibitem{tu2010improved}
S.-C. Tu, W.-K. Tai, M.~Isenburg, and C.-C. Chang, ``An improved data hiding
  approach for polygon meshes,'' \emph{The Visual Computer}, vol.~26, no.~9,
  pp. 1177--1181, 2010.

\bibitem{tu2010permutation}
S.~Tu, H.~Hsu, and W.~Tai, ``Permutation steganography for polygonal meshes
  based on coding tree,'' \emph{International Journal of Virtual Reality},
  vol.~9, no.~4, pp. 55--60, 2010.

\bibitem{tu2012high}
S.-C. Tu and W.-K. Tai, ``A high-capacity data-hiding approach for polygonal
  meshes using maximum expected level tree,'' \emph{Computers \& Graphics},
  vol.~36, no.~6, pp. 767--775, 2012.

\bibitem{wang2019breaking}
Y.~Wang, L.~Kong, Z.~Qian, G.~Feng, X.~Zhang, and J.~Zheng, ``Breaking
  permutation-based mesh steganography and security improvement,'' \emph{IEEE
  Access}, vol.~7, pp. 183\,300--183\,310, 2019.

\bibitem{kanai1998digital}
S.~Kanai, H.~Date, T.~Kishinami \emph{et~al.}, ``Digital watermarking for 3d
  polygons using multiresolution wavelet decomposition,'' in
  \emph{International Workshop on Geometric Modeling: Fundamentals and
  Applicattion}, vol.~5, 1998, pp. 296--307.

\bibitem{praun1999robust}
E.~Praun, H.~Hoppe, and A.~Finkelstein, ``Robust mesh watermarking,'' in
  \emph{Conference on Computer Graphics and Interactive Techniques}, 1999, pp.
  49--56.

\bibitem{ohbuchi2001watermarking}
R.~Ohbuchi, S.~Takahashi, T.~Miyazawa, and A.~Mukaiyama, ``Watermarking 3d
  polygonal meshes in the mesh spectral domain,'' in \emph{Graphics interface},
  vol. 2001.\hskip 1em plus 0.5em minus 0.4em\relax Citeseer, 2001, pp. 9--17.

\bibitem{wang2011robust}
K.~Wang, G.~Lavou{\'e}, F.~Denis, and A.~Baskurt, ``Robust and blind mesh
  watermarking based on volume moments,'' \emph{Computers \& Graphics},
  vol.~35, no.~1, pp. 1--19, 2011.

\bibitem{zafeiriou2005blind}
S.~Zafeiriou, A.~Tefas, and I.~Pitas, ``Blind robust watermarking schemes for
  copyright protection of 3d mesh objects,'' \emph{IEEE Transactions on
  Visualization and Computer Graphics}, vol.~11, no.~5, pp. 596--607, 2005.

\bibitem{wang2008new}
Y.-P. Wang and S.-M. Hu, ``A new watermarking method for 3d models based on
  integral invariants,'' \emph{IEEE Transactions on Visualization and Computer
  Graphics}, vol.~15, no.~2, pp. 285--294, 2008.

\bibitem{yang2014mesh}
Y.~Yang and I.~Ivrissimtzis, ``Mesh discriminative features for 3d
  steganalysis,'' \emph{ACM Transactions on Multimedia Computing,
  Communications, and Applications}, vol.~10, no.~3, p.~27, 2014.

\bibitem{li20163d}
Z.~Li and A.~G. Bors, ``3d mesh steganalysis using local shape features,'' in
  \emph{IEEE International Conference on Acoustics, Speech and Signal
  Processing}.\hskip 1em plus 0.5em minus 0.4em\relax IEEE, 2016, pp.
  2144--2148.

\bibitem{kim2017improved}
D.~Kim, H.-U. Jang, H.-Y. Choi, J.~Son, I.-J. Yu, and H.-K. Lee, ``Improved 3d
  mesh steganalysis using homogeneous kernel map,'' in \emph{International
  Conference on Information Science and Applications}.\hskip 1em plus 0.5em
  minus 0.4em\relax Springer, 2017, pp. 358--365.

\bibitem{li20183d}
Z.~Li, D.~Gong, F.~Liu, and A.~G. Bors, ``3d steganalysis using the extended
  local feature set,'' in \emph{IEEE International Conference on Image
  Processing}.\hskip 1em plus 0.5em minus 0.4em\relax IEEE, 2018, pp.
  1683--1687.

\bibitem{zhou2019feature}
H.~Zhou, K.~Chen, W.~Zhang, C.~Qin, and N.~Yu, ``Feature-preserving tensor
  voting model for mesh steganalysis,'' \emph{IEEE Transactions on
  Visualization and Computer Graphics}, 2019.

\bibitem{li2020steganalysis}
Z.~Li and A.~G. Bors, ``Steganalysis of meshes based on 3d wavelet
  multiresolution analysis,'' \emph{Information Sciences}, 2020.

\bibitem{li2016selection}
------, ``Selection of robust features for the cover source mismatch problem in
  3d steganalysis,'' in \emph{IEEE International Conference on Pattern
  Recognition}.\hskip 1em plus 0.5em minus 0.4em\relax IEEE, 2016, pp.
  4256--4261.

\bibitem{li2018selection}
------, ``Selection of robust and relevant features for 3-d steganalysis,''
  \emph{IEEE Transactions on Cybernetics}, 2018.

\bibitem{fridrich2002practical}
J.~Fridrich and M.~Goljan, ``Practical steganalysis of digital images: State of
  the art,'' in \emph{Security and Watermarking of Multimedia Contents IV},
  vol. 4675.\hskip 1em plus 0.5em minus 0.4em\relax International Society for
  Optics and Photonics, 2002, pp. 1--13.

\bibitem{hearst1998support}
M.~A. Hearst, S.~T. Dumais, E.~Osuna, J.~Platt, and B.~Scholkopf, ``Support
  vector machines,'' \emph{IEEE Intelligent Systems and their applications},
  vol.~13, no.~4, pp. 18--28, 1998.

\bibitem{kodovsky2012ensemble}
J.~Kodovsk{\`y}, J.~Fridrich, and V.~Holub, ``Ensemble classifiers for
  steganalysis of digital media,'' \emph{IEEE Transactions on Information
  Forensics and Security}, vol.~7, no.~2, pp. 432--444, 2012.

\bibitem{li20183dlevel}
Z.~Li, F.~Liu, and A.~G. Bors, ``3d steganalysis using laplacian smoothing at
  various levels,'' in \emph{International Conference on Cloud Computing and
  Security}.\hskip 1em plus 0.5em minus 0.4em\relax Springer, 2018, pp.
  223--232.

\bibitem{fridrich2002steganalysis}
J.~Fridrich, M.~Goljan, and D.~Hogea, ``Steganalysis of jpeg images: Breaking
  the f5 algorithm,'' in \emph{International Workshop on Information
  Hiding}.\hskip 1em plus 0.5em minus 0.4em\relax Springer, 2002, pp. 310--323.

\bibitem{kodovsky2009calibration}
J.~Kodovsk{\`y} and J.~Fridrich, ``Calibration revisited,'' in \emph{ACM
  Workshop on Multimedia and Security}.\hskip 1em plus 0.5em minus 0.4em\relax
  ACM, 2009, pp. 63--74.

\bibitem{taubin1995signal}
G.~Taubin, ``A signal processing approach to fair surface design,'' in
  \emph{International Conference on Computer Graphics and Interactive
  Techniques}.\hskip 1em plus 0.5em minus 0.4em\relax ACM, 1995, pp. 351--358.

\bibitem{bollobas2013modern}
B.~Bollob{\'a}s, \emph{Modern graph theory}.\hskip 1em plus 0.5em minus
  0.4em\relax Springer Science \& Business Media, 2013, vol. 184.

\bibitem{rugis2006scale}
J.~Rugis and R.~Klette, ``A scale invariant surface curvature estimator,'' in
  \emph{Pacific-Rim Symposium on Image and Video Technology}.\hskip 1em plus
  0.5em minus 0.4em\relax Springer, 2006, pp. 138--147.

\bibitem{sun2002triangle}
Y.~Sun, D.~L. Page, J.~K. Paik, A.~Koschan, and M.~A. Abidi, ``Triangle
  mesh-based edge detection and its application to surface segmentation and
  adaptive surface smoothing,'' in \emph{IEEE International Conference on Image
  Processing}, vol.~3.\hskip 1em plus 0.5em minus 0.4em\relax IEEE, 2002, pp.
  825--828.

\bibitem{lounsbery1997multiresolution}
M.~Lounsbery, T.~D. DeRose, and J.~Warren, ``Multiresolution analysis for
  surfaces of arbitrary topological type,'' \emph{ACM Transactions on
  Graphics}, vol.~16, no.~1, pp. 34--73, 1997.

\bibitem{dyn1990butterfly}
N.~Dyn, D.~Levine, and J.~A. Gregory, ``A butterfly subdivision scheme for
  surface interpolation with tension control,'' \emph{ACM transactions on
  Graphics}, vol.~9, no.~2, pp. 160--169, 1990.

\bibitem{chen2009benchmark}
X.~Chen, A.~Golovinskiy, and T.~Funkhouser, ``A benchmark for 3d mesh
  segmentation,'' in \emph{ACM Transactions on Graphics}, vol.~28, no.~3.\hskip
  1em plus 0.5em minus 0.4em\relax ACM, 2009, p.~73.

\bibitem{li2015strategy}
B.~Li, M.~Wang, X.~Li, S.~Tan, and J.~Huang, ``A strategy of clustering
  modification directions in spatial image steganography,'' \emph{IEEE
  Transactions on Information Forensics and Security}, vol.~10, no.~9, pp.
  1905--1917, 2015.

\bibitem{denemark2015improving}
T.~Denemark and J.~Fridrich, ``Improving steganographic security by
  synchronizing the selection channel,'' in \emph{Proceedings of the ACM
  Workshop on Information Hiding and Multimedia Security}, 2015, pp. 5--14.

\bibitem{zhang2016decomposing}
W.~Zhang, Z.~Zhang, L.~Zhang, H.~Li, and N.~Yu, ``Decomposing joint distortion
  for adaptive steganography,'' \emph{IEEE Transactions on Circuits and Systems
  for Video Technology}, vol.~27, no.~10, pp. 2274--2280, 2016.

\bibitem{DBLP:conf/ih/ZhangZCLLY18}
Y.~Zhang, W.~Zhang, K.~Chen, J.~Liu, Y.~Liu, and N.~Yu, ``Adversarial examples
  against deep neural network based steganalysis,'' in \emph{Proceedings of the
  6th {ACM} Workshop on Information Hiding and Multimedia Security},
  R.~B{\"{o}}hme, C.~Pasquini, G.~Boato, and P.~Sch{\"{o}}ttle, Eds.\hskip 1em
  plus 0.5em minus 0.4em\relax {ACM}, 2018, pp. 67--72.

\bibitem{DBLP:journals/tifs/TangLTBH19}
W.~Tang, B.~Li, S.~Tan, M.~Barni, and J.~Huang, ``Cnn-based adversarial
  embedding for image steganography,'' \emph{IEEE Transactions on Information
  Forensics and Security}, vol.~14, no.~8, pp. 2074--2087, 2019.

\bibitem{DBLP:conf/ih/BernardPBK19}
S.~Bernard, T.~Pevn{\'{y}}, P.~Bas, and J.~Klein, ``Exploiting adversarial
  embeddings for better steganography,'' in \emph{Proceedings of the {ACM}
  Workshop on Information Hiding and Multimedia Security}, R.~Cogranne,
  L.~Verdoliva, S.~Lyu, J.~R. Troncoso{-}Pastoriza, and X.~Zhang, Eds.\hskip
  1em plus 0.5em minus 0.4em\relax {ACM}, 2019, pp. 216--221.

\bibitem{ker2006batch}
A.~D. Ker, ``Batch steganography and pooled steganalysis,'' in
  \emph{International Workshop on Information Hiding}.\hskip 1em plus 0.5em
  minus 0.4em\relax Springer, 2006, pp. 265--281.

\bibitem{ker2007capacity}
------, ``A capacity result for batch steganography,'' \emph{IEEE Signal
  Processing Letters}, vol.~14, no.~8, pp. 525--528, 2007.

\bibitem{koch2020dna}
J.~Koch, S.~Gantenbein, K.~Masania, W.~J. Stark, Y.~Erlich, and R.~N. Grass,
  ``A dna-of-things storage architecture to create materials with embedded
  memory,'' \emph{Nature Biotechnology}, vol.~38, no.~1, pp. 39--43, 2020.

\bibitem{bajaj2003anisotropic}
C.~L. Bajaj and G.~Xu, ``Anisotropic diffusion of surfaces and functions on
  surfaces,'' \emph{ACM Transactions on Graphics}, vol.~22, no.~1, pp. 4--32,
  2003.

\bibitem{fleishman2003bilateral}
S.~Fleishman, I.~Drori, and D.~Cohen-Or, ``Bilateral mesh denoising,'' in
  \emph{ACM SIGGRAPH}, 2003, pp. 950--953.

\bibitem{schneider2001geometric}
R.~Schneider and L.~Kobbelt, ``Geometric fairing of irregular meshes for
  free-form surface design,'' \emph{Computer Aided Geometric Design}, vol.~18,
  no.~4, pp. 359--379, 2001.

\bibitem{eigensatz2008curvature}
M.~Eigensatz, R.~W. Sumner, and M.~Pauly, ``Curvature-domain shape
  processing,'' in \emph{Computer Graphics Forum}, vol.~27, no.~2.\hskip 1em
  plus 0.5em minus 0.4em\relax Wiley Online Library, 2008, pp. 241--250.

\bibitem{zhao2019normalnet}
W.~Zhao, X.~Liu, Y.~Zhao, X.~Fan, and D.~Zhao, ``Normalnet: Learning based
  guided normal filtering for mesh denoising,'' \emph{arXiv preprint
  arXiv:1903.04015}, 2019.

\bibitem{krizhevsky2012imagenet}
A.~Krizhevsky, I.~Sutskever, and G.~E. Hinton, ``Imagenet classification with
  deep convolutional neural networks,'' in \emph{Advances in Neural Information
  Processing Systems}, 2012, pp. 1097--1105.

\bibitem{hamilton2017inductive}
W.~Hamilton, Z.~Ying, and J.~Leskovec, ``Inductive representation learning on
  large graphs,'' in \emph{Advances in neural information processing systems},
  2017, pp. 1024--1034.

\bibitem{feng2019meshnet}
Y.~Feng, Y.~Feng, H.~You, X.~Zhao, and Y.~Gao, ``Meshnet: mesh neural network
  for 3d shape representation,'' in \emph{Proceedings of the AAAI Conference on
  Artificial Intelligence}, vol.~33, 2019, pp. 8279--8286.

\bibitem{DBLP:journals/spl/XuWS16}
G.~Xu, H.~Wu, and Y.~Shi, ``Structural design of convolutional neural networks
  for steganalysis,'' \emph{{IEEE} Signal Processing Letters}, vol.~23, no.~5,
  pp. 708--712, 2016.

\bibitem{DBLP:journals/tifs/YeNY17}
J.~Ye, J.~Ni, and Y.~Yi, ``Deep learning hierarchical representations for image
  steganalysis,'' \emph{IEEE Transactions on Information Forensics and
  Security}, vol.~12, no.~11, pp. 2545--2557, 2017.

\bibitem{DBLP:journals/tifs/BoroumandCF19}
M.~Boroumand, M.~Chen, and J.~J. Fridrich, ``Deep residual network for
  steganalysis of digital images,'' \emph{IEEE Transactions on Information
  Forensics and Security}, vol.~14, no.~5, pp. 1181--1193, 2019.

\end{thebibliography}

\begin{IEEEbiography}[{\includegraphics[width=1in,height=1.25in,clip,keepaspectratio]{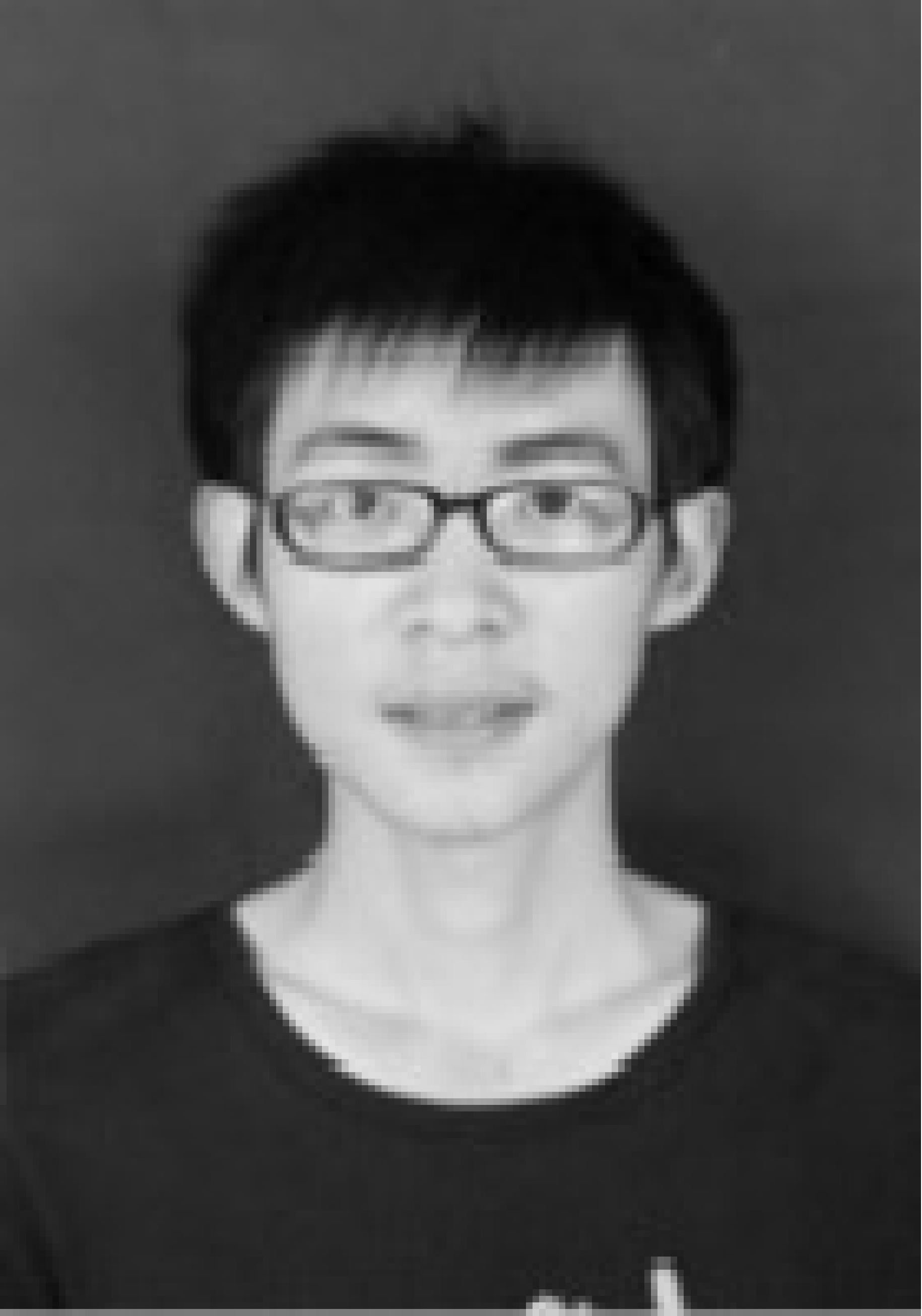}}]{Hang Zhou}
received his B.S. degree in 2015 from Shanghai University (SHU) and a Ph.D. degree in 2020 from the University of Science and Technology of China (USTC). Currently, he is a postdoctoral researcher at Simon Fraser University. His research interests include computer graphics, multimedia security and deep learning.
\end{IEEEbiography}

\begin{IEEEbiography}[{\includegraphics[width=1in,height=1.25in,clip,keepaspectratio]{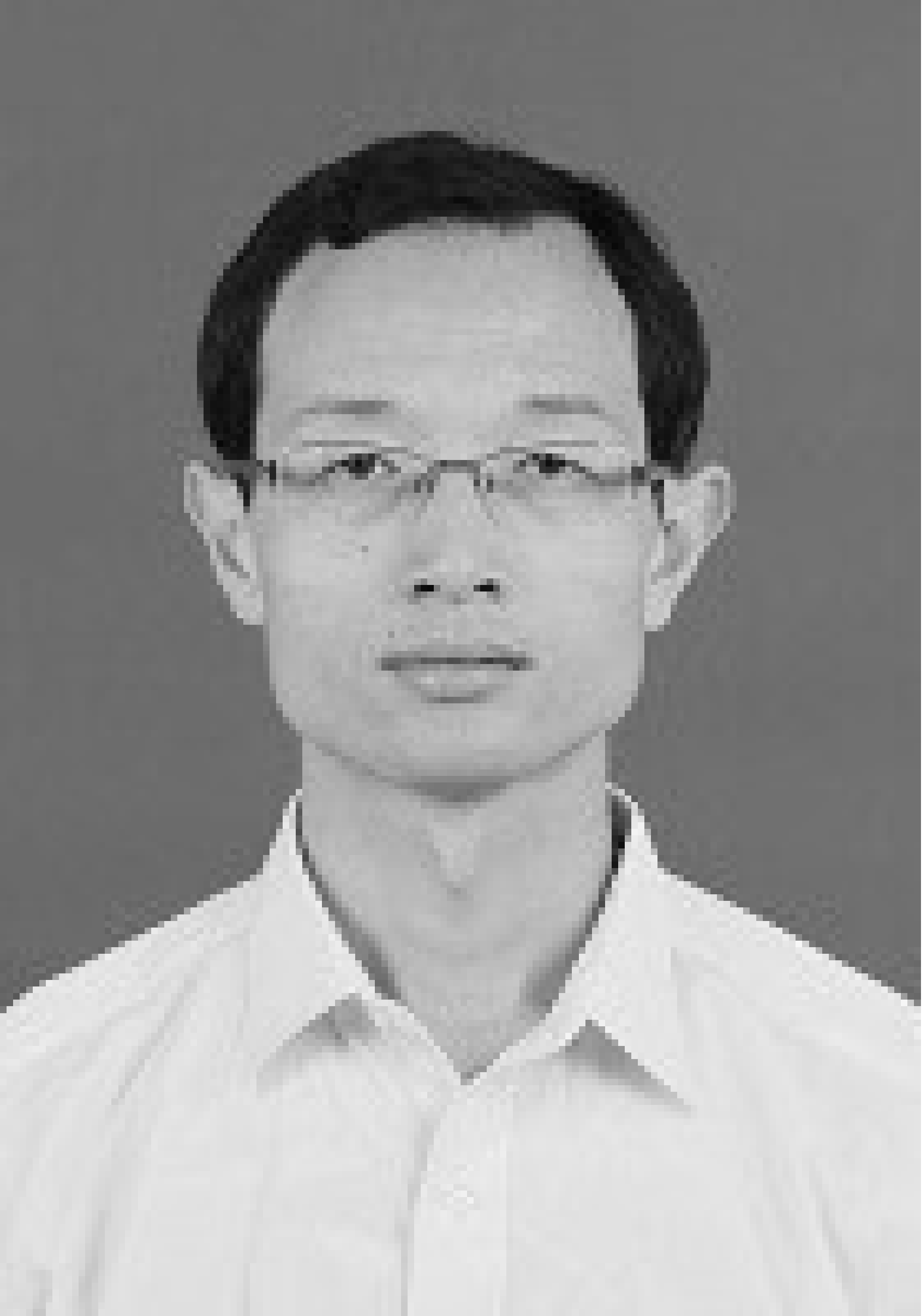}}]{Weiming Zhang}
received his M.S. degree and Ph.D. degree in 2002 and 2005, respectively, from the Zhengzhou Information Science and Technology Institute, P.R. China. Currently, he is a professor with the School of Information Science and Technology, University of Science and Technology of China. His research interests include information hiding and multimedia security.
\end{IEEEbiography}

\begin{IEEEbiography}[{\includegraphics[width=1in,height=1.25in,clip,keepaspectratio]{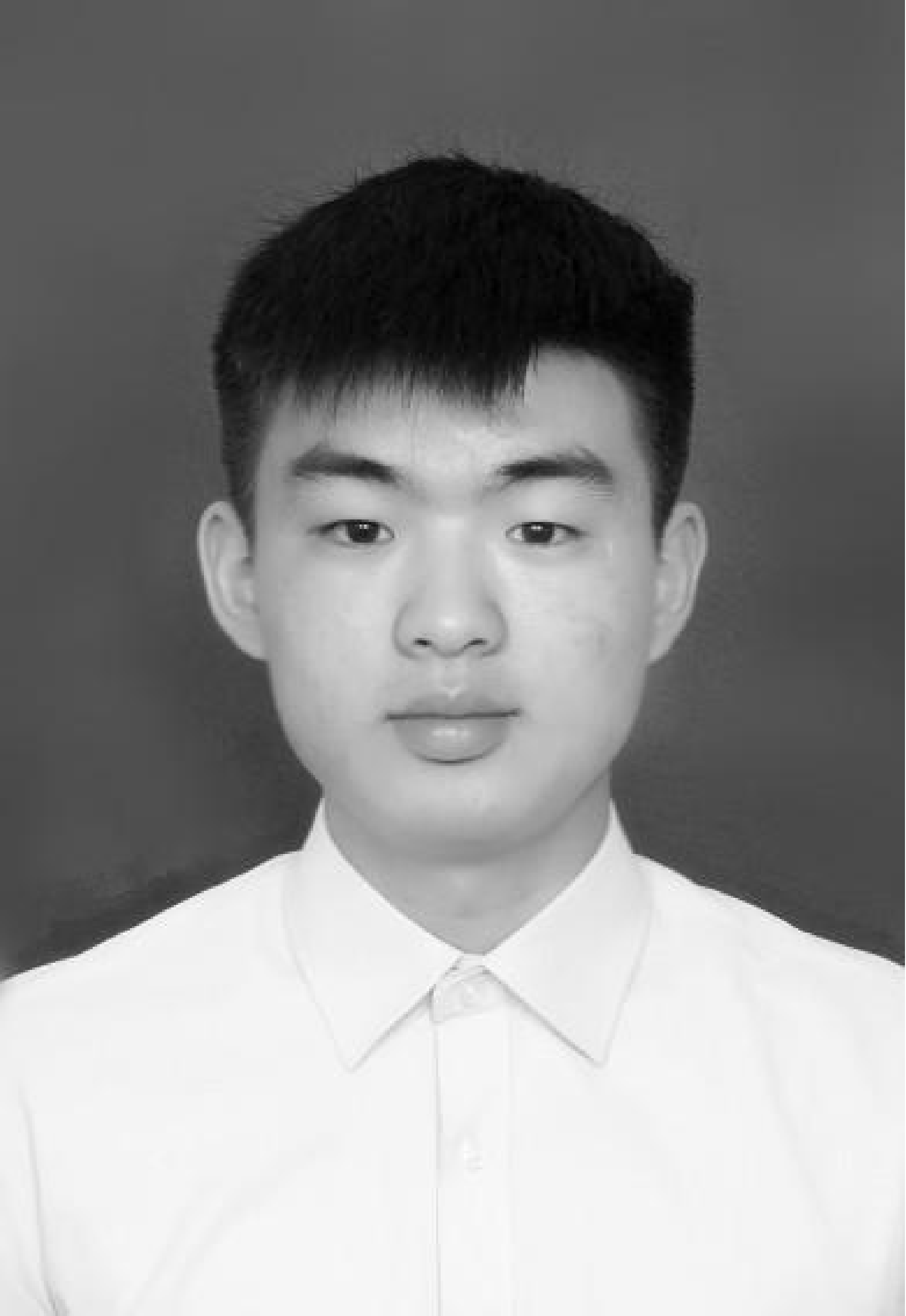}}]{Kejiang Chen}
received his B.S. degree in 2015 from Shanghai University (SHU) and a Ph.D. degree in 2020 from the University of Science and Technology of China (USTC). Currently, he is a postdoctoral researcher at the University of Science and Technology of China. His research interests include information hiding, image processing and deep learning.
\end{IEEEbiography}

\begin{IEEEbiography}[{\includegraphics[width=0.90in,height=1.30in,clip,keepaspectratio]{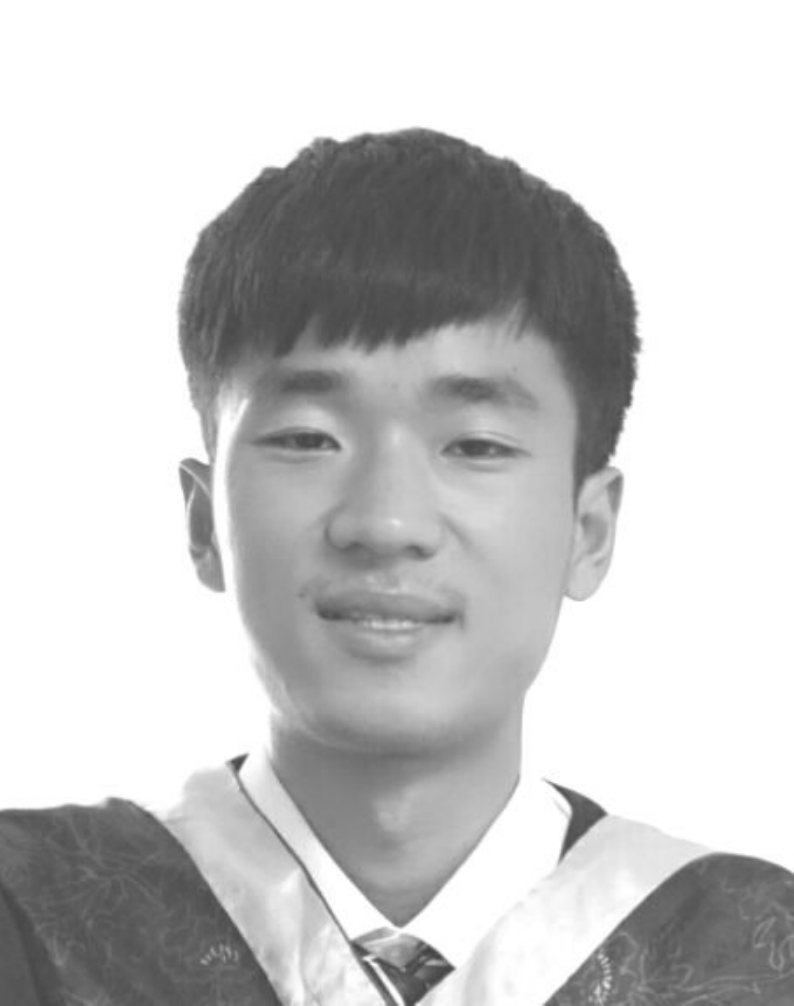}}]{Weixiang Li}
received a B.S. degree from Xidian University (XDU) in 2016. He is currently pursuing a Ph.D. degree from the University of Science and Technology of China (USTC). His research interests include image processing, steganography, and steganalysis. He received the Best Student Paper Award at the 6th ACM IH\&MMSec in 2018.
\end{IEEEbiography}

\begin{IEEEbiography}[{\includegraphics[width=1in,height=1.25in,clip,keepaspectratio]{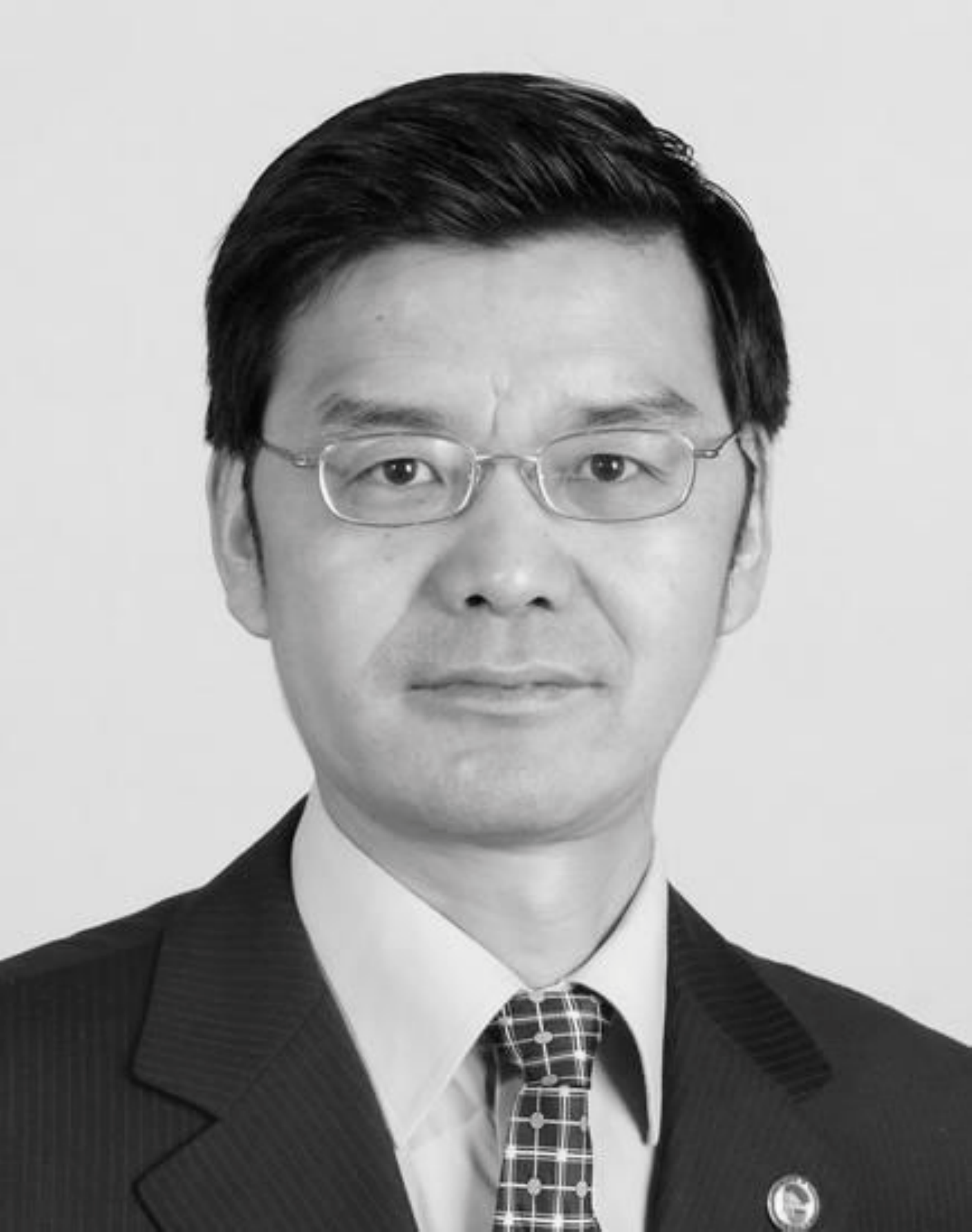}}]{Nenghai Yu}
received his B.S. degree in 1987 from Nanjing University of Posts and Telecommunications, an M.E. degree in 1992 from Tsinghua University and a Ph.D. degree in 2004 from the University of Science and Technology of China, where he is currently a professor. His research interests include multimedia security, multimedia information retrieval, video processing and information hiding.
\end{IEEEbiography}

\end{document}